\documentclass{aa}

\usepackage{graphicx}
\usepackage{txfonts}

\usepackage{natbib}
\bibpunct{(}{)}{;}{a}{}{,} 
\usepackage{color}
\usepackage{colortab}
\usepackage{pstricks}
\usepackage{enumerate}
\usepackage[normalem]{ulem}

\usepackage{amsfonts,amssymb}

\usepackage{longtable}

\begin{document}

\title{Multi-resonance orbital model of high-frequency quasi-periodic oscillations: possible high-precision determination of black hole and neutron star spin}

\author{Zden\v{e}k Stuchl\'{\i}k, Andrea Kotrlov\'a \and Gabriel T\"{o}r\"{o}k}
\institute{Institute of Physics, Faculty of Philosophy and Science, Silesian
  University in Opava, Bezru\v{c}ovo n\'{a}m. 13, CZ-74601 Opava,
  \\
  Czech Republic\\
  \email{andrea.kotrlova@fpf.slu.cz}
  }

\date{Received / Accepted}

\titlerunning{On the possibility of determination of black hole and neutron star spin}
\authorrunning{Z. Stuchl\'{\i}k et al.}

\abstract
{Using known frequencies of the~twin-peak high-frequency quasiperiodic oscillations (HF~QPOs) and known mass of the~central black hole, the~black-hole dimensionless spin $a$ can be determined by assuming a~concrete version of the~resonance model. However, a~wide range of observationally limited values of the~black hole mass implies low precision of the~spin estimates.}
{We discuss the~possibility of higher precision of the~black hole spin $a$ measurements in the~framework of a~multi-resonance model inspired by observations of more than two HF~QPOs in the~black hole systems, which are expected to occur at two (or more) different radii of the~accretion disc. This framework is also applied in a~modified form to the~neutron star systems.}
{We determine the~spin and mass dependence of the~twin-peak frequencies with a~general rational ratio $n:m$, assuming a~non-linear resonance of oscillations with the~epicyclic and Keplerian frequencies or their combinations. In the~multi-resonant model, the~twin-peak resonances are combined properly to give the~observed frequency set. For the~black hole systems we focus on the~special case of duplex frequencies, when the~top, bottom, or mixed frequency is common at two different radii where the~resonances occur giving triple frequency sets.}
{The~sets of triple frequency ratios and the~related spin $a$ are given. The~resonances are considered up to $n=5$ since excitation of higher order resonances is improbable. The~strong resonance model for ``magic'' values of the~black hole spin means that two (or more) versions of resonance could occur at the~same radius, allowing cooperative effects between the~resonances. For neutron star systems we introduce a~resonant switch model that assumes switching of oscillatory modes at resonant points.}
{In the~case of doubled twin-peak HF~QPOs excited at two different radii with common top, bottom, or mixed frequency, the~black hole spin $a$ is given by the~triple frequency ratio set. The~spin is determined precisely, but not uniquely, because the~same frequency set could correspond to more than one concrete spin $a$. The~black hole mass is given by the~magnitude of the~observed frequencies. The~resonant switch model puts relevant limits on the~mass and spin of neutron stars, and we expect a~strong increase in the~fitting procedure precision when different twin oscillatory modes are applied to data in the~vicinity of different resonant points. We expect the~multi-resonance model to be applicable to data from the~planned LOFT or similar X-ray satellite observatory.}



\maketitle

\section{Introduction}\label{1-uvod}

In the~black hole systems observed in both Galactic and extragalactic sources, strong gravity effects play a~crucial role in three phenomena related to the~accretion disc that is the~emitting source: the~spectral continuum, spectral profiled lines, and oscillations of the~disc. Moreover, it is clear that strong gravity should also play an~important role in the~binary systems containing neutron (quark) stars. Quasiperiodic oscillations (\emph{QPOs}) of X-ray brightness had been observed at {low}-(Hz) and {high}-(kHz) frequencies in many Galactic low mass X-ray binaries (LMXBs) containing neutron~stars \citep[see, e.g.,][]{Kli:2000:ARASTRA:,Kli:2006:CompStelX-Ray:,Bar-Oli-Mil:2005:MONNR:,Bel-Men-Hom:2005:ASTRA:,Bel-Men-Hom:2007:MONNR:BriNSQPOCor,Wan-etal:2013} or black holes \citep[see, e.g.,][]{McCli-Rem:2004:CompactX-Sources:,Rem:2005:ASTRN:,Rem-McCli:2006:ARASTRA:}. Some of the~HF~QPOs are in the~kHz range and often come in pairs of the~upper and lower frequencies ($\nu_{\mathrm{U}}$, $\nu_{\mathrm{L}}$) of {\it twin peaks}\footnote{Also called \emph{double peaks}.} in the~Fourier power spectra. Since the~peaks of high frequencies are close to the~orbital frequency of the~marginally stable circular orbit, representing the~inner edge of Keplerian discs orbiting black holes (or neutron~stars), the~strong gravity effects must be relevant in explaining HF~QPOs \citep{Ter-Abr-Klu:2005:ASTRA:QPOresmodel}.

Before the~twin-peak HF~QPOs were discovered in microquasars \citep[first by][]{Str:2001:ASTRJ2L:} and the~3\,:\,2 ratio pointed out, \citet{Klu-Abr:2000:ASTROPH:} suggested that these QPOs should have rational ratios, because of the~resonances in oscillations of nearly Keplerian accretion discs; see also \citet{Ali-Gal:1981:GENRG2:}. It seems that the~resonance hypothesis is now well supported by observations and that, in
particular, the~3\,:\,2 ratio ($2\nu_{\mathrm{U}} = 3\nu_{\mathrm{L}}$) is seen most often in twin-peak QPOs in the~LMXBs containing black holes (microquasars). In addition, there is even some evidence of the~same 3\,:\,2 ratio in the~X-ray spectra of the~Galaxy centre's black hole in Sgr\,A$^*$ \citep{Abr-etal:2004:ASTRJ2L:,Asch:2004:ASTRA:,Ter:2005:ASTRA:},  the~galactic nuclei MCG-6-30-15 and NGC 4051 \citep[]{Lac-Cze-Abr:2006:astro-ph0607594:}, and other active galactic nuclei \citep{Mid-Utt-Don:2011:,Mid-Don-etal:2009:}.

There is a~crucial difference between the~HF~QPOs in black hole (BH) and neutron star (NS) systems. In BH systems, the~peaks of HF~QPOs are detected at (nearly) constant frequencies that are characteristic of a~given source across long times and various observations. When more frequencies are detected (simultaneously or in various observations), they come in ratios of small numbers, usually in the~3\,:\,2 ratio \citep[][]{Rem-McCli:2006:ARASTRA:}. In NS systems, the~HF~QPOs often appear as twin QPOs consisting from two simultaneously observed peaks with distinct actual frequencies that substantially change over time (in one observational sequence); moreover, sometimes only one of the~frequencies is observed and has evolved \citep[][]{Bel-etal:2007:MONNR:RossiXTE,Bou-Bar-Lin-Tor:2010:MNRAS:}. Most of the~twin QPOs in the~so-called atoll sources \citep[][]{Kli:2006:CompStelX-Ray:} have been detected at lower frequencies, $600-800\,\mathrm{Hz}$ vs. upper frequencies $900-1200\,\mathrm{Hz}$, demonstrating a~clustering of the~twin HF~QPOs frequency ratio around 3\,:\,2 and thus indicating some analogy to the~BH case \citep[][]{Abr-etal:2003:ASTRA:,Bel-etal:2007:MONNR:RossiXTE,Tor-etal:2008:ACTA:DistrKhZ4U1636-53,Tor-Bak-Stu-Cec:2008:ACTA:TwPk4U1636-53,Tor-etal:2008:ACTA:Clustering4U1636-53,Bou-Bar-Lin-Tor:2010:MNRAS:}. The~amplitudes of twin HF~QPOs in NS systems are usually much stronger, and their coherence times are much longer than those of BH sources \citep[see, e.g.,][]{Rem-McCli:2006:ARASTRA:,Bar-Oli-Mil:2005:MONNR:,Bar-Oli-Mil:2006:MONNR:QPO-NS}. It is probable that a~3\,:\,2 resonance also plays a~significant role in the~LMXBs containing neutron stars. However, the~evidence is much more complicated, since the~frequency ratio is concentrated around 3\,:\,2, but has much wider range than in the~black hole systems \citep[][]{Bel-Men-Hom:2005:ASTRA:,Ter-Abr-Klu:2005:ASTRA:QPOresmodel,Abr-etal:2005:RAGtime6and7:CrossRef,Tor-etal:2008:ACTA:Clustering4U1636-53,Tor:2009:ASTRA:ReversQPOs,Bou-Bar-Lin-Tor:2010:MNRAS:}. It remains controversial whether the~peak in distribution of the~twin-peak frequency ratios $\nu_{\mathrm{U}}/\nu_{\mathrm{L}}$ at 3\,:\,2 is physical \citep{Mon-Zan:2012:MNRAS:}.

It has also been suggested that the~multi-peaked distribution in the~frequency ratios meant that more than one resonance could be realized if a~resonant mechanism is involved in generating the~neutron star \citep{Bel-Men-Hom:2005:ASTRA:,Stu-Kot-Tor:2011:ASTRA:ResRadKep} and black hole \citep{Stu-Tor:2005:RAGtime6and7:CrossRef} HF~QPOs. For a~given neutron star source, the~upper and lower HF~QPO frequencies can be traced over the~whole observed range, but the~probability of detecting both QPOs simultaneously increases when the~frequency ratio is close to the~ratio of small natural numbers, namely 3\,:\,2, 4\,:\,3, 5\,:\,4, in the~case of the~atoll sources \citep[][]{Tor:2009:ASTRA:ReversQPOs,Bou-Bar-Lin-Tor:2010:MNRAS:} or the~ratios that are close to 2\,:\,1, and 3\,:\,1 in the~Z-source Circinus~X-1 \citep[][]{Bou-etal:2006:ASTRJ2:,Tor-etal:2010:ASTRJ2:MassConstraints}.
The~analysis of root-mean-squared-amplitude evolution in a~group of six atoll sources (4U~1636$-$53, 4U~1608$-$52, 4U~0614$-$09, 4U~1728$-$34, 4U~1820$-$30, 4U~1735$-$44) shows that the~upper and lower HF~QPO amplitudes equal each other and alter their dominance, while passing rational frequency ratios corresponding to the~datapoint clustering \citep{Tor:2009:ASTRA:ReversQPOs}. Such an~``energy switch effect'' can be explained well in the~framework of non-linear resonant orbital models as shown in \citet{Hor-etal:2009:ASTRA:IntResQPOs}.

In the~orbital resonance models of HF~QPOs, the~resonance conditions applied to the~frequency ratio of oscillations entering the~resonances imply specific radii of accretion discs where the~resonances can occur. For any specific version of the~orbital resonance models, the~resonant radii at thin, Keplerian accretion discs \citep[][]{Ter-Abr-Klu:2005:ASTRA:QPOresmodel} are given by the~frequency ratio and the~parameters of the~central black hole or neutron star (mass $M$, spin $a$, and quadrupole moment $q$ in the~case of neutron stars).\footnote{For toroidal discs \citep{Rez-Ahm-Mil:1991:MONNR:,Rez-etal:2003:MNRAS:,Bla-etal:2006:ASTRJ2:,Str-Sra:2009:CLAQG:EpiOscNonSleKerrBH,Mon-Zan:2012:MNRAS:} or slightly charged discs around magnetic neutron stars \citep{Bak-etal:2012:CQG:,Pac-etal:2011:,Bak-etal:2010:CLAQG:MagIndNonGeo,Kov-Stu-Kar:2008:CLAQG:OffEqOrb}, additional parameters must enter the~resonance conditions.} The~frequencies of disc oscillation modes can be expressed in terms of the~geodetical orbital and epicyclic frequencies of the~test particle motion for both Keplerian discs and slender tori when gravitation is the~relevant restoring force \citep{Abr-etal:2003:PUBASJ:,Sra-Tor-Abr:2006:ASTRA:}. For non-slender tori, corrections to the~epicyclic frequencies have to be introduced, depending on the~thickness of the~tori, with differences to the~epicyclic frequencies approaching tens of percent \citep{Bla-etal:2006:ASTRJ2:,Sra:2005:ASTRN:,Str-Sra:2009:CLAQG:EpiOscNonSleKerrBH}.

When the~pressure gradients in thick tori become important as the~restoring force, the~radial oscillation frequency strongly differs from the~radial geodetical epicyclic frequency \citep{Rez-etal:2003:MNRAS:,Rez-etal:2003:MONNR:,Mon-Rez-Yos:2004:MONNR:}, similar to the~model of oscillating string loops where the~string tension due to magnetic field is the~main restoring force \citep{Stu-Kol:2012:JCAP:,Stu-Kol:2012:PHYSR4:,Jac-Sot:2009:PHYSR4:,Sem-Dya-Pun:2004:Science:,Spr:1981:AA:}.

We focus our attention mainly on the~black hole systems. According to the~resonance hypothesis \citep{Klu-Abr:2000:ASTROPH:}, the~two modes in resonance should have eigenfrequencies $\nu_{r}$ (equal to the~radial epicyclic frequency) and $\nu_{\mathrm{v}}$ \citep[equal to the~vertical epicyclic frequency $\nu_{\theta}$, or to the~Keplerian frequency $\nu_{\mathrm{K}}$; see also][]{Abr-etal:2004:ASTRJ2L:,Ter-Abr-Klu:2005:ASTRA:QPOresmodel}. While models based on the~\emph{parametric resonance} identify the~two observed frequencies of the~twin peak ($\nu_\mathrm{U}$, $\nu_\mathrm{L}$) directly with the~eigenfrequencies of resonant oscillations, models based on the~\emph{forced resonance} allow combinational (beat) frequencies of the~modes to be
observed. Both parametric and forced resonance models make clear and precise
predictions about the~values of observed frequencies in connection with the~spin and mass of the~observed object, at least in the~case of black holes -- see \citet{Ter-Abr-Klu:2005:ASTRA:QPOresmodel} where the~black hole specific (dimensionless) internal angular momentum (spin) is predicted by a~few possible models for microquasars with observed twin-peak HF~QPOs.

Although the~observed frequencies are consistent with several (but not all) resonance models, the~most probable and natural explanation for the~presence of a~3\,:\,2 ratio is the~3\,:\,2 parametric (or \emph{internal}) resonance
\citep{Abr-etal:2004:ASTRJ2L:,Ter-Abr-Klu:2005:ASTRA:QPOresmodel}. However, in all the~cases under consideration, a~relatively wide range of observationally allowed black hole masses implies rather unprecise determination of the~black hole spin \citep{Ter-Abr-Klu:2005:ASTRA:QPOresmodel}. Therefore, it is useful to look for some possibilities how to make the~spin estimate more precise \citep{Stu-Tor:2005:RAGtime6and7:CrossRef,Stu-Kot-Tor:2008:ACTA:BHadmStrResPhen}.

Of course, the~spin estimates based on various versions of the~resonance model have to be compared with the~spin estimates based on the~accretion disc spectral continuum fitting \citep{McCli-etal:2011:CLAQG:,Don-Dav:2008:,Don-Gie-Kub:2007:,McCli-etal:2006:astro-ph/0606076:,Mid-etal:2006:MONNR:,Sha:2005::astro-ph/0508302} and profiled spectral lines \citep{Laor:1991:ASTRJ2:,Bao-Stu:1992:ASTRJ2:,Kar-Vok-Pol:1992:ASTRJ2L:,Dov-Kar-Mar:2004:RAGtime4and5:CrossRef,Fab-Min:2005:XSpectraKerr:Book,Zak:2003:,Cad-Cal-Fan:2003:MEMSA1:XrFeProf,Fan-etal:1997:,Cad-Cal:2005:MNRAS:,Zak-Rep:2006:,Sch-Stu:2009:INTJMD:OpPheBraKerr,Sch-Stu:2009:GENRG2:ProEmRingBran,Mil-etal:2009:}. In the~case of Sgr\,A$^*$ the~orbital precession of some stars moving in close vicinity of the~central black hole could also give an~interesting restriction on the~black hole spin \citep[]{Kra:2005:,Kra:2007:}.

In some sources, more than two high-frequency peaks are observed. The~microquasar GRS~1915$+$105 reveals high-frequency QPOs appearing at four frequencies with the~lower and upper pairs in the~ratio close to 3\,:\,2 or 5\,:\,3 \citep{Rem-McCli:2006:ARASTRA:}, and even a~fifth frequency was reported \citep{Bel-Men-San:2001:ASTRA}. An~additional sixth frequency was mentioned \citep{Str:2001:ASTRJ2L:}, although not confirmed. In Sgr\,A$^*$, three frequencies were reported with ratios close to 3\,:\,2\,:\,1 \citep[][]{Asch:2004:ASTRA:,Asc-etal:2004:ASTRA:,Ter:2005:ASTRA:}. In the~source NGC~5408~X-1, oscillations with three frequencies of ratios close to 6\,:\,4\,:\,3 were observed \citep{Str-etal:2007:ASTRJ2:QuaPerVar}. In the~galactic nuclei MCG-6-30-15 and NGC 4051, two pairs of QPOs were reported with the~ratios close to 3\,:\,2 and 2\,:\,1, respectively \citep[]{Lac-Cze-Abr:2006:astro-ph0607594:}.

In the~microquasar GRS~1915$+$105, a~near-extreme Kerr black hole with $a \sim 1$ is expected according to the~spectral continuum fitting \citep[]{McCli-etal:2006:astro-ph/0606076:}, and all five (six) frequencies of QPOs can be explained in the~framework of the~extended resonance model with the~hump-induced oscillations \citep{Stu-Sla-Tor:2007:ASTRA:}. In the~extended resonance model, forced resonances of the~epicyclic oscillations are assumed with the~oscillations induced by the~humpy orbital velocity profile \citep[related to the~locally non-rotating frames,][]{Bar-Pre-Teu:1972:ASTRJ2:} that occurs in Keplerian discs orbiting Kerr black holes with $a > 0.9953$ \citep{Asc:2006:China:,Asch:2004:ASTRA:,Stu-etal:2004:PHYSR4:,Stu-Sla-Ter:2006:ASTRA:Humpy,Stu-Sla-Tor:2007:ASTRA:,Stu-Bla-Sla:2011:}. In the~extended ``humpy'' resonance model, all the~oscillations in resonance are related to the~exclusively defined ``humpy radius'' with extremal orbital velocity gradient within the~humpy profile \citep{Stu-Sla-Ter:2006:ASTRA:Humpy}. However, this model can only be relevant for near-extreme Kerr black holes with spins of $a > 0.9953$. Some concrete implications of a~rapid spin in GRS~1915$+$105 for other frequently quoted QPO models are discussed in \citet{Tor-Kot-Sra-Stu:2011:}.

When more than two resonant frequencies or, precisely, more than one resonant point corresponding to two resonant eigenfrequencies of the~system are observed, a~multi-resonant model of HF~QPOs is necessary in order to understand the~behaviour of a~concrete system generating such HF~QPOs. There are different types of the~multi-resonant model based on oscillations with the~orbital frequencies $\nu_{\mathrm{K}}$, $\nu_{\theta}$, and $\nu_{r}$, all of which allow higher precision in determining the~black hole parameters in comparison with situations where only one resonant point is observed.

First, for special (``magic'') values of the~black hole spin, the~resonance points share the~same radius in the~disc that enables cooperative phenomena between resonances of different origin. Of special interest is the~case of so-called strong resonant phenomena allowing direct resonances of oscillations having frequency ratio $\nu_{\mathrm{K}}\!:\!\nu_{\theta}\!:\!\nu_{r} = s\!:\!t\!:\!u$ (with $s,t,u$ being small natural numbers) at the~radius given by the~frequency ratio; for each triple frequency set ratio, the~black hole spin $a$ is given uniquely \citep[][]{Stu-Kot-Tor:2008:ACTA:BHadmStrResPhen}.

Second, two resonance points occur at two different radii. Generally, a~set of four frequencies is  observed. However, a~common frequency appears for special values of the~black hole spin, which is related to the~types of resonances involved and the~frequency ratio. Then the~ratio of the~three observed frequencies determines the~black hole spin uniquely, if the~resonance conditions at the~resonant points are given. Here, we concentrate our attention on these situations.

Third, one fixed type of resonance is enhanced at different radii, with different frequency ratios. This type is probably relevant in neutron star atoll and Z systems, where it is shown that the~observed datapoint clusters concentrated around frequency ratios 3\,:\,2, 4\,:\,3, 5\,:\,4 could be approximated by a~variety of models, e.g., the~relativistic precession model \citep{Ste-Vie:1999:PHYRL:,Ste-Vie:1998:ASTRJ2L:} or the~so-called total precession model, where the~Keplerian oscillations of frequency $\nu_{\mathrm{K}}$ are in resonance with the~total precession oscillations of frequency $\nu_{\mathrm{T}}=\nu_{\theta}-\nu_{r}$ \citep[][]{Stu-Tor-Bak:2007:arXiv:}. However, we observe a~bad quality of data fitting by the~HF~QPO models in neutron star LMXBs \citep{Tor-etal:2010:ASTRJ2:MassConstraints,Tor-etal:2012:ApJ:,Bou-Bar-Lin-Tor:2010:MNRAS:}. 
Therefore, we introduce a~new, resonant switch model that could be applied to some neutron star LMXBs exploring observational data spanning two resonant points. Such a~model could improve data fitting without using a~non-geodesic corrections to the~orbital and epicyclic frequencies.

In general, for two resonant points observed, we can expect the~oscillations to be excited at two different radii of the~accretion disc and to enter the~resonance in the~framework of different versions of the~resonance model. In such situations we are making the~black hole spin estimate within two properly chosen versions of the~resonance model, thus in principle obtaining a~more precise determination of the~spin in comparison with situations where only one resonant point is observed. In special cases where a~common upper, lower, or mixed frequency is observed in the~two frequency pairs (i.e., only three different frequencies are observed), the~triple frequency set is precisely given independently of the~black hole mass for appropriately fixed values of the~black hole spin \citep{Stu-Tor:2005:RAGtime6and7:CrossRef}. Then, the~spin is in principle determined precisely (within the~precision of the~frequency measurements), but not uniquely, because the~same frequency set could occur for different values of the~spin within different versions of the~resonance model. It is clear that in such situations the~black hole spin estimates coming from the~spectral continuum fitting and the~line profile models have to be relevant for determining the~proper versions of the~resonant model.\footnote{Very promising seems to be modelling of the~energy dependencies of high-frequency QPO, i.e., the~QPO spectra generated at the~QPO radius \citep{Zyc-Nie-Sob:2007:MNRAS}, as they could be credited by the~multi-resonant models giving exact values of the~black hole spin and the~QPO radii.} The~recently discussed method based on the~frequency shift interval of radiation emitted from matter excited at the~resonant radii can also be very efficient in restricting the~spin of the~black hole or some other compact object \citep{Mur-etal:2009:ApJ:,Stu-Sch:2010:CLAQG:AppKepDiOrKerrSSp}. When the~black hole spin is found, the~black hole mass can be determined from the~magnitude of the~observed frequencies. We expect the~multi-resonance model of HF~QPOs to be applicable to both black hole and neutron star systems for precise data obtained by the~planned LOFT \citep{Fer-etal:2012:ExA:} or a~similar X-ray satellite observatory.

\section{Digest of orbital resonance model}\label{2-modely}

The~standard orbital resonance model \citep{Abr-Klu:2001:ASTRA:,Ter-Abr-Klu:2005:ASTRA:QPOresmodel,Ali-Gal:1981:GENRG2:}
assumes non-linear resonance between oscillation modes of an~accretion disc orbiting a~central object, here considered to be a~rotating Kerr black hole (or a~neutron star with exterior described by the~Kerr geometry -- see \citet{Tor-etal:2010:ASTRJ2:MassConstraints} for details).\footnote{We consider here only the~Kerr spacetime as the~standard description for rotating black holes, although alternatives have been discussed in the~same context \citep[see][]{Kot-Stu-Ter:2008:CLAQG:,Stu-Kot:2009:GENRG2:OrResDiBraKBH,Ali-Tal:2009:PHYSR4:RotBraBH,Rah-Abd-Ahm:2011:,Ali-etal:2012:arXiv:,Hor-Ger:2012:}.} The~accretion disc can be a~thin disc with Keplerian angular velocity profile \citep{Nov-Tho:1973:BlaHol:} or a~thick toroidal disc with angular velocity profile given by the~distribution of the~specific angular momentum of the~fluid \citep{Koz-Jar-Abr:1978:ASTRA:,Abr-Jar-Sik:1978:ASTRA:,Stu-Sla-Hle:2000:ASTRA:}. The~frequency of the~oscillations is related to the~Keplerian frequency (orbital frequency of tori) or to the~radial and vertical epicyclic frequencies of the~circular test particle motion. The~epicyclic frequencies can be relevant both for the~thin, Keplerian discs with quasicircular geodetical motion \citep{Kat-Fuk-Min:1998:BHAccDis:,Now-Leh:1998:TheoryBlackHoleAccretionDisks:,Abr-etal:2003:PUBASJ:,Klu-etal:2007:REVHA:,Kat:2001:PUBASJ:b,Kat:2004:PUBASJ:QPOsmodel,Wag:1999:Discoseismology} and for slender toroidal discs \citep{Sch-Rez:2006:ASTRJ2:QPOsRelTor,Rez:2004:qpoblackhole,Rez:2004:RAGtime4and5:Proceedings,Rez-etal:2003:MNRAS:}. However, with the~thickness of an~oscillating toroid growing, the~eigenfrequencies of its radial and vertical oscillations deviate from the~epicyclic test particle frequencies \citep{Sra:2005:ASTRN:,Bla-etal:2006:ASTRJ2:,Str-Sra:2009:CLAQG:EpiOscNonSleKerrBH}.
Here we focus our attention on the~Keplerian thin discs.

A~variety of different versions of the~orbital resonance model exists that could be classified according to the~following criteria:
\begin{enumerate}[a)]
    \item the~type of the~resonance (parametric or forced),
    \item the~presence of beat, combinational frequencies,
    \item the~type of oscillations entering the~resonance.
\end{enumerate}
Thus, according to the~basical criterion a), two main groups of orbital resonance model versions exist, differing by the~type of the~resonance. In both of them, the~epicyclic frequencies of the~equatorial test particle circular motion play a~crucial role \citep{Ter-Abr-Klu:2005:ASTRA:QPOresmodel}.

\subsection{Parametric internal resonance}

The~internal resonance model is based on the~idea of \emph{parametric resonance} between vertical and radial epicyclic oscillations with the~frequencies $\nu_\theta=\omega_\theta/2\pi$ and $\nu_{r}=\omega_{r}/2\pi$. The~parametric resonance is described by the~Mathieu equation \citep{Lan-Lif:1976:Mech:}
\begin{equation}
\label{Mathieu} \delta \ddot \theta + \omega_{\theta}^2\,[ 1 + h
\cos (\omega_{r} t) ]\, \delta \theta = 0.
\end{equation}
Theory behind the~Mathieu equation implies that a~parametric resonance is excited when
\begin{equation}
\label{Equation6} {\frac{\omega_{r}} {\omega_{\theta}}} =
{\frac{\nu_{r}}  {\nu_{\theta}}} = {\frac{2}  {n}}, ~~~~n
=1, \,2, \,3,\,\dots
\end{equation}
and is strongest for the~lowest possible value of $n$ \citep{Lan-Lif:1976:Mech:}. Because there is $\nu_{r} < \nu_{\theta}$ near black holes, the~lowest possible value for the~parametric resonance in the~so-called \emph{epicyclic resonance model} is $n = 3$, which means that $2 \nu_{\theta} = 3 \nu_{r}$. This explains observed 3\,:\,2 ratio, assuming $\nu_{\mathrm{U}} = \nu_{\theta}$ and $\nu_{\mathrm{L}} = \nu_{r}$. The~internal resonance corresponds to a~system with conserved energy, as shown, e.g., in \citet{Hor-etal:2009:ASTRA:IntResQPOs}.

\subsection{Forced resonance}

Models based on the~\emph{forced resonance} come from the~idea of a~forced non-linear oscillator, when the~relation of the~latitudinal (vertical) and radial oscillations is given by the~formulae
\begin{equation}
\delta \ddot \theta + \omega_{\theta}^2\delta \theta +
\left[{\mathrm{non~linear~terms~in}}~\delta \theta \right] =
g (r)\cos (\omega_0\,t),
\end{equation}
\begin{equation}
\delta \ddot r + \omega_{{r}}^2\delta r +
\left[{\mathrm{non~linear~terms~in}}~\delta \theta, \delta r \right] =
h (r)\cos (\omega_0\,t),
\end{equation}
with
\begin{equation}
\omega_{\theta} = \left(\frac{k}{l}\right)\, \omega_{r},
\end{equation}
where $k,l$ are small natural numbers and $\omega_0$ is the~frequency of the~external force.\footnote{For example, the~gravitational perturbation forces are discussed in \citet{Stu-Hle:2005:RAGtime6and7:CrossRef} and \citet{Stu-Kon-Mil-Hle:2008:ASTRA:GravExc} for the~case of  a~neutron star with ``mountains'' and of a~binary partner of the~neutron star or a~black hole.} The~non-linear terms allow combinational (beat) frequencies in resonant solutions for $\delta \theta (t)$ and $\delta r(t)$ \citep[see, e.g.,][]{Lan-Lif:1976:Mech:}, which in the~simplest case give
\begin{equation}
\omega_- = \omega_{\theta} - \omega_{r}, ~~~\omega_+ =\omega_{\theta} + \omega_{r}.
\end{equation}
Such resonances can produce the~observable frequencies in the~3\,:\,2 ratio, as well as in other rational ratios. (One of the~cases that give 3\,:\,2 observed ratio is also the~``direct'' case of $k\!:\!l=3\!:\!2$ corresponding to the~same frequencies and radius as in the~case of 3\,:\,2 parametric resonance.) Therefore, we consider direct and simple combinational resonances.

The \emph{``Keplerian'' resonance} model assumes parametric or forced resonances between oscillations with radial epicyclic frequency $\nu_{r}$ and Keplerian orbital frequency $\nu_\mathrm{K}$. Of course, there are many additional possibilities for composing the~resonance conditions. We systematically treat the~possibilities in the~following.

For the~parametric and direct forced resonances, the~resonance conditions related to the~frequency ratio of oscillations are common, however, physical details, such as the~time evolution of the~resonance, the~dependence of the~resonance strength and the~resonant frequency width on the~order of the~resonance, are different \citep[see][]{Lan-Lif:1976:Mech:,Nay-Moo:1979:NonOscilations:,Stu-Kot-Tor:2008:ACTA:BHadmStrResPhen}. Here we concentrate on the~frequency-ratio resonant conditions, leaving the~physical details aside for
investigation in the~concrete situations expected to be discovered in future thanks to development of the~observational techniques \citep[e.g., the~proposed LOFT observatory,][]{Fer-etal:2012:ExA:}.

\section{Determination of the~black hole spin from the~resonance model}\label{3-spiny}

It is well known that the~formulae for the~vertical epicyclic frequency $\nu_{\theta}$ and the~radial epicyclic frequency $\nu_{r}$ of the~geodetical motion take in the~gravitational field of a~rotating Kerr black hole (with the~mass $M$ and spin $a$) the~form
\begin{equation}
\label{frequencies}
\nu_{\theta}^2 = \alpha_\theta\,\nu_\mathrm{K}^2\,,
\qquad
\nu_{r}^2 = \alpha_{r}\,\nu_\mathrm{K}^2
\end{equation}
\citep[e.g.,][]{Ali-Gal:1981:GENRG2:,Kat-Fuk-Min:1998:BHAccDis:,Ste-Vie:1998:ASTRJ2L:,Ter-Stu:2005:ASTRA:}
where the~Keplerian frequency $\nu_\mathrm{K}$ and related epicyclic frequencies are given by the~formulae
\begin{eqnarray}
\nu_{\mathrm{K}}&=&\frac{1}{2\pi}\left(\frac{\mathrm{G}M}{r_\mathrm{G}^{~3}}\right)^{1/2}\frac{1}{x^{3/2} + a} =
\frac{1}{2\pi}\left(\frac{\mathrm{c}^3}{\mathrm{G}M}\right)
\frac{1}{x^{3/2}+a}\,,\\
\alpha_\theta&=& 1-\frac{4\,a}{x^{3/2}}+\frac{3\,a^2}{x^{2}}\,,\\
\alpha_{r}&=&1-\frac{6}{x}+\frac{8\,a}{x^{3/2}}-\frac{3\,a^2}{x^{2}}\,.
\end{eqnarray}
Here $x = r/(\mathrm{G}M/\mathrm{c}^2)$ is the~dimensionless radius, expressed in terms of the~gravitational radius of the~black hole, and $a$ is the~dimensionless spin.

It is known that the~Keplerian frequency $\nu_{\mathrm{K}}(x,a)$ is a~monotonically decreasing function of the~radial coordinate for any value of the~black hole spin. On the~other hand, the~radial epicyclic frequency has the~global maximum for any Kerr black hole. However, the~vertical epicyclic frequency is also not monotonic if the~spin is sufficiently high \citep[see, e.g.,][]{Kat-Fuk-Min:1998:BHAccDis:,Per-etal:1997:ASTRJ2:}.\footnote{In the~model of p-mode (axisymmetric) oscillations of thick tori with the~fundamental frequency and its first overtone, both the~frequencies decrease monotonically with increasing dimension of the~oscillating torus \citep{Rez-etal:2003:MNRAS:,Rez-etal:2003:MONNR:}. In the~model of oscillating string loops, the~radial profiles of the~radial and vertical frequencies cross each other near the~black hole horizon \citep{Stu-Kol:2012:JCAP:}. In both these models the~resonant phenomena could be irrelevant, as for the~model of shock waves \citep{Mon-Chak:2006:MNRAS:,Mon-etal:2009:MNRAS:,Mon-Choi:2013:NewA:}.}

For the~Kerr black-hole spacetimes, the~locations $\mathcal{R}_{r}\,(a),~\mathcal{R}_\theta\,(a)$ of maxima of the~epicyclic frequencies $\nu_{r},~\nu_\theta$ are implicitly given by the~conditions \citep{Ter-Stu:2005:ASTRA:}
\begin{eqnarray}
\label{implicitcondition}
\beta_\mathrm{j}(x,a)&=&\frac{1}{2}\frac{\sqrt{x}}{x^{3/2}+a}\,\alpha_\mathrm{j}(x,a)\,,\quad \mathrm{where}\ ~\mathrm{j}\in\{{r},\theta\}\,,\\
\beta_{{r}}(x,a)&\equiv&\frac{1}{x^{2}}-\frac{2\,a}{x^{5/2}}+ \frac {a^2}
{x^3}\,,\\
\beta_{\theta}(x,a)&\equiv&\frac{a}{x^{5/2}}-\frac{a^2}{x^3}\,.
\end{eqnarray}

For any black hole spin, the~extrema of the~radial epicyclic frequency $\mathcal{R}_{r}\,(a)$ must be located above the~marginally stable orbit. On the~other hand, the~latitudinal extrema $\mathcal{R}_\theta\,(a)$ are located above the~photon (marginally bound or marginally stable) circular orbit only if the~limits on the~black hole spin $a>0.748$ (0.852, 0.952) are satisfied \citep{Ter-Stu:2005:ASTRA:}. In the~Keplerian discs, with the~inner boundary $x_{\mathrm{in}} \sim x_{\mathrm{ms}}$, the~limiting value $a=0.952$ is relevant.

For a~particular resonance with frequency ratio $n:m$, the~equation
\begin{equation}
\label{ratios} n \nu_{r} = m \nu_{\mathrm{v}};
\qquad\nu_\mathrm{v}\in\{\nu_\theta,\, \nu_\mathrm{K}\}
\end{equation}
determines the~dimensionless resonance radius $x_{n:m}$ as a~function of the~spin $a$ in the~case of direct resonances. This can be easily extended to the~resonances with combinational frequencies, as
discussed in detail later.

The known mass of the~central black hole, the~observed twin-peak frequencies ($\nu_{\mathrm{U}}$, $\nu_{\mathrm{L}}$), the~equations (\ref{frequencies})\,--\,(\ref{ratios}) and a~concrete type of resonance, assumed to be direct or to have the~combinational frequencies, enable to determine the~black hole spin. This procedure was first applied to the~microquasar GRO~1655--40 by \citet{Abr-Klu:2001:ASTRA:}, to the~other three microquasars \citep{Ter-Abr-Klu:2005:ASTRA:QPOresmodel}, and also to the~Galaxy centre black hole Sgr\,A$^*$ \citep{Ter:2005:ASTRA:}.

More complex resonant phenomena in HF~QPOs have to be expected in the~field of Kerr superspinars (naked singularities), as shown in \citet{Ter-Stu:2005:ASTRA:} and \citet{Stu-Sch:2012:CQG:ObsPhenKerrSSp:}. In the~superspinar (or naked singularity) backgrounds, the~optical effects also demonstrate considerable differences as compared with those generated in the~field of black holes \citep{Stu-Sch:2010:CLAQG:AppKepDiOrKerrSSp,Stu-Sch:2012:CQG:CountRotKerrSSp:,Stu-Sch:2012:CQG:ObsPhenKerrSSp:,Vir-Ell:2002:,Vir-Kee:2008:PHYSR4:TimDelGrLens,Tak-Har:2010:CLAQG:ObsTestKerr}.

\section{Multiple resonances and resonance conditions}\label{4-multiplres}

The~very probable interpretation of twin-peak frequencies observed in microquasars is the~3\,:\,2 parametric resonance of epicyclic oscillations; however, generally it is possible that more than one resonance could be excited in the~disc simultaneously (or at different times) under different internal conditions. Indeed, observations of the~HF~QPOs in the~microquasar GRS~1915$+$105 \citep{Rem:2005:ASTRN:}, in extragalactic sources NGC 4051, MCG-6-30-15 \citep[]{Lac-Cze-Abr:2006:astro-ph0607594:}, NGC~5408~X-1 \citep[]{Str-etal:2007:ASTRJ2:QuaPerVar}, and the~Galaxy centre Sgr\,A$^*$ \citep[]{Asc-etal:2004:ASTRA:} show a~variety of QPOs with frequency ratios differing from (or additional to) the~3\,:\,2 ratio.

The~resonances could be parametric or forced and have different versions according to the~epicyclic (Keplerian) frequencies entering the~resonance directly, or in some combinational form. In principle, for any case of the~orbital resonance model version, one can determine both the~spin and mass of the~black hole just from the~eventually observed set of frequencies. However, the~obvious difficulty would be to identify the~right combination of resonances and its relation to the~observed frequency set. Within the~range of black hole mass allowed by observations, each set of twin-peak frequencies puts a~limit on the~black hole spin. The~resonance model versions are consistent with observations if the~allowed spin ranges overlap. Two or more twin peaks (resonant oscillations) then generally make the~spin measurement more precise.

Here we consider a~special case of two different resonances, determined by a~doubled ratio of natural numbers $n$\,:\,$m$ and $n'$\,:\,$m'$. The~resonances occur at the~corresponding radii of the~disc $x_{n:m}$, $x_{n':m'}$ that can be determined from the~observed set of frequencies using the~relevant orbital resonance models. Thus, the~generic relations $n$\,:\,$m$\,, $n'$\,:\,$m'$ put restrictions on the~spin of the~central black hole. We consider general frequency ratios of small integers, with the~order of the~resonances $n+m \leq 9$ ($n'+m' \leq 9$); only for $n \leq 5$ are the~resonant phenomena realistic \citep[see][for details]{Lan-Lif:1976:Mech:,Nay-Moo:1979:NonOscilations:}. In some sources (e.g., NGC~5408~X-1), the~ratio $4$\,:\,$3$ occurs for the~strongest QPOs \citep[]{Str-etal:2007:ASTRJ2:QuaPerVar}, supporting the~necessity of considering the~general ratios at both radii. In special cases, the~resonant radii could coincide, i.e., $x_{n:m}=x_{n':m'}$. Then we can expect strongest resonances due to possible direct interactions of the~relevant
oscillatory modes \citep[see, e.g.,][]{Nay-Moo:1979:NonOscilations:,Stu-Kot-Tor:2008:ACTA:BHadmStrResPhen}.

We first discuss how the~frequency ratio of twin oscillation modes in a~resonance implies the~resonance conditions, i.e., dependence of the~radius of the~accretion disc where the~resonance occurs on the~spin parameter $a$ of the~Kerr geometry and the~frequency ratio $n$\,:\,$m$.

\subsection{Resonance conditions}

We study radial coordinates determining positions where the~rational frequency ratios occur for all possible versions of the~resonance between the~radial and vertical epicyclic and the~Keplerian oscillations ($\nu_{\mathrm{K}}>\nu_{\theta}>\nu_{r}$ for $1 \geq a \geq 0$), taking both the~direct and simple combinational resonances into account. For all possible resonances of the~epicyclic and Keplerian oscillations, the~resonance condition is given in terms of the~frequency ratio parameter
\begin{equation}
p=\left(\frac{m}{n}\right)^2.
\end{equation}
All the~resonant conditions that implicitly determine the~resonant radius $x_{n:m}(a,p)$ must be related to the~radius of the~innermost stable circular geodesic $x_{\mathrm{ms}}(a)$ giving the~inner edge of Keplerian discs. Therefore, for all the~relevant resonance radii, we require $x_{n:m}(a,p) \geq x_{\mathrm{ms}}(a)$, where $x_{\mathrm{ms}}(a)$ is implicitly given by \citep{Bar-Pre-Teu:1972:ASTRJ2:}
\begin{equation}
a=a_{\mathrm{ms}}\equiv\frac{\sqrt{x}}{3}\left(4-\sqrt{3x-2}\right).
\end{equation}

The~results are summarized in such a~way as to relate the~black hole spin $a$ and the~dimensionless resonance radius $x_{n:m}(a,p)$ determined for the~corresponding resonance model version and the~frequency ratio parameter $p$. The resonance functions are denoted as $a^{\nu_\mathrm{U}/\nu_\mathrm{L}}(x,p)$.

\subsubsection{Direct resonances}

We consider first the~direct resonances concerning the~epicyclic and Keplerian orbital frequencies giving the~resonance conditions that have to be fulfilled:

\begin{figure*}
\begin{center}
\begin{minipage}[h]{.48\hsize}
\includegraphics[width=\hsize]{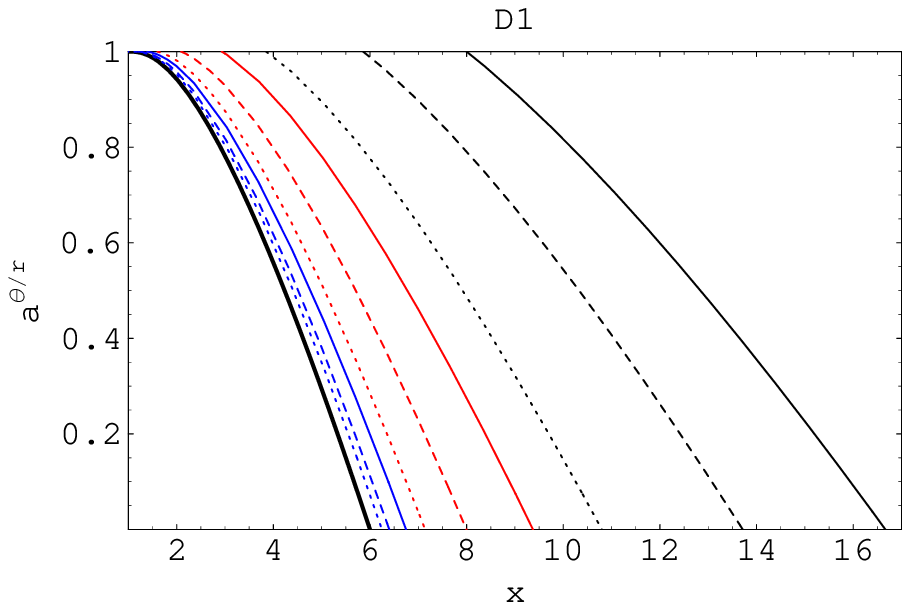}
\end{minipage}
\begin{minipage}[h]{.48\hsize}
\includegraphics[width=\hsize]{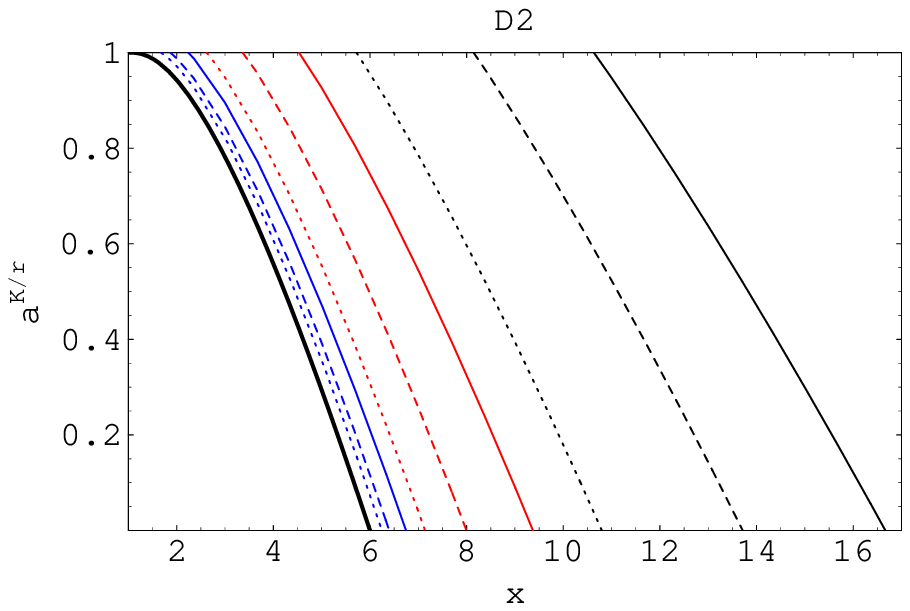}
\end{minipage}
\end{center}
\begin{center}
\includegraphics[width=.93\hsize]{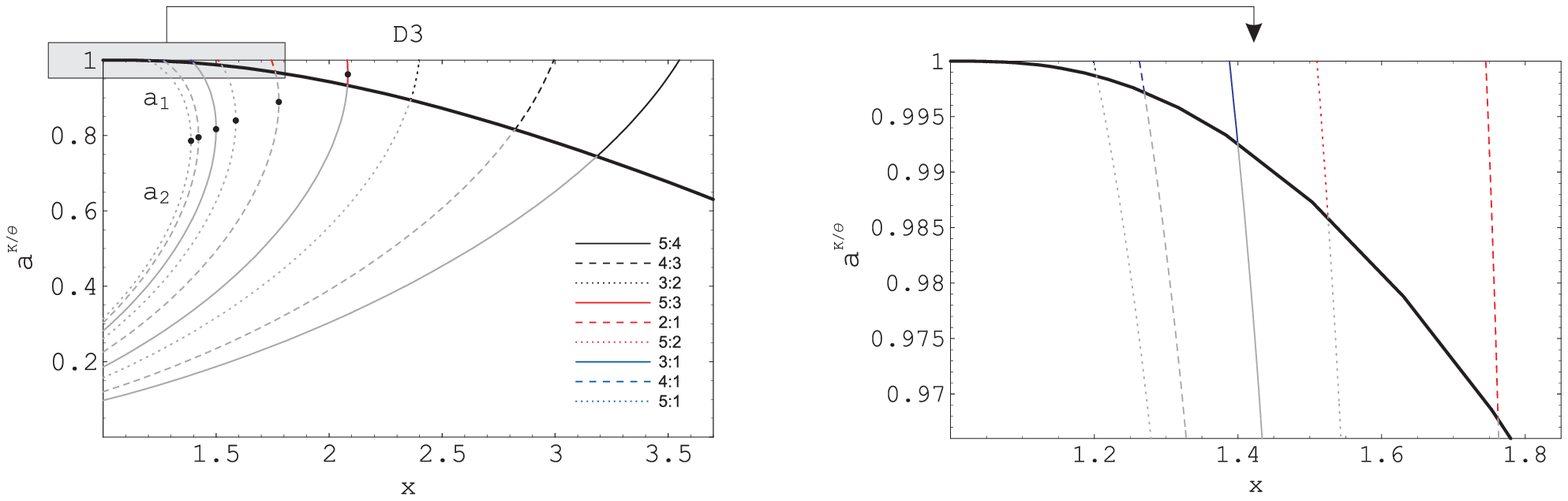}
\end{center}
\caption{The~spin resonance functions $a^{\nu_\mathrm{U}/\nu_\mathrm{L}}(x,p)$ for the~direct
resonances D1 -- D3 and the~frequency ratios
$n$\,:\,$m$\,=\,5\,:\,4 (black solid line), 4\,:\,3 (black dashed
line), 3\,:\,2 (black dotted line), 5\,:\,3 (red solid line),
2\,:\,1 (red dashed line), 5\,:\,2 (red dotted line), 3\,:\,1
(blue solid line), 4\,:\,1 (blue dashed line), 5\,:\,1 (blue
dotted line). Black thick line represents $a_{\mathrm{ms}}(x)$, which
implicitly determines the~radius of the~marginally stable orbit
$x_{\mathrm{ms}}$ giving a~natural restriction on the~validity of
the~resonance model versions involving radial epicyclic
oscillations. In the~case D3, the~black points determine turning points of the~functions $a^{\nu_\mathrm{U}/\nu_\mathrm{L}}(x,p)$ joining two separate branches of the~solution of the~resonance condition equation. Of course, only those parts of the~resonance functions with $x>x_{\mathrm{ms}}(a)$ are acceptable.\label{aDirect}}
\end{figure*}

\begin{enumerate}[{D}1.]\setlength{\itemsep}{1.5ex plus 1ex}
    \item 
          $\displaystyle \nu_\mathrm{U} = \nu_{\theta},\,\nu_\mathrm{L} = \nu_{r}\,,$
          \begin{eqnarray}\label{aD1}
          a&=&a^{\theta/{r}}(x,p)\equiv\frac{\sqrt{x}}{3(p+1)}\Bigg\{2(p+2)\nonumber\\
          && -\sqrt{(1-p)\left[3x(p+1)-2(2p+1)\right]}\Bigg\}\,,
          \end{eqnarray}
   \item 
         $\displaystyle \nu_\mathrm{U} = \nu_{\mathrm{K}},\, \nu_\mathrm{L} = \nu_{r}\,,$
          \begin{equation}\label{aD2}
           a=a^{\mathrm{K}/{r}}(x,p)\equiv\frac{\sqrt{x}}{3}\left[4-\sqrt{3x(1-p)-2}\right]\,,
          \end{equation}
   \item 
         $\displaystyle \nu_\mathrm{U} = \nu_{\mathrm{K}},\, \nu_\mathrm{L} = \nu_{\theta}\,,$
           \begin{equation}\label{aD3}
           a=a^{\mathrm{K}/\theta}(x,p)\equiv\frac{\sqrt{x}}{3}\left[2\pm\sqrt{4-3x(1-p)}\right]\,.
           \end{equation}
\end{enumerate}

The~results are illustrated in Fig.~\ref{aDirect}. In cases D1 and D2, the~resonance lines $a^{\nu_\mathrm{U}/\nu_\mathrm{L}}(x,p)$ are monotonous functions of $x$ and do not cross the~stability line $a_{\mathrm{ms}}(x)$. They are given by a~single branch of the~solutions of the~resonance conditions. In case D3, the~resonance lines cross the~stability line and in some cases are determined by two branches of the~resonance condition solutions, joined at a~turning point. The~solutions are acceptable at the~stability domain, i.e., for $a>a^{\nu_\mathrm{U}/\nu_\mathrm{L}}_{\mathrm{ms}}(p)$, where $a^{\nu_\mathrm{U}/\nu_\mathrm{L}}_{\mathrm{ms}}(p)$ denotes crossing of the~resonance line $a^{\nu_\mathrm{U}/\nu_\mathrm{L}}(x,p)$ with $a_{\mathrm{ms}}(x)$ and is called ``stability point'' of the~resonance function.

\subsubsection{Simple combinational resonances}

We study the~simple combinational resonances in Appendix~\ref{apendix-A} for combinations of frequency pairs, and in Appendix~\ref{apendix-B} for combinations of all three frequencies  $\nu_{\mathrm{K}}$, $\nu_{\theta}$, $\nu_{r}$.

\subsubsection{Double-beat combinational resonances}

Resonances of beat frequencies (four frequencies combined from $\nu_{\mathrm{K}}$, $\nu_{\theta}$, $\nu_{r}$) constitute family of 13 cases. The behavior of the~resonance functions is similar to the~case of the~simple combinational resonances. The double-beat combinational resonances are treated in Appendix~\ref{apendix-C}.

\subsection{Warped disc oscillations}

In the~framework of the~warped disc oscillations models \citep{Kat:2004:PUBASJ:QPOsmodel,Wag:1999:Discoseismology}, where the~frequencies of the~disc oscillations are given by combinations of the~(multiples of the) Keplerian and epicyclic frequencies, resonant phenomena could be relevant too. Usually, the~inertial-acoustic and g-mode oscillations and their resonances are relevant \citep{Kato:2007:PUBASJ:FreqCorr}. We give as an~example the~frequency relation $\nu_{\mathrm{L}}=\nu_{\mathrm{K}}-\nu_{r}$; $\nu_{\mathrm{U}}=2 \nu_{\mathrm{K}}-\nu_{\theta}$ which can be considered as an~acceptable model of HF~QPOs in some atoll sources \citep[e.g., 4U~1636$-$53,][]{Bar-Oli-Mil:2005:MONNR:} and typical Z-source Circinus~X-1 \citep{Bou-etal:2006:ASTRJ2:,Tor-etal:2010:ASTRJ2:MassConstraints}.

Assuming
\begin{equation}
    \frac{2
    \nu_{\mathrm{K}}-\nu_{\theta}}{\nu_{\mathrm{K}}-\nu_{r}}=\frac{n}{m}\,,
\end{equation}
we find that the~resonance points are determined by the~resonance condition
\begin{equation}
        a = a^{(2\mathrm{K}-\theta)/(\mathrm{K}-r)}(x,p)
\end{equation}
where the~resonance function $a^{(2\mathrm{K}-\theta)/(\mathrm{K}-r)}(x,p)$ is given by the~equation
\begin{equation}\label{warp}
    \left(p \alpha_{\theta}-\alpha_{{r}}\right)^2 - 2 \left(p \alpha_{\theta}+\alpha_{{r}}\right)
     \left(2\sqrt{p}-1\right)^2+\left(2\sqrt{p}-1\right)^4=0\,.
\end{equation}
Results of solutions of (\ref{warp}) are given in Fig.~\ref{aDC2-WD} in Appendix~\ref{apendix-C}.\footnote{In the~realistic models of warped disc
oscillations that could be used to explain QPOs observed in the~black hole systems, and due to precession of the~disc also in the~neutron star systems \citep{Kato:2007:PUBASJ:FreqCorr}, the~higher-harmonic oscillations are usually considered up to the~third order being thus allowed up to the~frequencies $3\nu_{\mathrm{K}}-\nu_{r}$, etc.}

%
%
%
%
%
%
%
%
%
%
%

\section{Characteristic frequency sets with a~duplex frequency}\label{5-charsets}

In some specific situations, for some specific values of the~black hole spin, two twin-peak QPOs observed at the~radii $x_{n:m}$ and $x_{n':m'}$ have the~bottom, top, or mixed (the~bottom at the~inner radius and the~top in the~outer radius, or vice versa) frequencies identical. Such situations can be characterized by sets of three frequencies (upper $\nu_{\mathrm{U}}$, middle $\nu_{\mathrm{M}}$ and lower $\nu_{\mathrm{L}}$ frequency) with ratio $\nu_{\mathrm{U}}:\nu_{\mathrm{M}}:\nu_{\mathrm{L}} = s:t:u$, given by the~$n:m$ and $n':m'$ ratios, the~relevant versions of the~resonance, and the~type of the~duplex (common) frequency.

When only direct resonances of the~epicyclic oscillations are allowed, the~first case with ``bottom identity'' can be realized when two resonances have a~common radial epicyclic frequency, while the~second case with ``top identity'' can be realized when two resonances have a~common vertical epicyclic frequency. These two possibilities only are in principle allowed by the~non-monotonicity of the~epicyclic frequencies (\ref{frequencies}) discussed in detail by \citet{Ter-Stu:2005:ASTRA:}. When the~Keplerian oscillations and the~combinational frequencies are allowed, all the~mixed, bottom, and top identities are possible.

From the~point of view of the~observational consequences, it is important to know for which frequency ratios $n\!:\!m$ the~resonant frequency $\nu_{\theta}(a,n\!:\!m)$, which is considered as a~function of the~black hole spin $a$ for a~given frequency ratio $n\!:\!m$, has a~non-monotonic character. A~detailed analysis \citep[]{Ter-Stu:2005:ASTRA:} shows that $\nu_{\theta}(a,n\!:\!m)$ has a~local maximum for $n\!:\!m > 11\!:\!5$; i.e., in physically relevant situations ($n$, $m$ small enough for the~resonance), it occurs for the~ratios $\nu_{\theta}:\nu_{r}=$ 5\,:\,2, 3\,:\,1, 4\,:\,1, 5\,:\,1. This means that while the~``bottom identity'' could happen for any black hole spin $a$, the~``top identity'' can only arise for $a\sim1$ if only the~epicyclic oscillations are considered.

\begin{figure*}
\begin{minipage}[h]{.465\hsize}
(a)
\begin{center}
\includegraphics[width=\hsize]{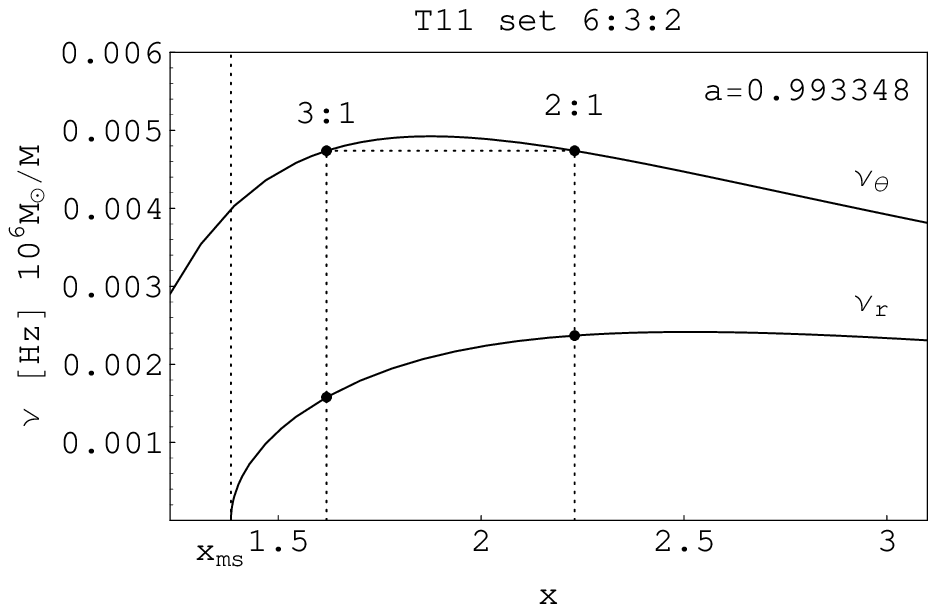}
\end{center}
\end{minipage}
\hfill
\begin{minipage}[h]{.465\hsize}
(b)
\begin{center}
\includegraphics[width=\hsize]{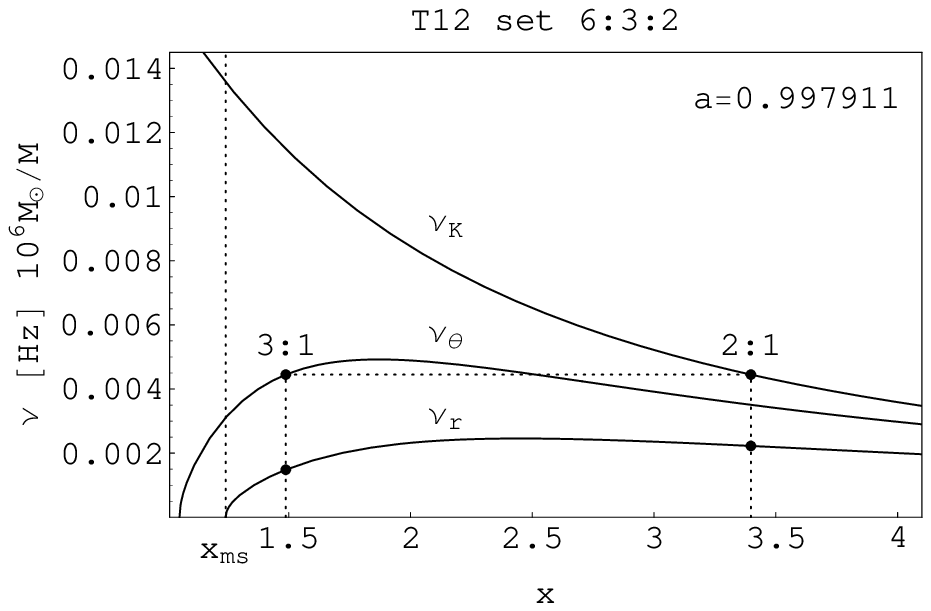}
\end{center}
\end{minipage}
\begin{minipage}[h]{.465\hsize}
(c)
\begin{center}
\includegraphics[width=\hsize]{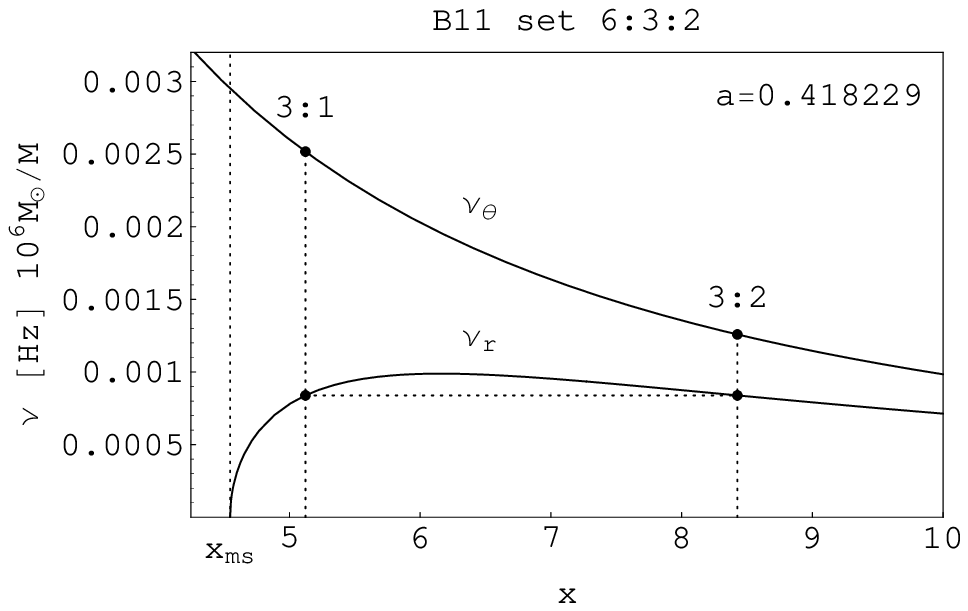}
\end{center}
\end{minipage}
\hfill
\begin{minipage}[h]{.465\hsize}
(d)
\begin{center}
\includegraphics[width=\hsize]{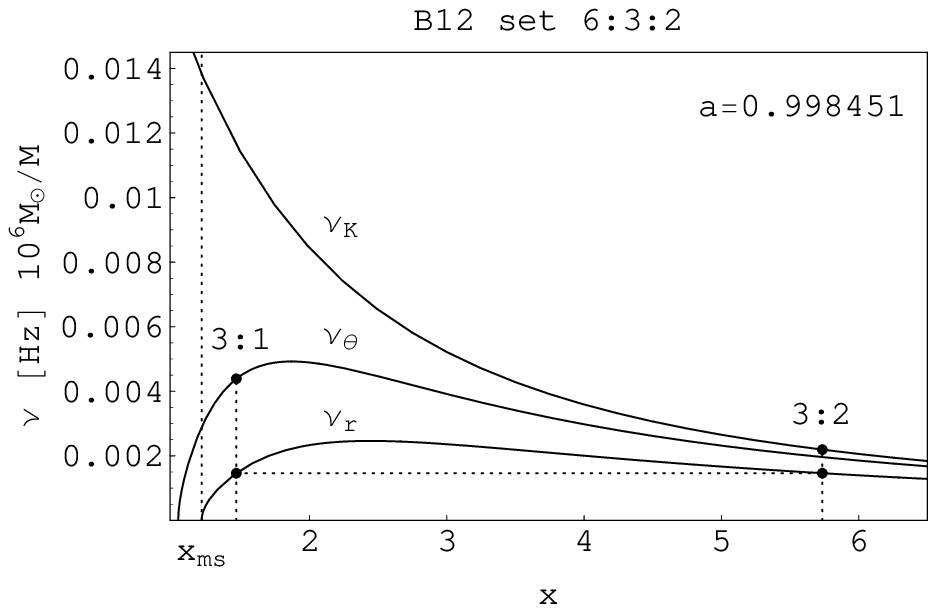}
\end{center}
\end{minipage}
\begin{minipage}[h]{.465\hsize}
(e)
\begin{center}
\includegraphics[width=\hsize]{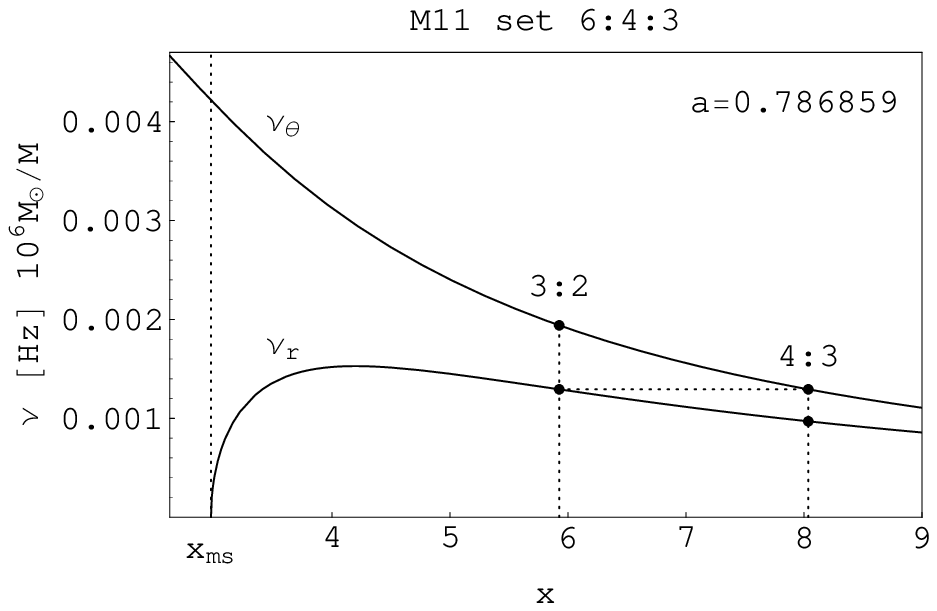}
\end{center}
\end{minipage}
\hfill
\begin{minipage}[h]{.465\hsize}
(f)
\begin{center}
\includegraphics[width=\hsize]{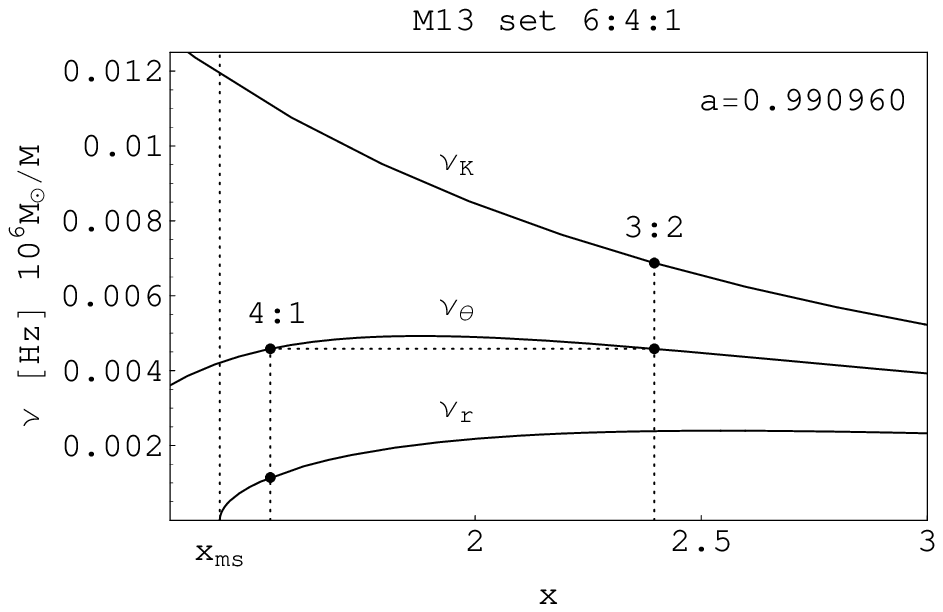}
\end{center}
\end{minipage}
\begin{minipage}[h]{.465\hsize}
(g)
\begin{center}
\includegraphics[width=\hsize]{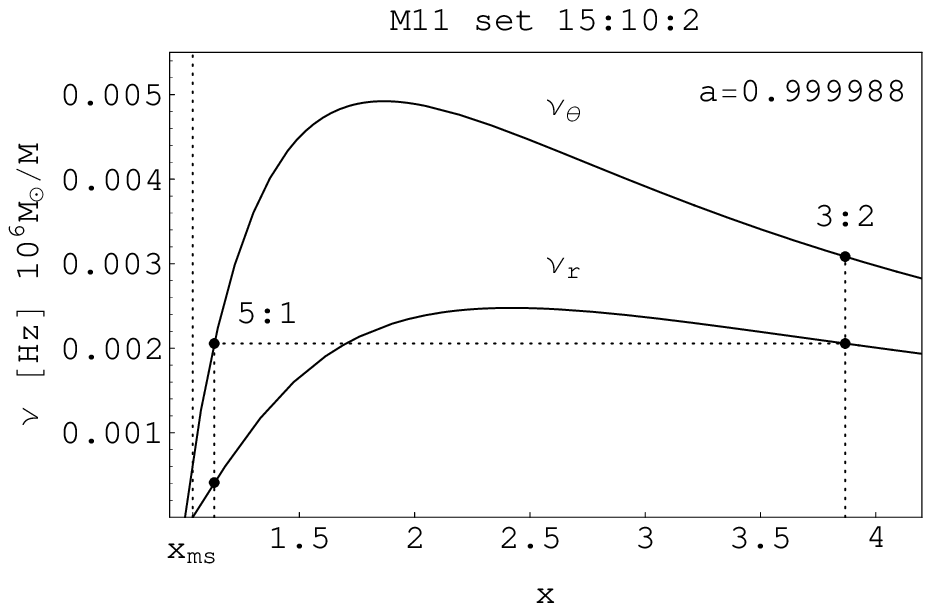}
\end{center}
\end{minipage}
\hfill
\begin{minipage}[h]{.465\hsize}
(h)
\begin{center}
\includegraphics[width=\hsize]{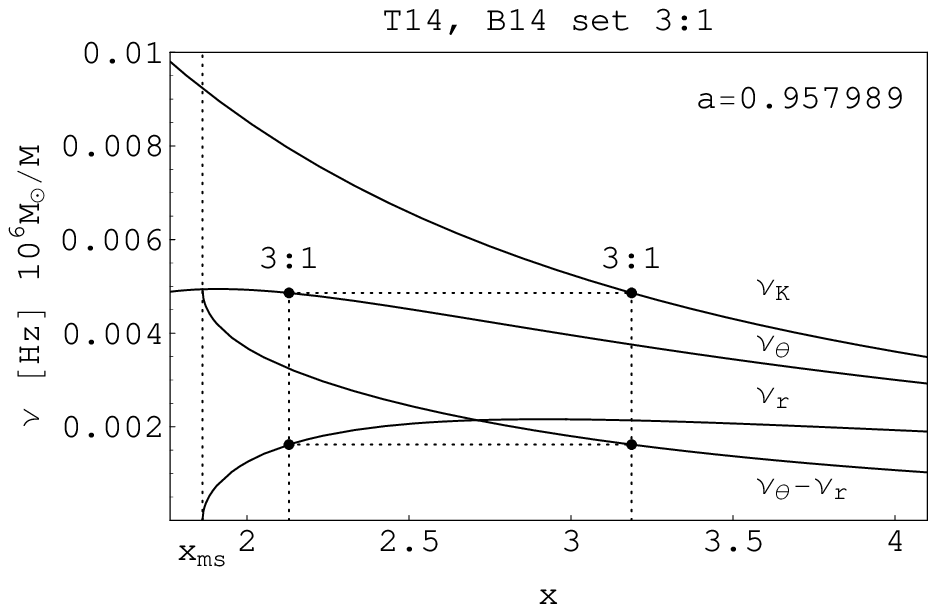}
\end{center}
\end{minipage}
\caption{The~typical cases of the~frequency triples with top
(a), (b), bottom (c), (d), and both types of mixed identities
(e), (f), (g). Also shown is the~interesting exceptional case where two oscillations of frequencies with the~same magnitude and the~same ratio are in resonance at two different
radii (h).\label{typical}}
\end{figure*}

The~typical cases of the~frequency triple sets with bottom, top, and both types of mixed identities containing duplex frequencies are illustrated in Fig.~\ref{typical}. Another interesting exceptional case occurs, e.g., for the~spin
$a=0.958$, when two oscillations of frequencies with the~same magnitude (and ratio) are in resonance at two different radii, see Fig.~\ref{typical}h.

We use the~following system of notation throughout this paper: $\mathrm{T}(X)(Y)$, $\mathrm{B}(X)(Y)$, and $\mathrm{M}(X)(Y)$ where T stands for top, B for bottom, and M for mixed identity. The~$(X)(Y)$ corresponds to the~given types of the~doubled resonances $(X)$ and $(Y)$ with identical top, bottom, or mixed frequencies (for the~case of mixed identity $(X)$ always denotes the~resonance with the~top frequency and $(Y)$ the~resonance with the~identical bottom frequency). The~concrete types of resonances $(X)(Y)$ can be easily deduced from the~notation presented in Appendix~\ref{apendix-D} (Tables~\ref{top-id-znaceni} and \ref{top-id-D-DC-znaceni}).

\subsection{Triple frequency sets and black hole spin}

We consider a~simple situation with the~``top identity'' of the~upper frequencies in two resonances between the~radial and vertical epicyclic oscillations at the~radii $x_{p}, x_{p'}$ with $p^{1/2}=m:n$, $p'^{1/2}=m':n'$. The~condition
$\nu_{\theta}(a,x_p)=\nu_{\theta}(a,x_{p'})$ is then transformed to the~relation
\begin{equation}
\frac{\alpha_{\theta}^{1/2}(a,x_{p})}{x_{p}^{3/2}+a} =
\frac{\alpha_{\theta}^{1/2}(a,x_{p'})}{x_{p'}^{3/2}+a}\,,
\end{equation}
which uniquely determines the~black hole spin $a$. When two different resonances are combined, we proceed in the~same manner. For example, the~case of ``bottom identity'' in the~resonance between the~radial and vertical epicyclic oscillations at $x_{p}$ and the~resonance between the~Keplerian oscillations with $\nu_{\mathrm{K}}$ and total precession oscillations with $\nu_{\mathrm{T}}=\nu_{\theta}-\nu_{r}$ at $x_{p'}$ implies the~condition
$\nu_{r}(a,x_p)=\left(\nu_{\theta}-\nu_{r}\right)(a,x_{p'})$ that leads to the~relation
\begin{equation}
\frac{\alpha_{{r}}^{1/2}(a,x_{p})}{x_{p}^{3/2}+a} =
\frac{\left(\alpha_{\theta}-\alpha_{{r}}\right)(a,x_{p'})}{x_{p'}^{3/2}+a}\,,
\end{equation}
which uniquely determines the~spin $a$, since in Eq. (\ref{aD1}) (and (\ref{aCT1})), the~radii $x_{p}$ and $x_{p'}$ are related to the~spin $a$ by the~resonance conditions for $a^{\theta/{r}}(x,p)$ and $a^{\mathrm{K}/(\theta-{r})}(x,p')$, respectively.

Therefore, for given types of the~doubled resonances, the~ratios $n:m$ and $n':m'$ determine the~ratio in the~triple frequency set $s:t:u$. The~black hole spin $a$ is given by the~types of the~two resonances and the~ratios $p$, $p'$, quite independently of the~black hole mass $M$.

Since the~radial and vertical epicyclic frequencies and the~Keplerian frequency have the~same dependence on the~black hole mass $M$, the~above arguments hold in the~same way for any kind of the~three frequency sets, for any of the~bottom, top, or mixed frequency identity with any two resonances containing any combination of the~frequencies $\nu_{\mathrm{K}}, \nu_{\theta}, \nu_{r}$. Therefore, the~triple frequency sets with the~``duplex'' frequencies can be used to determine the~black hole spin with very high precision, independently of the~uncertainties in determining the~black hole mass $M$: the~parameter $M$ can be addressed by the~magnitude of the~measured frequencies. However, the~relation between the~black hole spin and the~triple frequency ratios is not unique in general. For a~given frequency ratio set, several values of $a$ are allowed, and some other methods of the~spin measurement (spectral continuum fitting, profiled spectral lines) must be involved.

\begin{figure*}[t]
\begin{minipage}[h]{.48\hsize}
\begin{center}
\includegraphics[width=\hsize]{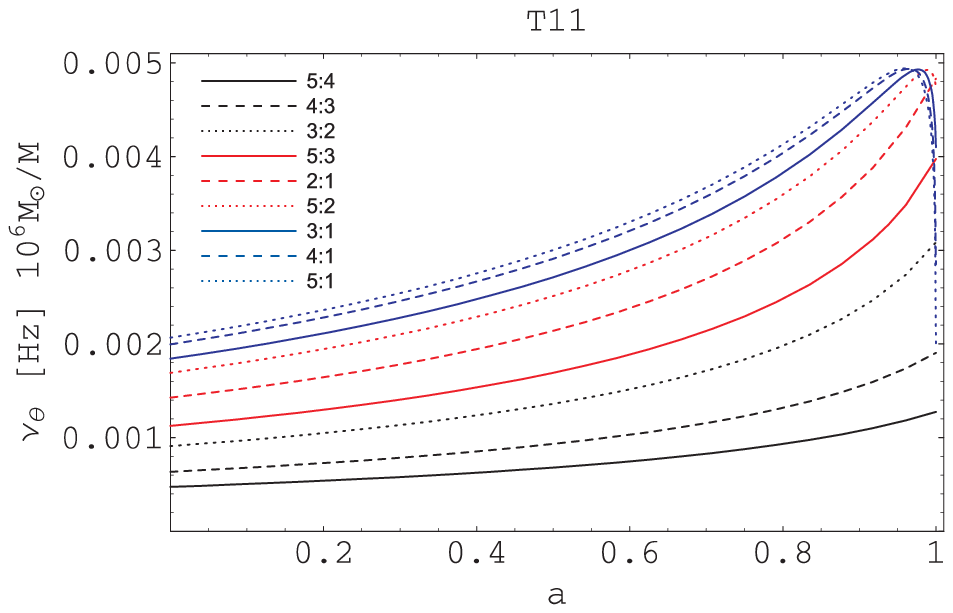}
\mbox{}
\end{center}
\end{minipage}
\hfill
\begin{minipage}[h]{.48\hsize}
\begin{center}
\includegraphics[width=\hsize]{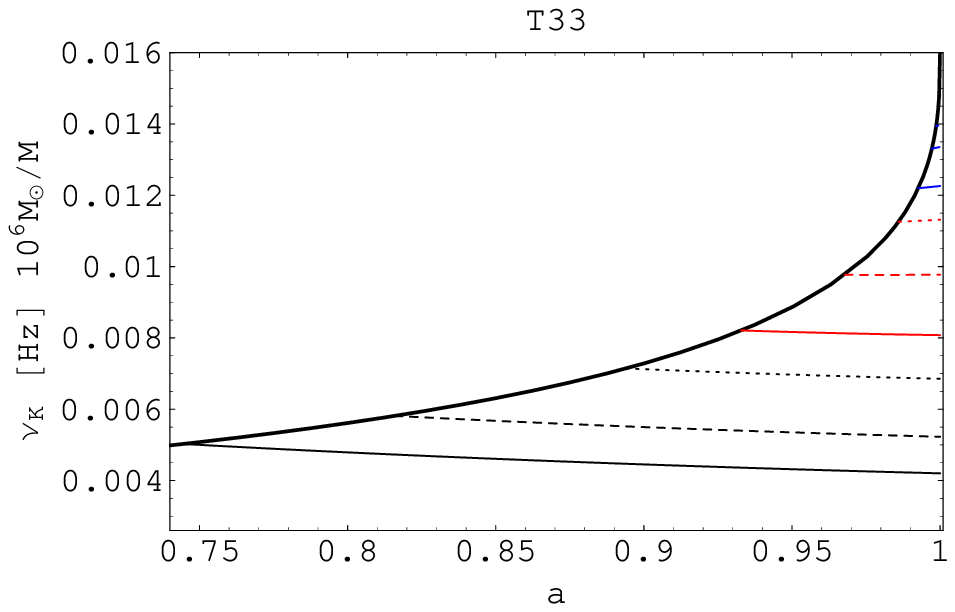}
\mbox{}
\end{center}
\end{minipage}
\begin{minipage}[h]{.48\hsize}
\begin{center}
\includegraphics[width=\hsize]{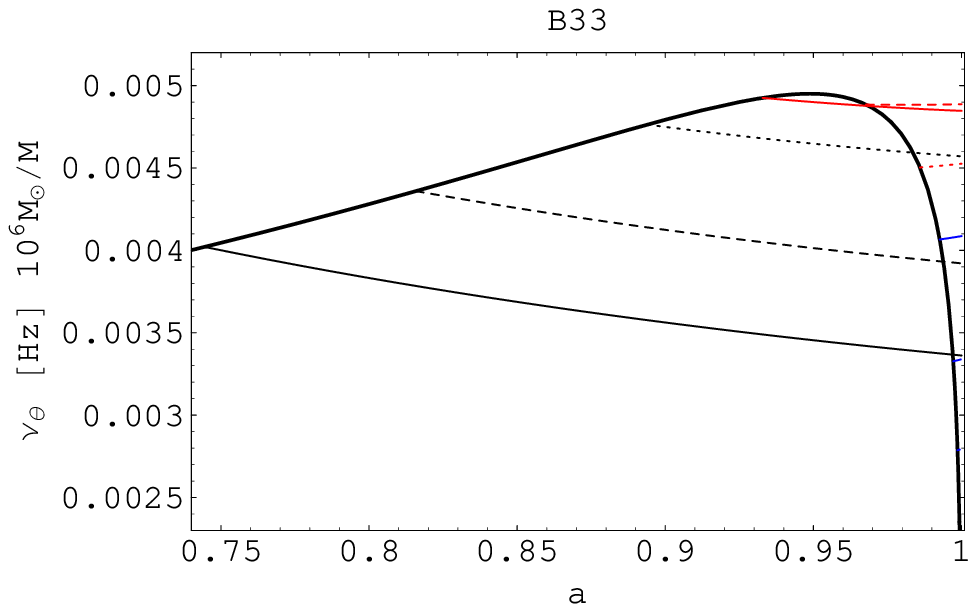}
\mbox{}
\end{center}
\end{minipage}
\hfill
\begin{minipage}[h]{.48\hsize}
\begin{center}
\includegraphics[width=\hsize]{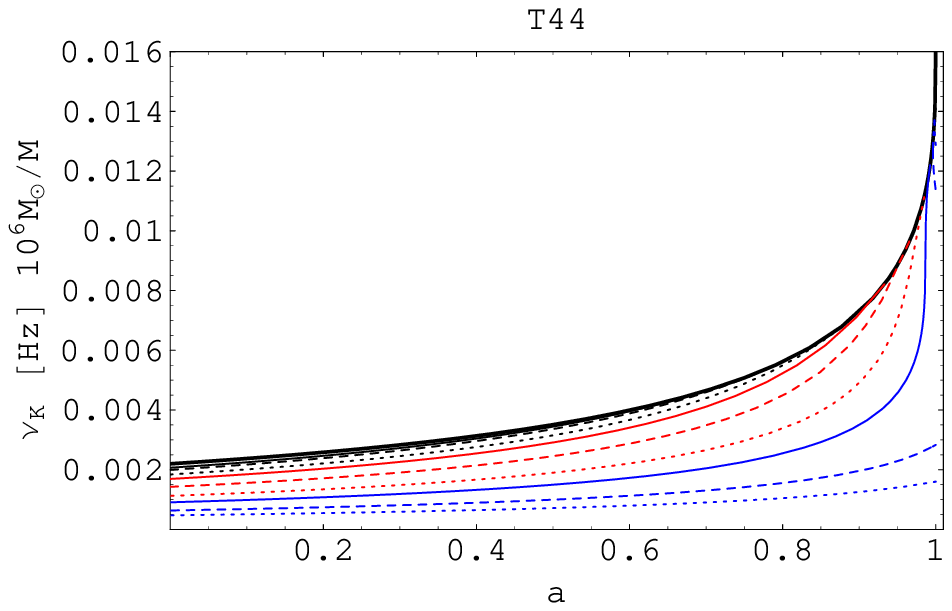}
\mbox{}
\end{center}
\end{minipage}
\begin{minipage}[h]{.48\hsize}
\begin{center}
\includegraphics[width=\hsize]{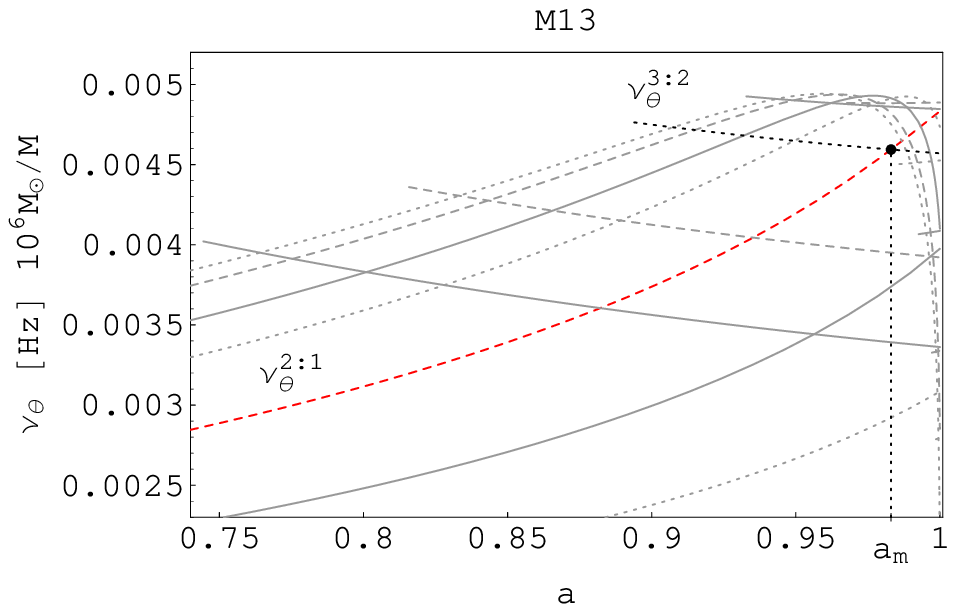}
(a)
\end{center}
\end{minipage}
\hfill
\begin{minipage}[h]{.48\hsize}
\begin{center}
\includegraphics[width=\hsize]{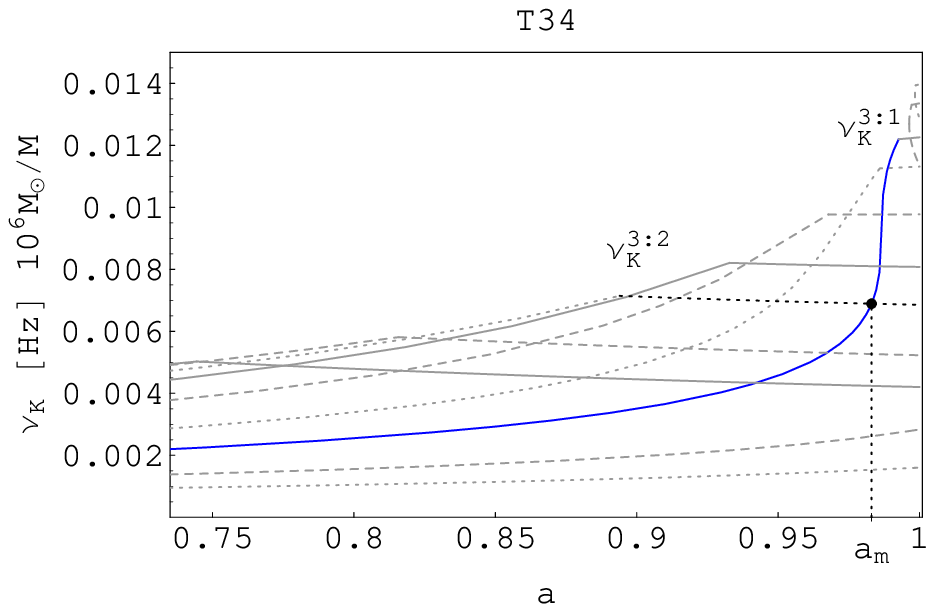}
(b)
\end{center}
\end{minipage}
\caption{Determination of black hole spin for several cases of doubled resonances. The~functions
$\nu_{\mathrm{v}}=\nu_{\mathrm{v}}^{n:m}(a,x_{n:m}(a))$,
$\mathrm{v}\in\{\theta,\mathrm{K}\}$ are drawn for the~frequency ratios $n$\,:\,$m$\,=\,5\,:\,4 (black
solid line), 4\,:\,3 (black dashed line), 3\,:\,2 (black dotted
line), 5\,:\,3 (red solid line), 2\,:\,1 (red dashed line),
5\,:\,2 (red dotted line), 3\,:\,1 (blue solid line), 4\,:\,1
(blue dashed line), 5\,:\,1 (blue dotted line). Black thick line
in T33, T44, and B33 represents
$\nu_{\mathrm{v}}=\nu_{\mathrm{v}}(a,x_{\mathrm{ms}})$, which
denotes the~frequency on the~marginally stable
orbit. Here we show how to determine the~``magic'' spin $a_{\mathrm{m}}=0.983$ in the~case of (a) ``mixed identity'' M13 with $\nu_{\theta}:\nu_{r}=2:1$ and $\nu_{\mathrm{K}}:\nu_{\theta}=3:2$ giving the~triple frequency set $\nu_{\mathrm{K}}^{3:2}:\nu_{\theta}:\nu_{r}^{2:1} = 3:2:1$ and (b) ``top identity'' T34 with $\nu_{\mathrm{K}}:\nu_{\theta}=3:2$ and $\nu_{\mathrm{K}}:\left(\nu_{\theta}-\nu_{r}\right)=3:1$ giving the~triple frequency set $\nu_{\mathrm{K}}:\nu_{\theta}^{3:2}:\left(\nu_{\theta}-\nu_{r}\right)^{3:1} = 3:2:1$.\label{Tab-examples}}
\end{figure*}

The~triple frequency set ratios are directly given by the~versions of resonance that occur in the~two twin-peak QPOs under consideration, while the~relevant spin $a$ can be easily determined by seeking common points of the~relevant
frequency functions for a~fixed mass $M$. The~schemes for treating the~situations with duplex frequencies are given in
Figs.~\ref{typical} and \ref{Tab-examples}.

The~method of finding the~spin $a$ is illustrated in Fig.~\ref{Tab-examples} for several cases of resonances. We can see from Fig.~\ref{Tab-examples} that there are no solutions for the~cases T33, T44, and B33, but for T11, T34, and M13 there are some specific values of the~black hole spin when the~top (or mixed) frequencies are identical and give the~characteristic triple frequency ratio sets. The~complete results of detailed analysis are given in Appendix~\ref{apendix-D} (Tables~\ref{top-id} -- \ref{mid-id}). We give schemes of the~direct resonances (D1\,--\,D3) and some of the~triple combinational resonances (CT1, CT2) that cannot be deduced from the~direct resonances.

For the~simple combinational resonances containing only two of the~three relevant frequencies, the~results can be easily determined from the~direct resonances, because at a~given radius $x_{p}$, where a~direct resonance occurs, the~combinational resonances of the~two relevant frequencies occur, and the~relevant ratio of the~combinational resonances is given by the~relations expressed explicitly for the~direct resonance. In all of the~simple combinational resonances occurring at the~same radius as the~corresponding direct resonance, the~related black hole spin $a$ remains unchanged, and only the~triple of the~frequency ratios is different and is determined by the~relations for the~combinational frequencies.

In presenting the~final results for the~direct and simple combinational resonances in the~tables in Appendix~\ref{apendix-D}, we have considered all the~ratios with $n,m = 1,2,3,4$. Of course, owing to the~theory \citep{Lan-Lif:1976:Mech:}, the~highest resonance order that should be considered here is $n+m = 9$, corresponding to the~highest order resonances with $n:m = 5:4$ that are observed in some black hole systems \citep[see][]{Rem-McCli:2006:ARASTRA:,Str-etal:2007:ASTRJ2:QuaPerVar} and in some neutron star systems \citep[see][]{Bel-Men-Hom:2005:ASTRA:,Bel-Men-Hom:2007:MONNR:BriNSQPOCor,Bar-Oli-Mil:2005:MONNR:,Tor:2009:ASTRA:ReversQPOs,Stu-Kot-Tor:2011:ASTRA:ResRadKep}. However, such tables are too extended to be presented here, and that is the~reason we restrict ourselves to the~limited range of $n,m = 1,2,3,4$. The~complete tables with $n,m = 1,2,3,4,5$ can be found for some important cases in \citet{Stu-Kot-Tor:2007:RAGtime8and9CrossRef:MrmQPO}.

In the~case of the~doubled combinational (and warped disc) frequency ratios, we restrict our attention to triple frequency sets that contain their combinations with the~most frequently expected direct resonance of the~radial and vertical epicyclic oscillations and to the~ratios involving $n,m = 1,2,3,4$. The~results are presented in Appendix~\ref{apendix-D} (Tables~\ref{top-id-D-DC} -- \ref{mid-id-D-DC}). The~other combinations can be constructed in an~analogous manner. An~explicit form of tables representing all the~possible combinations of doubled beat frequencies is very extended, and that is why we do not present them explicitly. Using our results they could be constructed in a~straightforward way. The~resonance ``guide'' book can be thus completed easily.

\subsection{Strong resonant radii and related black hole spin}

\begin{figure}[t]
\includegraphics[width=\hsize]{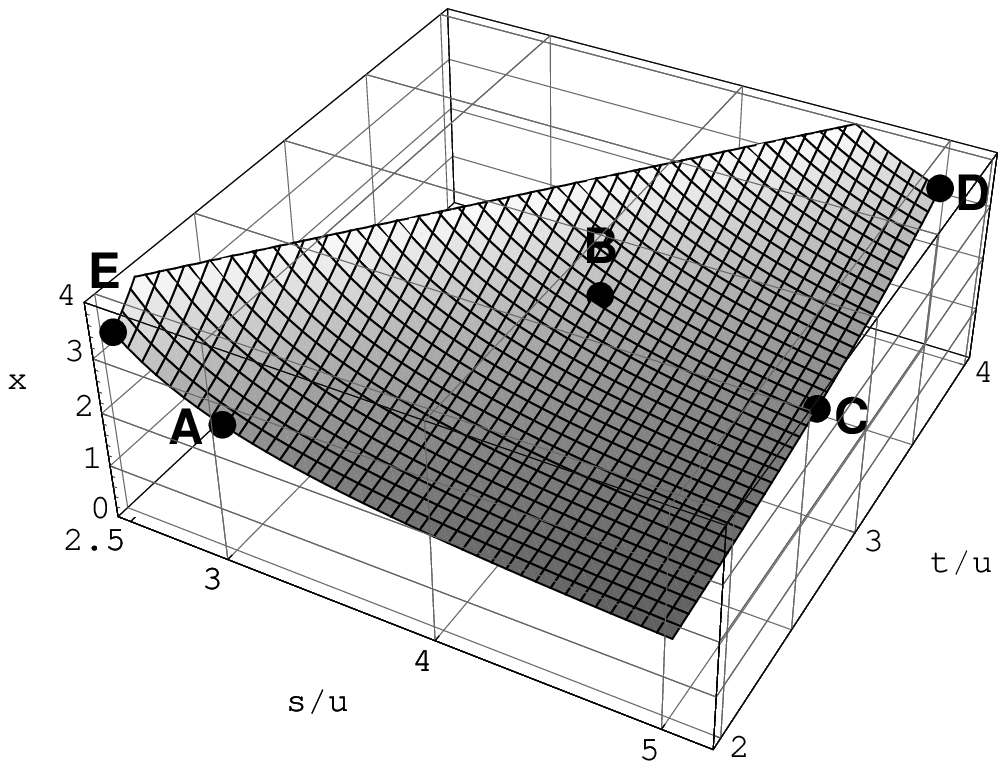}
\caption{The~function $x(s/u,t/u)$ determining the~triple frequency ratio set $s:t:u$ at the~properly given radius of a~black hole with a~``magic'' spin. The~points
represent the~ratio: $s$\,:\,$t$\,:\,$u=$ 3\,:\,2\,:\,1 (A),
4\,:\,3\,:\,1 (B), 5\,:\,3\,:\,1 (C), 5\,:\,4\,:\,1 (D),
5\,:\,4\,:\,2 (E).\label{x-stu}}
\end{figure}

The~presented results show that usually the~triple frequency sets fixing the~black hole spin $a$ occur at two different radii. However, there are important cases when the~triple frequencies occur at the~same (shared) radius. Then one could expect an~intuitively higher probability that the~resonant phenomena will arise, especially in the~cases of ratios of very low integers
\begin{equation}
3:2:1,\ 4:3:2,\ 6:3:2,\,\dots
\end{equation}
because a~causally related cooperation of the~resonances at the~given radius should come into play. A~crucial role is expected for direct resonances of oscillations with all three orbital frequencies characterized by a~triple frequency ratio set ($s$, $t$, $u$ being small natural numbers)
\begin{equation}
\nu_{\mathrm{K}} : \nu_{\theta} : \nu_{r} = s:t:u
\end{equation}
when strong resonant phenomena are possible \citep{Stu-Kot-Tor:2008:ACTA:BHadmStrResPhen}. Assuming two resonances with ratios $\nu_{\mathrm{K}}:\nu_{\theta}=s:t$ and $\nu_{\mathrm{K}}:\nu_{r}=s:u$ sharing the~same radius $x$, we can determine the~radius giving $s:t:u$ ratio from the~equation \citep{Stu-Kot-Tor:2008:ACTA:BHadmStrResPhen} {\setlength\arraycolsep{2pt}
    \begin{eqnarray}
&&x\left(s/u,t/u\right)\equiv 6\left(s/u\right)^2{X}^{-1}\,,\\
&&X=6 \left(s/u\right)^2\pm 2 \sqrt{2} \sqrt{\left(t/u-1\right)
\left(t/u+1\right) \left[3 \left(s/u\right)^2-\left(t/u\right)^2-2
\right]}\nonumber\\
&&- \left[\left(t/u\right)^2+5\right]\,,
    \end{eqnarray}}
and the~related black hole ``magic'' spin is given by
   \begin{equation}
a\left(x\left(s/u,t/u\right),u/s\right)\equiv\frac{\sqrt{x}}{3}\left(4\pm\sqrt{-2+3x\left[1-\left(u/s\right)^2\right]}\,\right).
   \end{equation}
A~detailed discussion of the~black holes that admit strong resonant phenomena is for small integers ($s\leq 5$) given in \citet{Stu-Kot-Tor:2008:ACTA:BHadmStrResPhen}. All the~results (denoted by an~asterisk) are contained in Appendix~\ref{apendix-D} (Tables~\ref{top-id} -- \ref{mid-id} and \ref{top-id-D-DC} -- \ref{mid-id-D-DC}). Since $x(s,t,u)$ can be considered as a~two-parameter ($s/u,t/u$) family of solutions for the~shared resonant radii, we are able to give the~results of finding the~strong resonance radii and the~corresponding spin in Fig.~\ref{x-stu}. The~radius and spin at these points are $x_{\mathrm{A}} = 2.395$, $a_{\mathrm{A}} = 0.983$; $x_{\mathrm{B}} = 2.880$, $a_{\mathrm{B}} = 0.866$; $x_{\mathrm{C}} =2.083$, $a_{\mathrm{C}} =0.962$; $x_{\mathrm{D}} = 3.240$, $a_{\mathrm{D}} = 0.775$; $x_{\mathrm{E}} = 3.407$, $a_{\mathrm{E}} = 0.882$.

\begin{figure*}
\begin{center}
\begin{minipage}[h]{.48\hsize}
\includegraphics[width=\hsize]{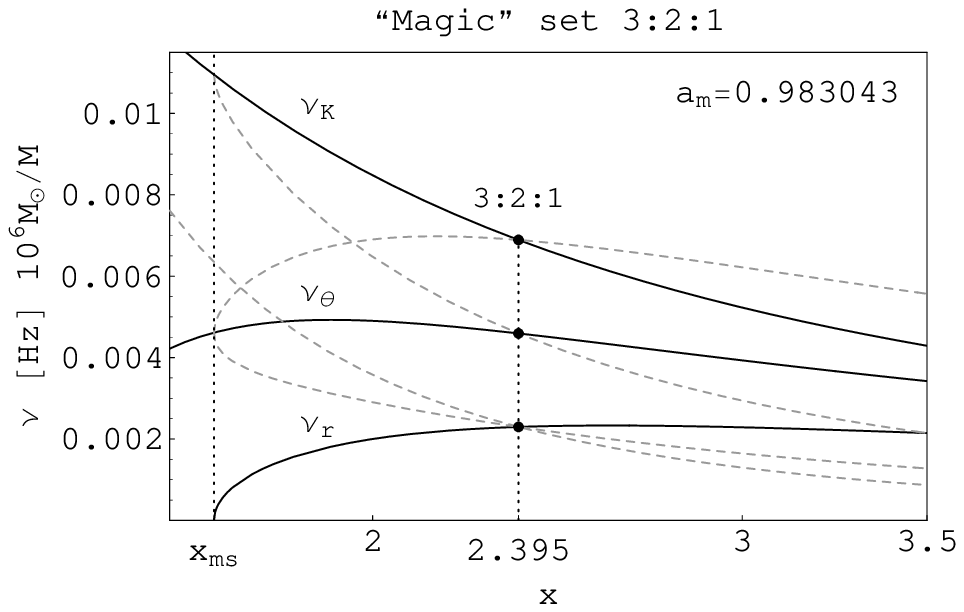}
\end{minipage}
\begin{minipage}[h]{.48\hsize}
\includegraphics[width=\hsize]{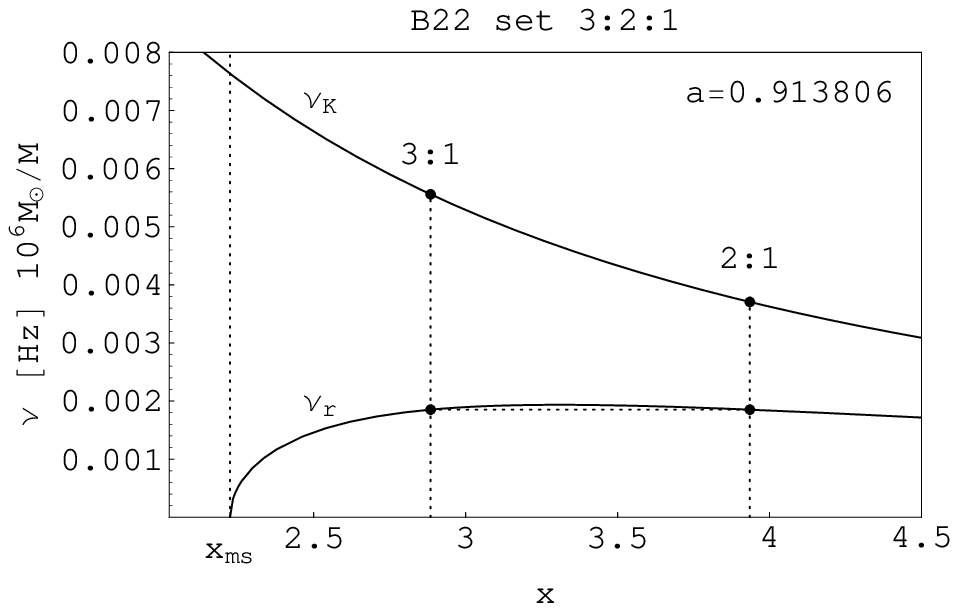}
\end{minipage}
\end{center}
\caption{\label{3-2-1-sets}Left: The~case of a~``magic'' spin, where the~strongest resonances
could occur at the~same radius. For completeness we present the~relevant simple combinational frequencies
$\nu_{\theta}-\nu_{r}$, $\nu_{\theta}+\nu_{r}$,
$\nu_{\mathrm{K}}-\nu_{\theta}$,
$\nu_{\mathrm{K}}-\nu_{r}$ (grey dashed lines). The~``magic'' spin $a_{\mathrm{m}}=0.983$ represents
the~only 
case where the~combinational and direct orbital frequencies
coincide at the~shared resonance radius. Right:
Another example of the~characteristic frequency ratio set 3\,:\,2\,:\,1 that
appears at different radii.}
\end{figure*}

Of special interest seems to be the~case of the~``magic'' spin $a=0.983$, when the~Keplerian and epicyclic frequencies are in the~ratio $\nu_{\mathrm{K}}:\nu_{\theta}:\nu_{r} = 3:2:1$ at the~common radius $x_{3:2:1} = 2.3947$ (see
Fig.~\ref{3-2-1-sets}a). In fact, this case involves rather extended structure of resonances with $\nu_{\mathrm{K}}:\nu_{r}=3:1$, $\nu_{\mathrm{K}}:\nu_{\theta}=3:2$, $\nu_{\theta}:\nu_{r}=2:1$. Also the~simple combinational frequencies could be in this small integer ratio \citep{Stu-Kot-Tor:2008:ACTA:BHadmStrResPhen}. The~strongest possible resonances occur when  the~beat frequencies enter the~resonance satisfying the~conditions
            \begin{equation}
            \frac{\nu_{\theta}+\nu_{r}}{\nu_\mathrm{K}}=
            \frac{\nu_{\theta}}{\nu_\mathrm{K}-\nu_{r}}=
            \frac{\nu_{r}}{\nu_\mathrm{K}-\nu_{\theta}}=
            \frac{\nu_{\theta}-\nu_{r}}{\nu_{r}}=1.
            \end{equation}
It should be stressed that beside the~case of strong resonances between oscillations with $\nu_{\mathrm{K}}$, $\nu_{\theta}$, $\nu_{r}$ sharing the~same radius, the~characteristic set 3\,:\,2\,:\,1 could also appear because of resonances at different radii (see Fig.~\ref{3-2-1-sets}b). All the~relevant versions of the~multi-resonant model with 3\,:\,2\,:\,1 frequency ratio set are given in Appendix~\ref{apendix-E} by Table~\ref{3/2/1}.

\section{Resonant switch model}\label{6-RSmodel}

In the~case of LMXB systems containing neutron or quark stars, we propose a~new model of twin-peak HF~QPOs. It is based on the~idea that the~twin oscillatory modes with changing frequencies that create sequences of the~lower and upper HF~QPOs switch at a~resonant point -- we call it resonant switch (RS) model. We assume the~resonant point occurs at a~radius of the~oscillating disc where frequencies of the~twin oscillatory modes become commensurable. We expect that at the~resonant point non-linear resonant phenomena will cause excitation of a~new oscillatory mode (or two new oscillatory modes) and dumping (vanishing) of one of the~previously acting modes (or both the~previously acting modes), i.e., switching from one pair of the~oscillatory modes to the~other pair of them that will be acting up to the~next relevant resonant point.

The~observational X-ray data obtained for both the~atoll and Z-sources indicate the~feasibility of the~RS model in some of the~observed sources. In the~atoll sources (e.g., in 4U~1636$-$53), the~twin-peak HF~QPOs span a~wide range of frequencies crossing the~frequency ratios $3\!:\!2$, $4\!:\!3$ and finishing at $5\!:\!4$, just near the~inner edge of the~accretion disc \citep{Bou-Bar-Lin-Tor:2010:MNRAS:,Tor:2009:ASTRA:ReversQPOs}. In the~case of the~4U~1636$-$53 source, the~behaviour of these observed oscillations, reflected by the~energy switch effect occurring at the~related frequency ratios \citep{Tor:2009:ASTRA:ReversQPOs}, indicates the~resonant effects at disc radii where the~frequencies are in the~ratios 3\,:\,2 and 5\,:\,4. The~energy switch effect, reflecting the~change of the~dominance of the~lower frequency oscillations to dominance of the~upper frequency oscillations while the~resonant point is crossed, can be naturally explained in the~framework of non-linear resonant phenomena \citep{Hor-etal:2009:ASTRA:IntResQPOs}. Therefore, it is probably reasonable to assume that the~frequencies of oscillations at the~energy reverse point determine the~eigenfrequencies of the~oscillations at the~resonance point.

The~energy switch effect allows the~frequencies to be determined at the~resonant points that can then be used to determine (or significantly restrict) parameters of the~central neutron (quark) star for various models of the~HF~QPOs. A~similar, but slightly modified, situation occurs for Z-sources. In the~case of the~typical Z-source Circinus~X-1, we again observe a~wide range of HF~QPO frequencies, but these are crossing frequency ratios 2\,:\,1 and 3\,:\,1 which are substantially higher than in the~atoll sources, indicating that the~QPOs occur at different parts of the~accretion discs in comparison with the~atoll sources. We expect that in the~atoll sources HF~QPOs occur in the~innermost parts of the~disc \citep[we can even assume that the~resonant point with frequency ratio 5\,:\,4 represents the~inner boundary of the~disc,][]{Stu-Kot-Tor:2011:ASTRA:ResRadKep}, while in the~Z-sources we expect HF~QPOs generated at more distant parts of the~disc, lying behind the~radius where the~maximum of the~radial epicyclic frequency occurs \citep{Tor-etal:2010:ASTRJ2:MassConstraints}. We can determine the~resonant points (and their observational extension) by both (i)~the~switch energy effect, i.e., due to the~frequency where the~difference in amplitude of the~lower and upper frequency oscillations vanishes, and (ii)~the~region of observational data in the~resonant frequency ratio. We have checked that these methods agree in the~case of the~4U~1636$-$53 source.

In the~(simplest version of) RS model, we assume two resonant points at the~disc radii $r_{\mathrm{out}}$ and $r_{\mathrm{in}}$, with observed frequencies $\nu_{\mathrm{U}}^{\mathrm{out}}$, $\nu_{\mathrm{L}}^{\mathrm{out}}$ and $\nu_{\mathrm{U}}^{\mathrm{in}}$, $\nu_{\mathrm{L}}^{\mathrm{in}}$, being in commensurable ratios
$p^{\mathrm{out}} = n^{\mathrm{out}}:m^{\mathrm{out}}$ and $p^{\mathrm{in}} = n^{\mathrm{in}}:m^{\mathrm{in}}$. These resonant frequencies are determined by the~energy switch method \citep{Tor:2009:ASTRA:ReversQPOs,Hor-etal:2009:ASTRA:IntResQPOs}. The observations put the~restrictions $\nu_{\mathrm{U}}^{\mathrm{in}} > \nu_{\mathrm{U}}^{\mathrm{out}}$ and $p^{\mathrm{in}} < p^{\mathrm{out}}$. In the~region covering the~resonant point at $r_{\mathrm{out}}$ we assume the~twin oscillatory modes with the~upper (lower) frequency determined by the~function $\nu_{\mathrm{U}}^{\mathrm{out}}(x;M,a)$ ($\nu_{\mathrm{L}}^{\mathrm{out}}(x;M,a)$). In the~region of the~inner resonant point we assume (generally) different oscillatory modes with the~upper and lower frequency functions $\nu_{\mathrm{U}}^{\mathrm{in}}(x;M,a)$ and $\nu_{\mathrm{L}}^{\mathrm{in}}(x;M,a)$. We assume all the~frequency functions to be given by combinations of the~orbital and epicyclic frequencies of the~geodesic motion in the~Kerr backgrounds, and such a~simplification is correct with high precision for neutron (quark) stars with high mass. The~rotating neutron stars are described quite well by the~Hartle--Thorne geometry characterized by three parameters: mass $M$, internal angular momentum $J$, and quadrupole moment $Q$ \citep{Har-Tho:1968:ASTRJ2:SlowRotRelStarII}. It is convenient to use dimensionless parameters $a = J/M^2$ \citep[spin -- denoted $j$ in][]{Tor-etal:2010:ASTRJ2:MassConstraints} and $q = QM/J^2$ (dimensionless quadrupole moment). In the~special case where $q \sim 1$, the~Hartle--Thorne geometry reduces to the~Kerr geometry that is very convenient for relativistic calculations in a~strong gravitational field regime because of the~simplicity of relevant formulae. It has been shown that near-maximum-mass neutron (quark) star Hartle--Thorne models, constructed for any given realistic equation of state, imply $q \sim 1$, and the~simple Kerr geometry is quite correctly applicable in such situations instead of the~complicated Hartle--Thorne geometry \citep{Tor-etal:2010:ASTRJ2:MassConstraints}. High neutron star masses can be expected in the~LMXBs because their mass grows owing to the~accretion.

The~frequency functions have to meet the~observationally given resonant frequencies. In the~framework of the~simple RS model this requirement enables determination of the~parameters of the~Kerr background describing the~exterior of the~neutron (quark) star. The~``shooting'' of the~frequency functions to the~resonant points, giving the~neutron star parameters, can be realized efficiently in two steps. First, we can apply our method based on the~frequency ratio's independence of the~mass parameter $M$. Therefore, the~conditions
\begin{eqnarray}
      \nu_{\mathrm{U}}^{\mathrm{out}}(x;M,a) &:& \nu_{\mathrm{L}}^{\mathrm{out}}(x;M,a) = p^{\mathrm{out}} ,\nonumber\\
      \nu_{\mathrm{U}}^{\mathrm{in}}(x;M,a) &:& \nu_{\mathrm{L}}^{\mathrm{in}}(x;M,a) = p^{\mathrm{in}}
\end{eqnarray}
imply the~relations for the~spin $a$ in terms of the~dimensionless radius $x$ and the~resonant frequency ratio $p$. We can express them in the~form $a^{\mathrm{out}}(x,p^{\mathrm{out}})$ and $a^{\mathrm{in}}(x,p^{\mathrm{in}})$ as determined in Sect.~\ref{4-multiplres} for a~concrete version of the~twin oscillatory modes, or in an~inverse form $x^{\mathrm{out}}(a,p^{\mathrm{out}})$ and $x^{\mathrm{in}}(a,p^{\mathrm{in}})$. At the~resonant radii, the~conditions
\begin{equation}
        \nu^{\mathrm{out}}_{\mathrm{U}} = \nu^{\mathrm{out}}_{\mathrm{U}}(x;M,a) , \quad   \nu^{\mathrm{in}}_{\mathrm{U}} = \nu^{\mathrm{in}}_{\mathrm{U}}(x;M,a)
\end{equation}
are satisfied along the~functions $M^{\mathrm{out}}_{p_{\mathrm{out}}}(a)$ and $M^{\mathrm{in}}_{p_{\mathrm{in}}}(a)$, which are obtained by using the~functions $a^{\mathrm{out}}(x,p^{\mathrm{out}})$ and $a^{\mathrm{in}}(x,p^{\mathrm{in}})$. The~parameters of the~neutron (quark) star are then given by the~condition
\begin{equation}
        M^{\mathrm{out}}_{p_{\mathrm{out}}}(a) = M^{\mathrm{in}}_{p_{\mathrm{in}}}(a) .    \label{RS}
\end{equation}
Usually, condition (\ref{RS}) determines $M$ and $a$ uniquely, if the~resonant frequencies are determined precisely. If an~error occurs in determining the~resonant frequencies, as expected naturally, our method gives corresponding intervals of the~acceptable values of the~mass and spin parameter of the~neutron (quark) star.

Predictions of the~RS model have to be tested in three ways:
\begin{enumerate}[a)]
  \item by the~range of values of the~mass and spin parameters of the~central neutron (quark) star allowed by theoretical models of the~neutron (quark) star structure,
  \item by the~fitting of the~twin-peak HF~QPO data sequences around the~resonant points by the~twin oscillation modes used near the~resonant points,
  \item by the~observational limits on mass and spin given by different phenomena observed in X-rays coming from the~source, e.g., by the~profiled spectral lines.
\end{enumerate}

The~method of finding the~$M(a)$ relation in the~framework of the~RS model can be directly used for the~black hole systems that provide four QPO frequencies at two resonant points located at two different radii. Of course, for black holes the~four frequencies are exactly determined from precise observations, and there are generally no restrictions on the~models applied to the~two resonant points since we expect no relation between the~two resonances, contrary to the~RS model related to the~systems containing a~neutron star.

\section{Discussion}\label{7-diskuse}

We discuss the~applicability and precision of the~multi-resonance model in black hole or neutron star systems. As an~illustration we offer a~detailed discussion of the~HF~QPO phenomena for triple frequency set with ratio $3:2:1$ in terms of the~multi-resonance model, including estimates of the~error in determining the~black hole spin. The RS model for neutron star systems exhibiting two resonant points is tested for data from the~atoll source 4U~1636$-$53.

\subsection{Black holes}

The~relation between the~tripled frequency ratios and the~black hole spin is presented in Appendix~\ref{apendix-D}, which gives all the~possible cases of the~bottom, top, and mixed frequency identities. Of course, there is potentially a~real difficulty in choosing the~proper combination of the~resonance model versions in analysing the~frequency data from concrete sources. Then all the~relevant data on HF~QPOs have to be used, and the~results have to be compared with results from the~other methods of measuring black hole spin, e.g., those based on the~spectral continuum and profiled spectral lines.
\begin{figure*}
\begin{minipage}[h]{.48\hsize}
\begin{center}
\includegraphics[width=\hsize]{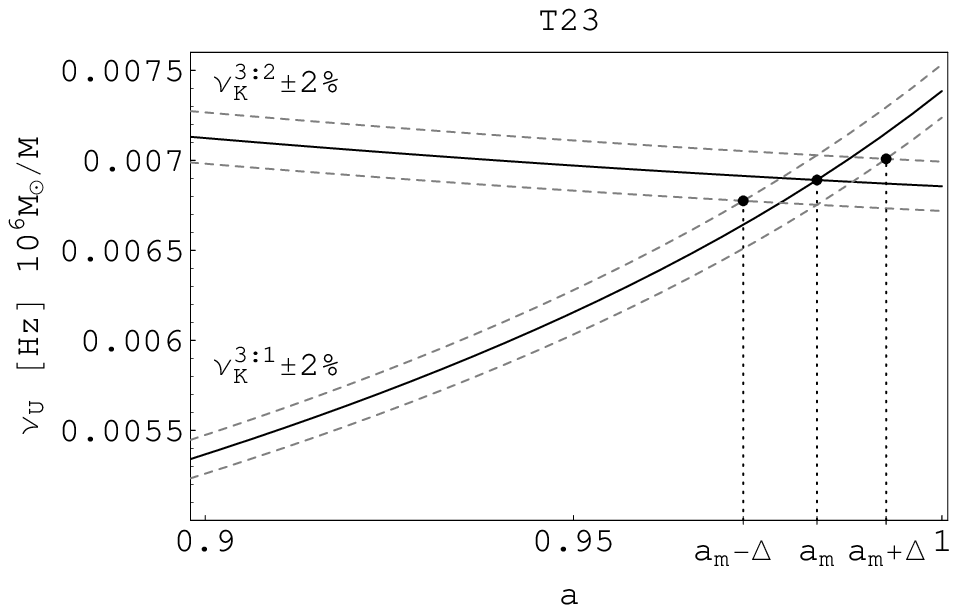}
\mbox{}
\end{center}
\end{minipage}
\hfill
\begin{minipage}[h]{.48\hsize}
\begin{center}
\includegraphics[width=\hsize]{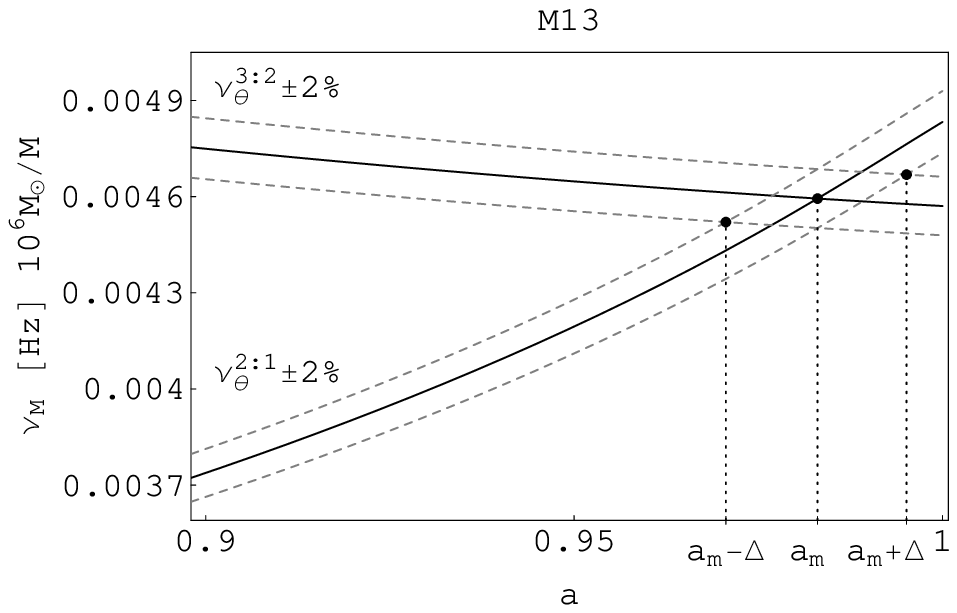}
\mbox{}
\end{center}
\end{minipage}
\begin{minipage}[h]{.48\hsize}
\begin{center}
\includegraphics[width=\hsize]{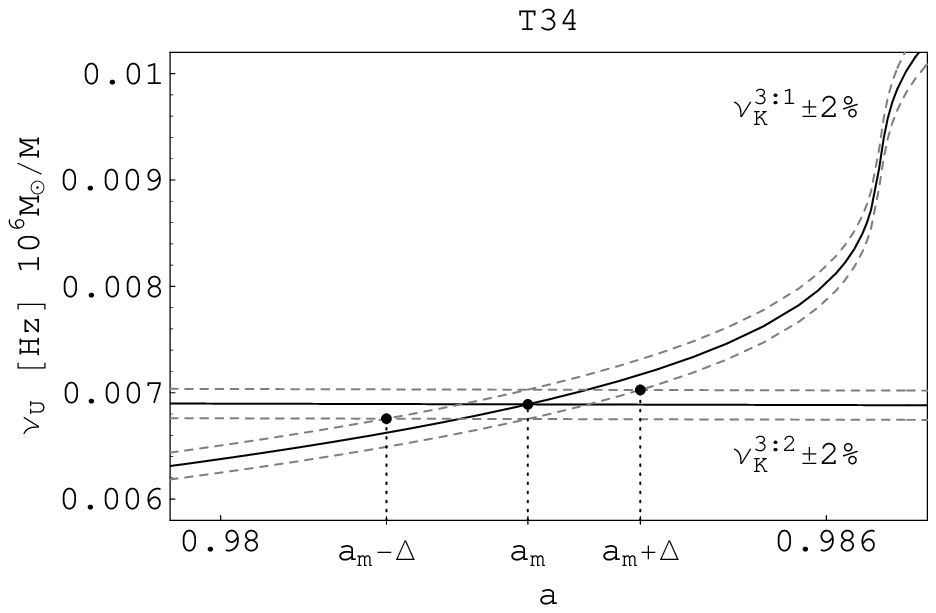}
\mbox{}
\end{center}
\end{minipage}
\hfill
\begin{minipage}[h]{.48\hsize}
\begin{center}
\includegraphics[width=\hsize]{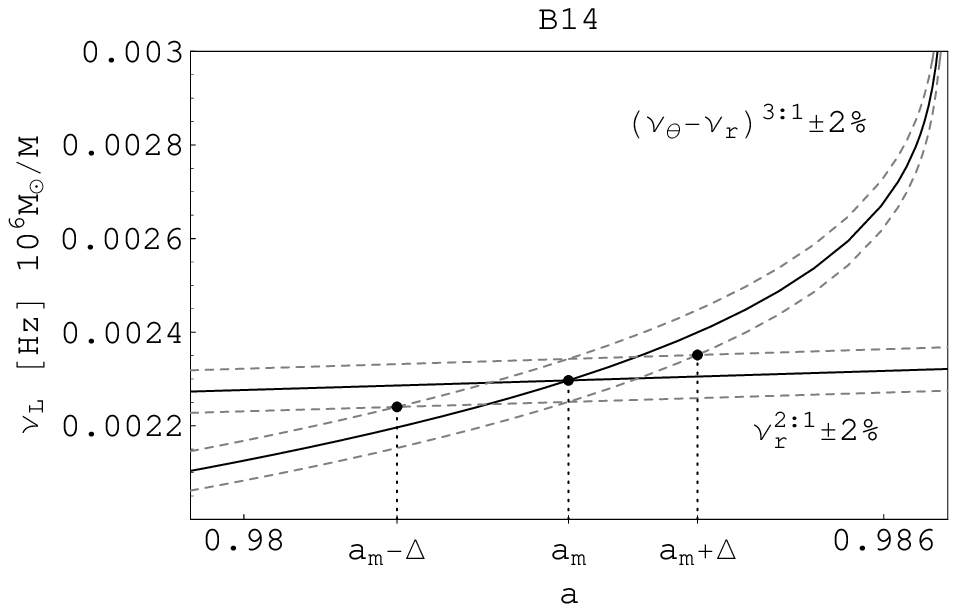}
\mbox{}
\end{center}
\end{minipage}
\begin{minipage}[h]{.48\hsize}
\begin{center}
\includegraphics[width=\hsize]{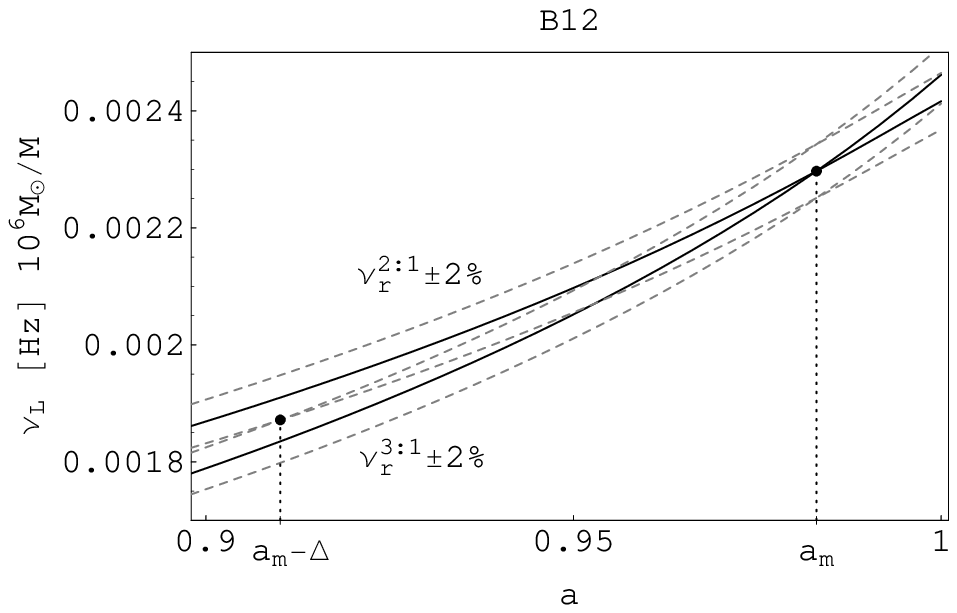}
\mbox{}
\end{center}
\end{minipage}
\hfill
\begin{minipage}[h]{.48\hsize}
\begin{center}
\includegraphics[width=\hsize]{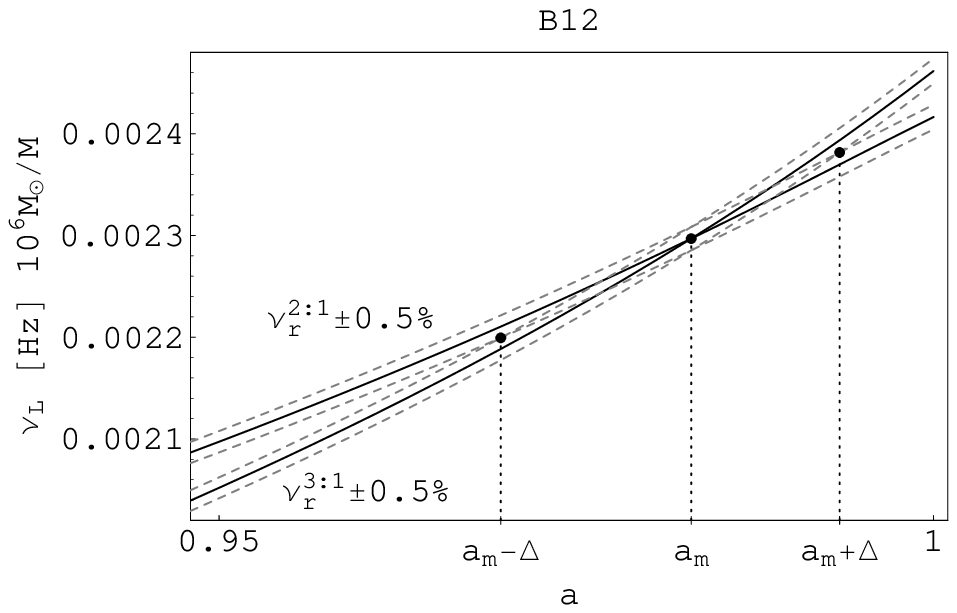}
\mbox{}
\end{center}
\end{minipage}
\caption{Error in determining the~``magic'' spin $a_\mathrm{m}=0.983043$ allowing
$\nu_{\mathrm{K}}$\,:\,$\nu_{\theta}$\,:\,$\nu_{r}=$
3\,:\,2\,:\,1 at the~same radius. The~interval of allowed values of the~black hole spin $a\in
\langle a_\mathrm{m}-\Delta;a_\mathrm{m}+\Delta \rangle$ related to the~$2\%$ error in frequency measurements is for T23: $a\in
\langle 0.973038; 0.992436 \rangle$; for M13: $a\in \langle
0.970599; 0.995125\rangle$; for T34: $a\in \langle 0.981643;
0.984157 \rangle$; for B14: $a\in \langle 0.981436;
0.984253\rangle$. For the~case of B12, only the~$0.5\%$ measurement
error gives a~reasonable restriction on black hole spin: $a\in
\langle 0.969714; 0.993428\rangle$ (for $2\%$ measurement error there is
$a\in \langle 0.91012; 1.01155\rangle$).\label{presnost}}
\end{figure*}

The~method of triple frequency sets must be treated very carefully because of uncertainties in the~frequency measurement in HF~QPOs \citep[]{Rem-McCli:2006:ARASTRA:}. In fact, in some versions of the~multi-resonance model, even a~relatively high precision of the~frequency measurements could imply rather high scatter in the~estimated black hole spin. The~triple frequency set method could be expected to work quite efficiently in the~case of ratios of small integers, such as 3\,:\,2\,:\,1, 4\,:\,3\,:\,2, etc. The~other possible frequency sets have to be taken seriously, but analysis must be very careful for high integers in the~triple frequency sets. It is increasingly difficult to distinguish different triple frequency sets when the~order of the~resonance increases, since the~uncertainties of the~frequency measurements are relatively high. Nevertheless, we can expect that data from a~new generation of X-ray satellites, especially from the~LOFT project aimed at extremely precise timing measurements \citep{Fer-etal:2012:ExA:}, could improve the~situation due to an~expected strong increase in the~frequency measurement precision.

We expect that analysis of each concrete source will need an~appropriate combination of different methods. Clearly, a~detailed analysis of assumed resonance phenomena, both parametric and forced, and their excitation by both external and internal causes must be taken into consideration. In particular, the~expected resonance strength and allowed range of resonant frequencies must be treated very carefully \citep{Stu-Kot-Tor:2008:ACTA:BHadmStrResPhen,Nay-Moo:1979:NonOscilations:}. When the~equations of motion for non-linear oscillations are solved by successive approximations, higher harmonics and combinational frequencies occur in oscillating systems corresponding to higher approximations. As the~degree of approximation increases, the~strength of the~resonances and the~resonant frequency width decrease rapidly as shown by \citet{Lan-Lif:1976:Mech:} and \citet{Nay-Moo:1979:NonOscilations:}.

The~most promising example of the~triple frequency ratios arising in a~single radius (or in its close vicinity) occurs for the~``magic'' black hole spin $a=0.983$, when at the~radius $x_{3:2:1}=2.395$, the~frequency ratio is $\nu_{\mathrm{K}}\!:\!\nu_{\theta}\!:\!\nu_{r} = 3\!:\!2\!:\!1$. Then the~resonant phenomena should be strongest, and all the~resonances, including those with beat frequencies, could cooperate efficiently even for frequencies scattered from the~exact resonant eigenfrequencies. For the~triple frequency ratio set $3\!:\!2\!:\!1$, all the~allowed combinations of the~oscillatory modes are given in Appendix~\ref{apendix-E} (Table~\ref{3/2/1}) and demonstrate a~significant complexity.

It is necessary for us to compare the~results of the~multi-resonance model to the~results of the~other methods of black hole spin measurements, namely those related to the~optical phenomena in the~vicinity of the~black hole \citep{Sch-Stu:2009:INTJMD:OpPheBraKerr,Bin-Nun:2010:PRD:SgrA:,Bin-Nun:2010:PRD:,Hor-Ger:2012:}. For Sgr\,A$^*$, considered to be a~candidate for the~test object of the~resonant phenomena with frequency ratio $3:2:1$ \citep{Asch:2004:ASTRA:,Ter:2005:ASTRA:}, the~relativistic precession of the~nearby star orbits is also very promising \citep[]{Kra:2005:,Kra:2007:}.

It is quite instructive to determine how the~frequency measurement precision could influence the~black hole spin estimates in this very special case.\footnote{A~similar estimate must be realized in analysing data from all sources that could be considered as realistic candidates of a~system for which the~triple frequency method is applicable.} We give the~range of the~black hole spin for which the~3\,:\,2\,:\,1 resonant phenomena could be relevant because of the~precision of frequency measurements that is usually reported in the~period of $0.5-2\,\%$ \citep{Rem-McCli:2006:ARASTRA:}. The~scatter of the~black hole spin related to the~frequency measurement errors is shown in Fig.~\ref{presnost}. We illustrate the~situation for some representative versions of oscillatory modes in resonance, namely T23, T34, B12, B14, M13. We can see that, except for the~case of B12, all the~cases give a~strong restriction on the~spin even for the~two percent error in the~frequency measurements. On the~other hand, for B12, only the~$0.5\,\%$ measurement error gives a~reasonable restriction on $a$, while even for $1\,\%$ error, values of $a > 1$ are allowed. Clearly, if the~relevant frequency curves cross in a~large (small) relative angle, the~spin is determined with high (low) precision. The~same rule is relevant in analysing any resonance triple frequency set. In the~case of T23, the~detailed results are illustrated in the~Table~\ref{T23-presnost}. We can see that the~one percent error in frequency measurement implies an~error of $0.005$ in spin determination, while a~three percent frequency error implies a~$0.015$ error in spin determination. In the~latter case, the~interval of allowed values for the~spin reads as $0.9678 < a_{\mathrm{m}} < 0.9969$, and the~upper limit enters the~region of applicability of the~extended resonance model \citep{Stu-Sla-Ter:2006:ASTRA:Humpy}. The~LOFT project promises high enough precision for extended application of our multi-resonance model. The assumed precision of the~LOFT measurements of HF~QPOs is expected to be at least one order higher than for recent satellites \citep{Fer-etal:2012:ExA:}, implying precision of the~spin determination at the~level of $0.0001$ (see Table~\ref{T23-presnost}).

\renewcommand{\arraystretch}{1.5}
\begin{table}[t]
\caption{\label{T23-presnost}The~scatter of the~black
hole spin $a\in
\langle a_\mathrm{m}-\Delta;a_\mathrm{m}+\Delta \rangle $ related to the~frequency measurement
errors for the~resonance of the~type T23 with the~``magic'' spin $a_\mathrm{m}=0.983043$ allowing
$\nu_{\mathrm{K}}$\,:\,$\nu_{\theta}$\,:\,$\nu_{r}=$
3\,:\,2\,:\,1 at the~same radius (see also Fig.~\ref{presnost}).}
\begin{center}
\begin{tabular}{cccc}
    \hline\hline
  $\Delta \nu_{\mathrm{U}}\, [\%]$ & $a_\mathrm{m}-\Delta$  & $a_\mathrm{m}+\Delta$ & $\sim\Delta a\, [\%]$ \\
    \hline
 $\pm 3.0$ & $0.967806$ & $0.996902$ & $\pm 1.5$ \\
 $\pm 2.0$ & $0.973038$ & $0.992436$ & $\pm 1.0$ \\
 $\pm 1.5$ & $0.975596$ & $0.990146$ & $\pm 0.7$ \\
 $\pm 1.0$ & $0.978117$ & $0.987817$ & $\pm 0.5$ \\
 $\pm 0.5$ & $0.980599$ & $0.985449$ & $\pm 0.2$ \\
    \hline
  \multicolumn{4}{l}{Expected precision of LOFT:} \\
 $\pm 0.05$ & $0.982801$ & $0.983286$ & $\pm 0.025$ \\
 $\pm 0.01$ & $0.982995$ & $0.983092$ & $\pm 0.005$ \\
    \hline
  \end{tabular}
\end{center}
\end{table}

In the~framework of the~multi-resonance model, a~detailed guide book of triple frequency sets and related black hole spins has been developed (see Appendix~\ref{apendix-D}). However, this is only an~introductory book based solely on the~frequency ratios of the~observed oscillations. Additional case-dependent information has to be related to the~multi-resonance model in order to restrict variants of the~oscillatory modes in resonance or to falsify all of them. The most important information is connected to the~amplitude of the~oscillatory modes, simultaneity of their appearance, their time dependence, and frequency scatter.

Properties of the~observed oscillations are related to the~magnitude of the~accretion flow \citep[see, e.g.,][]{For-etal:2000:ApJ:}. The multi-resonance model is unaffected by the~flow magnitude if the~gravitation remains the~main restoring force of the~oscillations. In those cases where the~restoring forces of non-gravitational origin becomes relevant, the~model has to be modified.

\subsection{Neutron stars}

In LMXB systems containing neutron (quark) stars, the~simple resonance model assuming resonance of
oscillations with radial $\nu_r$ and vertical $\nu_{\theta}$ epicyclic frequencies can be excluded with high probability, or requires substantial non-geodesic corrections to the~formulae of the~epicyclic frequencies \citep{Urb-etal:2010:ASTRA:DiscOscNS32,Abr-Klu-Yu:2011:AcA:}. The~evolution of the~lower and upper HF~QPOs frequencies (across a~single measurement sequence) in such sources suggests (very rough) agreement of the~data distribution with the~relativistic precession (RP), hot spot model \citep{Ste-Vie:1999:PHYRL:,Ste-Vie:1998:ASTRJ2L:}. In rough agreement with the~data are also other models based on the~assumption of the~oscillatory motion of hot spots or accretion disc oscillations, such as the~total precession model \citep{Stu-Tor-Bak:2007:arXiv:}, the~tidal disruption model \citep{Cad-etal:2008:AA:,Kos-etal:2009:tidal:,Ger-etal:2009:tidal:}, or the~warped disc oscillations model \citep{Kat:2001:PUBASJ:,Kat:2004:PUBASJ:QPOsmodel,Kat:2004:PUBASJ:mass,Kato:2007:PUBASJ:FreqCorr,Kato:2008:b:PASJ:,Kato:2009:PASJ:}. In all of these models the~frequency difference $\nu_{\mathrm{U}} - \nu_{\mathrm{L}}$ decreases with increase in the~magnitude of the~lower and upper frequencies in accord with trends given by the~observational data \citep{Bar-etal:2005:MONNR:HiCohQPO,Bel-etal:2007:MONNR:RossiXTE,Bou-Bar-Lin-Tor:2010:MNRAS:}. An~alternative explanation of HF~QPOs in neutron star systems provides the~sonic-point and spin-resonance models \citep{Lam-Mil:2001:,Lam-Mil:2003:}, higher-order nonlinearity model \citep{Muk:2009:ApJ:}, shock-wave model \citep{Chak-etal:2009:,Deb-Chak-Nan:2010:}, Rayleigh--Taylor gravity wave model \citep{Osh-Tit:1999:,Tit:2003:,Tit-Sha:2008:}, MHD Alfv\'{e}n wave oscillation model \citep{Zha:2004:,Zha-etal:2006:MONNR:kHzQPOFrCorr,Zha-etal:2007:MNRAS:,Shi:2011:}, and MHD model \citep{Shi-Li:2009:}. 

The~$\nu_\mathrm{U} - \nu_\mathrm{L}$ frequency relation, given by a~variety of the~relevant frequency-relation models mentioned above, can be fitted to the~observational data for some properly chosen neutron star sources. Rather surprisingly, the~fitting procedure applied implies a~mass-spin relation $M(a) = M_{0}[1 + k(a + a^2)]$ for both the~atoll and Z sources rather than concrete values of the~neutron star parameters $M$ and $a$. Both the~parameters $M_0$ and $k$ depend on the~source and the~combination of the~twin oscillatory modes \citep[for details see][]{Tor-etal:2010:ASTRJ2:MassConstraints,Tor-etal:2012:ApJ:}. The quality of the~fitting procedure is very poor for the~atoll source 4U~1636$-$53 as demonstrated in  \citet{Tor-etal:2012:ApJ:}. Similar, very bad fitting of observational data with the~frequency relation models has been found in \citet{Lin-etal:2011:ApJ:} for the~atoll source 4U~1636$-$53 and Sco~X-1. This disagreement of the~data distribution with their fitting by the~frequency-relation models is based on the~assumption of the~geodesic character of the~oscillatory frequencies. It implied some attempts to find a~non-geodesic correction reflecting additional physical ingredients (such as an~influence of the~magnetic field) that could make the~fitting procedure much better \citep{Tor-etal:2012:ApJ:}.

For the~RP model, a~strong improvement can be reached by a~small shift in the~geodetical radial epicyclic frequency and a~related small change in the~marginally stable orbit location \citep{Tor-etal:2010:ASTRJ2:MassConstraints,Tor-etal:2012:ApJ:}. The~non-geodesic corrections to the~radial epicyclic frequencies can be generated on the~level of tens of percent for oscillating non-slender tori \citep{Sra:2005:ASTRN:,Bla-etal:2006:ASTRJ2:,Str-Sra:2009:CLAQG:EpiOscNonSleKerrBH}, or can do so by assuming a~slightly charged internal part of the~accretion disc interacting with the~external dipole magnetic field of the~neutron star when very strong effects can be obtained, on the~level of hundred percent or higher \citep{Bak-etal:2012:CQG:,Pac-etal:2011:,Bak-etal:2010:CLAQG:MagIndNonGeo}. We can consider several diverse possibilities to explain the~behaviour of charged accretion discs. First, the~disc can be treated as a~``dielectric'' structure, with the~effective charge fixed to the~moving matter, as discussed in \citet{Kov-etal:2011:PHYSR4:}. Second, there might be an~inverse possibility, when the~electric charge happens to be concentrated in a~hot spot of a~solitonic character that could be a~cosmic object similar to the~well known and poorly understood ``ball lightning''. Third, an~interesting new and open possibility is represented by oscillations of coronal structures distributed in the~vicinity of the~off-equatorial circular orbits of charged particles discovered and discussed in \citet{Kov-Stu-Kar:2008:CLAQG:OffEqOrb}, \citet{Stu-Kov-Kar:2009:IAU:}, \citet{Kov-etal:2010:CLAQG:OffEquatOrbsII} and \citet{Kop-etal:2010:ASTRJ2:TraRegToChao}. Preliminary studies indicate that the~non-geodesic corrections make the~fits much better, however they do introduce additional free parameters into the~HF~QPOs models. Therefore, it is useful to also test the~possibility of improving the~fitting procedure by using proper combinations of the~models that assume purely geodesic origin of the~observed frequencies. Our multi-resonant approach can be applied to the~neutron star systems thanks to the~possible occurrence of one or two resonant points along the~observed twin-peak HF~QPOs \citep{Bel-Men-Hom:2007:MONNR:BriNSQPOCor,Tor-etal:2008:ACTA:Clustering4U1636-53,Tor-etal:2008:ACTA:DistrKhZ4U1636-53,Tor-Bak-Stu-Cec:2008:ACTA:TwPk4U1636-53,
Tor:2009:ASTRA:ReversQPOs,Bou-Bar-Lin-Tor:2010:MNRAS:}.

The~twin-peak frequencies are widely scattered, but concentrated in the~vicinity of the~resonance point with the~frequency ratio close to 3\,:\,2 in most of the~atoll sources. The~scatter of the~twin-peak frequencies around this 3\,:\,2 resonant point is correlated and approximated by the~linear fits $\nu_{\mathrm{U}}= A\nu_{\mathrm{L}} + B$, and an~anticorrelation of the~parameters $A$, $B$ was predicted by the~resonance theory and indicated from observational data related to twelve atoll neutron star X-ray binary systems \citep[]{Abr-etal:2005:RAGtime6and7:CrossRef,Abr-etal:2005:ASTRN:}.
However, analysis of the~neutron star source 4U~1636$-$53 indicates resonance at two resonance points with frequency ratios 3\,:\,2 and 5\,:\,4 \citep{Tor-etal:2008:ACTA:Clustering4U1636-53,Tor-etal:2008:ACTA:DistrKhZ4U1636-53,
Tor-Bak-Stu-Cec:2008:ACTA:TwPk4U1636-53,Tor:2009:ASTRA:ReversQPOs}; moreover, two resonance points with the~same frequency ratios are
demonstrated by analysis of the~source 4U~1608$-$52 \citep{Tor:2009:ASTRA:ReversQPOs}. This means that the~HF~QPO frequency linear fits and the~anticorrelation data \citep[]{Abr-etal:2005:ASTRN:} only related to the~3\,:\,2 frequency ratio are contaminated with the~data connected to the~5\,:\,4 resonance point in the~case of at least these two sources.

Recent observational data obtained for the~atoll sources indicate convincingly that the~resonance points are located very close to the~innermost stable circular geodesics (ISCO) of the~spacetime, which are usually assumed to represent the~inner edge of the~accretion disc \citep{Bar-Oli-Mil:2006:MONNR:QPO-NS,Bar-Oli-Mil:2005:MONNR:,Bar-Oli-Mil:2005:AstrNachr:}. Therefore, it is quite natural to expect that the~resonant oscillations could be excited by inhomogeneities on the~surface of the~neutron star, and these inhomogeneities strongly influence the~innermost parts of the~accretion disc \citep{Stu-Kon-Mil-Hle:2008:ASTRA:GravExc}. In fact, the~twin-peak QPOs' amplitude analysis of four atoll sources (4U~1636$-$53, 4U~1608$-$52, 4U~1820$-$30, and 4U~1735$-$44) demonstrates a~cut-off of HF~QPOs at the~resonant radii corresponding to the~frequency ratios 5\,:\,4 or 4\,:\,3 and indicating thus a~possibility that in such sources the~accretion discs are strongly influenced by resonant phenomena and that the~inner edge is shifted from the~ISCO to the~corresponding resonant radius \citep{Stu-Kot-Tor:2011:ASTRA:ResRadKep}.

A~non-standard way to improve the~data fitting can be realized in the~neutron star systems demonstrating two resonant points in HF~QPOs when assumption of a~unique frequency relation applicable along the~full twin-peak HF~QPOs frequency range can be abandoned thanks to the~RS model.

\subsection{Resonant switch model applied to the~atoll source 4U~1636$-$53}

We tested our RS model in the~case of the~atoll 4U~1636$-$53 source that seems to be the~best possibility owing to the~character of the~observational data clearly demonstrating two ``resonant points'' where the~energy switch effect occurs. Using the~results of \citet{Tor:2009:ASTRA:ReversQPOs}, the~resonant frequencies determined by the~energy switch effect are given in the~outer resonant point with frequency ratio $\nu_{\mathrm{U}}/\nu_{\mathrm{L}}=3/2$ by the~frequency intervals
\begin{eqnarray}
        \nu_{\mathrm{U}}^{\mathrm{out}} &=& \nu_{\mathrm{U0}}^{\mathrm{out}} \pm \Delta \nu^{\mathrm{out}} = (970 \pm 30) \,\mathrm{Hz}\,,\nonumber\\
        \nu_{\mathrm{L}}^{\mathrm{out}} &=& \nu_{\mathrm{L0}}^{\mathrm{out}} \pm \Delta \nu^{\mathrm{out}} = (647 \pm 20) \,\mathrm{Hz}\,,  \label{freu}
\end{eqnarray}
and at the~inner resonant point with frequency ratio $\nu_{\mathrm{U}}/\nu_{\mathrm{L}}=5/4$ there is
\begin{eqnarray}
        \nu_{\mathrm{U}}^{\mathrm{in}} &=& \nu_{\mathrm{U0}}^{\mathrm{in}} \pm \Delta \nu^{\mathrm{in}} =  (1180 \pm 20) \,\mathrm{Hz}\,,\nonumber\\
        \nu_{\mathrm{L}}^{\mathrm{in}} &=& \nu_{\mathrm{L0}}^{\mathrm{in}} \pm \Delta \nu^{\mathrm{in}} =   (944 \pm 16) \,\mathrm{Hz}\,. \label{frei}
\end{eqnarray}
The~resonant points determined by using the~energy switch effect accord with observational data points crossing the~lines of constant frequency ratios 3\,:\,2 and 5\,:\,4 as given in the~standard papers \citep{Bar-Oli-Mil:2005:MONNR:,Bel-etal:2007:MONNR:RossiXTE}.

\begin{figure*}
\begin{minipage}[h]{.48\hsize}
\begin{center}
\includegraphics[width=\hsize]{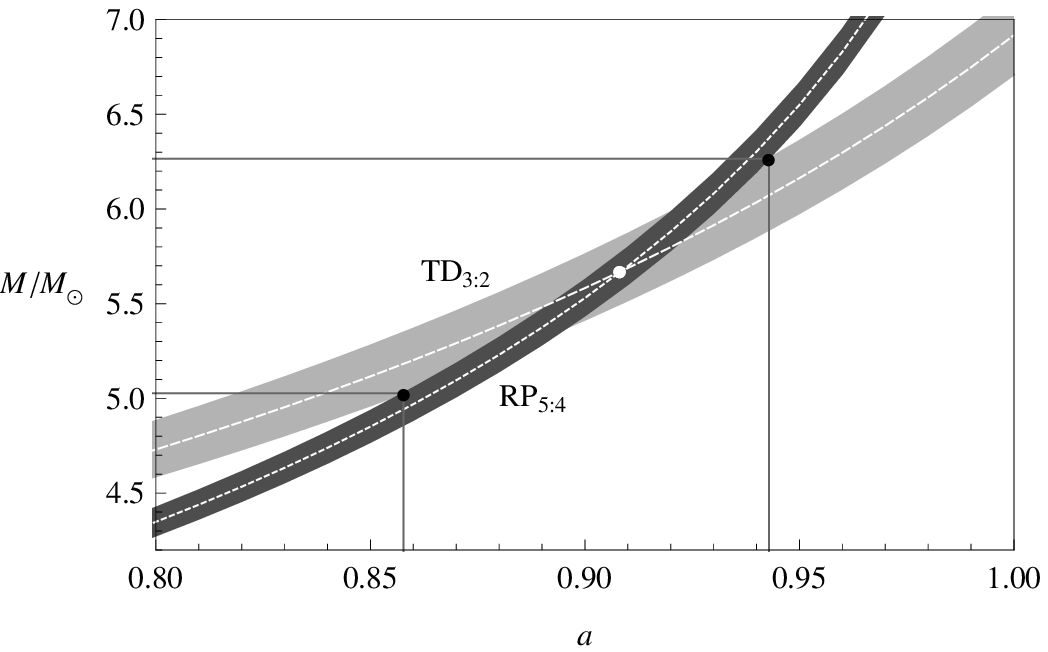}
\end{center}
\end{minipage}
\hfill
\begin{minipage}[h]{.48\hsize}
\begin{center}
\includegraphics[width=\hsize]{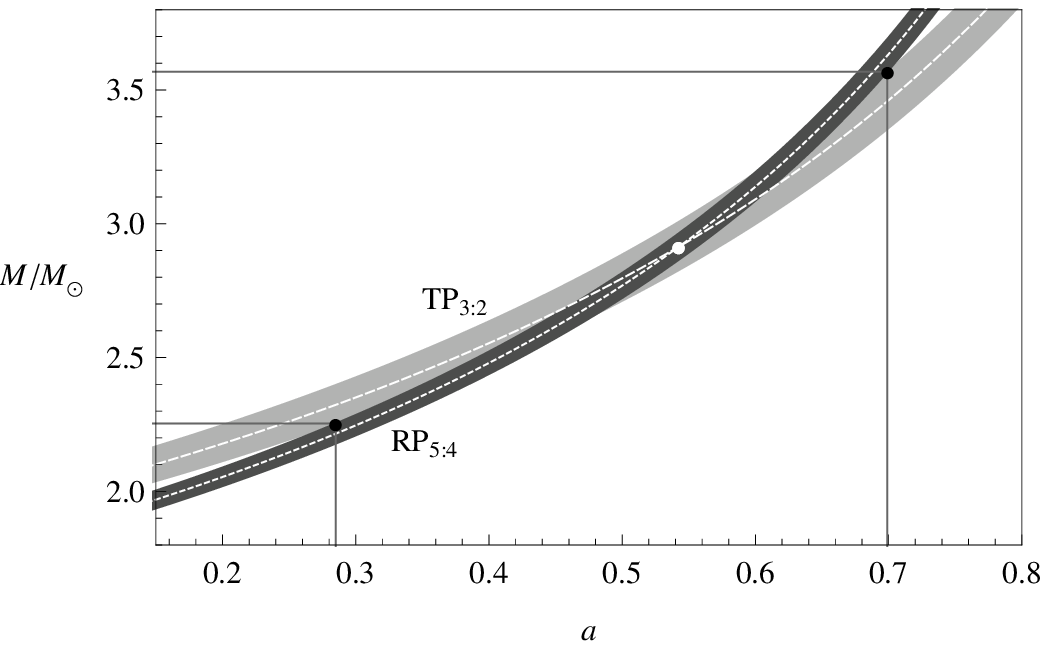}
\end{center}
\end{minipage}
\caption{\label{grafy-M-a-RS-model}
Intervals of acceptable values of mass and spin of the~neutron star in the~atoll source 4U~1636$-$53 predicted by the~resonant switch model for the~combination of RP--TD models (left) and RP--TP models (right) using the~scatter of the~resonant frequencies at each of the~two resonant points where $\mathrm{RP} \equiv \nu_\mathrm{K}:(\nu_\mathrm{K}-\nu_{r})=5:4$, $\mathrm{TD} \equiv (\nu_\mathrm{K}+\nu_{r}):\nu_\mathrm{K} = 3:2$ and $\mathrm{TP} \equiv \nu_\theta :(\nu_\theta-\nu_{r}) = 3:2$.}
\end{figure*}


We tested the~RS model determining the~parameters $M$ and $a$ of the~neutron (quark) star at the~4U~1636$-$53 source for the~standard relativistic precession (RP) model \citep{Ste-Vie:1999:PHYRL:} combined with the~tidal distortion (TD) model
\citep{Kos-etal:2009:tidal:}, and the~total precession (TP) model \citep{Stu-Tor-Bak:2007:arXiv:}. 
In the~case of the~RP model, the~frequency ratio of two oscillatory modes reads as
\begin{equation}
           \frac{\nu_{\mathrm{U}}}{\nu_{\mathrm{L}}} = \frac{\nu_\mathrm{K}}{\nu_\mathrm{K} - \nu_{r}}\,,
\end{equation}
for the~TD model we have to consider the~frequency ratio
\begin{equation}
           \frac{\nu_{\mathrm{U}}}{\nu_{\mathrm{L}}} = \frac{\nu_\mathrm{K} + \nu_{r}}{\nu_\mathrm{K}}\,,
\end{equation}
while in the~case of the~TP model there is
\begin{equation}
           \frac{\nu_{\mathrm{U}}}{\nu_{\mathrm{L}}} = \frac{\nu_{\theta}}{\nu_{\theta} - \nu_{r}}\,.
\end{equation}

\subsubsection{Combination of the~RP and TD models}

For the~combination of RP and TD models, the~RP model has to be related to the~inner resonant point, while the~TD model has to be related to the~outer resonant point. We applied the~method of determining the~neutron star parameters $M$ and $a$ presented in the~Sect.~\ref{6-RSmodel}, and the~results are presented in Fig.~\ref{grafy-M-a-RS-model}. For the~frequency scatter at the~resonant points, given by relations (\ref{freu}) and (\ref{frei}), the~range of allowed values for the~mass and spin of the~neutron star is given by
\begin{equation}
             0.86 < a < 0.94\, ,\quad
             5.03 < \frac{M}{\mathrm{M}_{\odot}} < 6.27 \,.
\end{equation}
The~central point of the~frequency ranges implies that the~central estimates of the~neutron star parameters are $a=0.91$ and $M = 5.67\,\mathrm{M}_{\odot}$.

Clearly, for the~combination of the~RP and TD oscillatory modes, the~RS model implies parameters of the~neutron star that are totally out of the~ranges accepted by the~recent theory of the~structure of the~neutron stars both for their spin and mass. It is well known that the~neutron star mass surely cannot exceed the~critical  value of $M \sim 3\,\mathrm{M}_{\odot}$ \citep{Rho-Ruf:1974:PHYRL:} and the~models based on the~realistic equations of state give maximal masses going up to $M_{\mathrm{max}} \sim 2.8\,\mathrm{M}_{\odot}$ \citep{Pos-etal:2010:LoveNum:}. In fact, the~extremal maximum $M_{\mathrm{max}} \sim 2.8\,\mathrm{M}_{\odot}$ is predicted by the~field theory \citep{Mul-Ser:1996:}. The~limit of $M_{\mathrm{max}} \sim 2.5\,\mathrm{M}_{\odot}$ is predicted by the~Dirac--Brueckner--Hartree--Fock approach in some special cases \citep{Mut-etal:1987:}, in the~variational approaches \citep{Akm-Pan:1997:,Akm-Pan:1998:}, and in other approaches \citep{Urb-Bet-Stu:2010:ACTA:ObsTestNSMeanField} allow for $M_{\mathrm{max}} \sim 2.25\,\mathrm{M}_{\odot}$. For the~maximum spin of neutron stars, the~limit $a < a_{\mathrm{max}} \sim 0.7$ has been demonstrated recently by \citet{Lo-Lin:2011:ApJ:}, independently of the~equation of state; on the~other hand, they have shown the~possible existence of quark stars with spin slightly exceeding the~black hole limit $a = 1$. Therefore, the~RP--TD combination can be excluded as a~realistic explanation of the~observed data in the~atoll source 4U~1636$-$53 because of the~high values of the~predicted neutron star parameters.

\subsubsection{Combination of the~RP and TP models}

For the~combination of RP and TP models, the~RP model has to be related to the~inner resonant point, while the~TP model has to be related to the~outer resonant point. The~method of determining the~neutron star parameters $M$ and $a$ implies the~results presented in Fig.~\ref{grafy-M-a-RS-model}. The~range of allowed values of the~mass and spin of the~neutron star is given by
\begin{equation}
             0.29 < a < 0.70 \,,\quad
             2.25 < \frac{M}{\mathrm{M}_{\odot}} < 3.57\,.
\end{equation}
The~central point of the~frequency ranges implies that the~central estimates of the~neutron star parameters are $a=0.54$ and $M = 2.91\,\mathrm{M}_{\odot}$.

The~RP--TP combination of the~oscillatory modes in the~RS model implies parameters of the~neutron star that are quite acceptable at the~lower end of the~allowed ranges of spin and mass. Owing to the~rotational frequency of the~central neutron star in the~atoll source 4U~1636$-$53, observed to be $f_{\mathrm{NS}} = 580\,\mathrm{Hz}$ (or $f_{\mathrm{NS}} = 290\,\mathrm{Hz}$), the~spin estimated by the~Hartle--Thorne models of rotating neutron stars puts limits in the~range $0.2<a<0.4$ on near-extreme masses predicted by models with realistic equations of state (Urbanec et al., in prep.). The related restriction on the~neutron star mass reads as $M<2.4\,\mathrm{M}_{\odot}$. The~physical explanation of the~RP--TP special model could be very simple. A~hot spot is oscillating in both vertical and radial directions, and its radiation is modified by the~frequencies $\nu_{\theta}$ and $\nu_{\theta} - \nu_r$; approaching the~3\,:\,2 resonant point, oscillations in the~vertical direction are successively damped by non-linear (e.g., tidal) effects and the~radial oscillations are enforced. Then the~Keplerian and the~radial epicyclic frequencies become important. This model requires a~detailed description that is planned in future.

The~RP--TP combination of the~RS model should be explicitly tested by a~fitting procedure realized for the~observed twin HF~QPO sequences related to the~resonant points. We plan to make such a~test in a~future work. Of course, we shall also test all the~possible combinations of the~RS model and estimate allowed values of the~spin and mass of the~neutron stars for these combinations.

\section{Conclusions}\label{8-zaver}

The~multi-resonance model of HF~QPOs assumes that both internal parametric resonance and/or forced resonance can occur in accretion discs rotating around black holes or neutron stars and that the~resonant non-linear phenomena can influence oscillations with the~vertical and radial epicyclic frequency and/or with the~orbital (Keplerian) frequency and their simple combinations. It is possible that the~resonances occur for different internal reasons, at different radii within the~accretion discs, and two pairs (or more complex combinations) of the~resonant frequencies could be exhibited in general situations. For simplicity, we considered two pairs of the~resonant frequencies, or their special reductions, but more complex situations can be treated in an~analogous way. The~spin and mass of the~black hole can then be found with higher precision than for individual twin peaks, but it could be rather difficult to identify any relevant combination of the~resonances.

For special values of the~black hole spin, the~bottom (top) epicyclic frequencies could be equal at different radii, since there are local extrema of the~radial profiles for both the~epicyclic frequencies in the~Kerr black hole spacetimes. If Keplerian frequency or beat frequencies can also enter the~resonance phenomena, top, bottom, or mixed coinciding frequencies in various versions of resonance are possible. In such situations, the~ratio of the~triples of the~resonant frequencies is directly related to the~black hole spin, independently of the~black hole mass. The related possibility of directly measuring the~black hole spin is very important because of relatively high uncertainties in observational estimates of the~black hole mass that are necessary for determining the~black hole spin in general resonant phenomena \citep{Ter-Abr-Klu:2005:ASTRA:QPOresmodel,Tor-Kot-Sra-Stu:2011:}, or in black hole spin determinations based on the~measurements of profiled spectral lines \citep[]{Laor:1991:ASTRJ2:,Kar-Vok-Pol:1992:ASTRJ2L:,Dov-Kar-Mar:2004:RAGtime4and5:CrossRef,Fab-Min:2005:XSpectraKerr:Book,
Zak:2003:,Zak-Rep:2006:,Zak-etal:2012:,Fan-etal:1997:,Cad-Cal-Fan:2003:MEMSA1:XrFeProf,Cad-Cal:2005:MNRAS:}.

The~multi-resonance model of HF~QPOs represents a~promising approach to understanding the~observational data from LMXBs containing both black holes and neutron stars and data from discs orbiting black holes in the~ULX sources or in active galactic nuclei. We focussed our attention on the~properties of the~frequency patterns implied by various versions of the~resonance orbital model, but not going into physical details because they are clearly case dependent.

The~special triple frequency set method determines the~black hole spin with high precision and is quite independent of the~measurement of the~black hole mass, but it could work only accidentally for special values of the~spin. Nevertheless, it is worth making a~detailed scan of all the~observational data for the~LMXB black hole systems or supermassive black holes in active galactic nuclei in order to look for some candidate systems, since any successful determination of the~spin could help very much in determining other physical parameters of the~system and in understanding a~wide scale of astrophysical phenomena. The~prepared new space X-ray mission LOFT \citep{Fer-etal:2012:ExA:} proposes a~sensitivity of the~observational instruments that is high enough to obtain data that could be both extended and precise enough to apply the~triple frequency set method in a~realistic way.

In~Appendix~\ref{apendix-D} 
 we present a~detailed guide to all the~possible triple frequency sets and related values of the~black hole spin $a$, shown for all possible double combinations of both the~direct and simple combinational resonances with the~order of individual resonances limited by $n \leq 4$.\footnote{Tables limited by $n \leq 5$ are much too extended to be published here; however, in relevant cases they can be found in \citet{Stu-Kot-Tor:2007:RAGtime8and9CrossRef:MrmQPO}.} It is clear that comparison of observational data with the~guide tables must be done extremely carefully, since different resonances can give the~same triple frequency ratio set and black hole spin. It is worth noting that the~triple frequency set resonant model can be approximately used also in situations where two resonant points are detected with two of the~four frequencies involved in the~resonances close to each other, and the~black hole spin then can be estimated independently of the~black hole mass.

The~efficiency of the~black hole spin determination by using the~triple frequency set ratios grows strongly with the~growing precision of the~frequency measurements. We believe that the~precision of the~planned LOFT mission will be high enough to enable application of the~multi-resonance methods for an~order of resonance up to $n=5$.

In the~LMXB neutron star sources exhibiting a~possible two resonant points in the~data of twin-peak HF~QPOs, the~RS model  can be relevant, since based on the~switching of the~twin oscillatory modes at the~outer resonant point. Of course, the~neutron star parameters given by the~RS model have to accord with those determined by fitting all the~observed data by the~frequency relations corresponding to the~oscillatory modes in action.

We expect differences between the~black hole and neutron star resonant phenomena to be attributed to the~presence of the~neutron star surface and related inhomogeneities influencing the~HF~QPOs in the~vicinity of the~star surface. A switch of the~frequency relations reflecting twin-peak HF~QPOs in neutron star systems from one pair to some other pair is not necessarily caused by the~resonant phenomena, but could be excited, e.g., by the~influence of the~neutron star magnetic field or by something else.

We can conclude that the~multi-resonant model of HF~QPOs based on the~orbital motion is able to explain a~wide range of HF~QPO phenomena observed in both black hole and neutron star X-ray binary systems, around the~supermassive galactic centre (Sgr\,A$^*$) black hole or in some other black hole systems in the~centre of active galactic nuclei, and in the~intermediate (NGC~5408~X-1) black hole systems.

A~lot of observational and theoretical research is necessary for deeper understanding of the~resonant phenomena indicated in the~black hole and neutron (quark) star systems, and we hope that the~multi-resonant model will turn out to be very useful in future research in connection with the~observational data provided by the~LOFT X-ray satellite or some similar X-ray observatory.

\begin{acknowledgements}
We would like to express our gratitude to the~Czech grants MSM~4781305903, GA\v{C}R~202/09/0772, GA\v{C}R~205/09/H033 and the~internal grants of the~Silesian University Opava FPF SGS/1,2/2010. The~authors further acknowledge the~project Supporting Integration with the~International Theoretical and Observational Research Network in Relativistic Astrophysics of Compact Objects, reg. no. CZ.1.07/2.3.00/20.0071, supported by Operational Programme \emph{Education for Competitiveness} funded by Structural Funds of the~European Union and the~state budget of the~Czech Republic.

\end{acknowledgements}


\providecommand{\uv}[1]{\glqq#1\grqq}


\Online
\appendix
\section{Simple combinational resonances of frequency pairs}\label{apendix-A}

The simple combinational resonances correspond to frequency relations of a~single orbital (epicyclic) frequency $\nu_{\alpha}$ and a~simple combinational (beat) frequency of another frequency $\nu_{\beta}$ with the~frequency $\nu_{\alpha}$, assuming $\nu_{\alpha}>\nu_{\beta}$. Considering the~pair $(\alpha,\beta)$ in the~combinations of $(\theta,r)$, $(\mathrm{K},r)$, $(\mathrm{K},\theta)$ and introducing the~new frequency ratio parameters
\begin{eqnarray}
        p^{\mathrm{I}}&=&\left(\frac{n-m}{m}\right)^2=\frac{\left(1-\sqrt{p}\right)^2}{p}\,,\\
        p^{\mathrm{II}}&=&\left(\frac{n-m}{n}\right)^2=\left(1-\sqrt{p}\right)^2\,,\\
        p^{\mathrm{III}}&=&\left(\frac{m}{n-m}\right)^2=\frac{p}{\left(\sqrt{p}-1\right)^2}\,,\\
        p^{\mathrm{IV}}&=&\left(\frac{m}{n+m}\right)^2=\frac{p}{\left(\sqrt{p}+1\right)^2}\,,\\
        p^{\mathrm{V}}&=&\left(\frac{n}{n+m}\right)^2=\frac{1}{\left(\sqrt{p}+1\right)^2}\,,\\
        p^{\mathrm{VI}}&=&\left(\frac{n-m}{n+m}\right)^2=\left(\frac{\sqrt{p}-1}{\sqrt{p}+1}\right)^2\,,
\end{eqnarray}
we arrive at the~following resonance conditions expressed in the~general form:
        \begin{enumerate}[{CS.}1.]\setlength{\itemsep}{1.5ex plus 1ex}
          \item $\displaystyle\nu_\mathrm{U}=\nu_{\alpha}+\nu_{\beta},\,\nu_\mathrm{L}=\nu_{\alpha}\,,$
\begin{equation}
              a=a^{(\alpha+\beta)/\alpha}(x,p)\equiv
              a^{\alpha/\beta}(x,p^{\mathrm{I}})\,,
\end{equation}
 \item $\displaystyle\nu_\mathrm{U}=\nu_{\alpha},\,\nu_\mathrm{L}=\nu_{\alpha}-\nu_{\beta}\,,$
\begin{equation}
              a=a^{\alpha/(\alpha-\beta)}(x,p)\equiv
              a^{\alpha/\beta}(x,p^{\mathrm{II}})\,,
\end{equation}
\item $\displaystyle\nu_\mathrm{U}=\nu_{\alpha}+\nu_{\beta},\, \nu_\mathrm{L}=\nu_{\beta}\,,$
\begin{equation}
              a=a^{(\alpha+\beta)/\beta}(x,p)\equiv
              a^{\alpha/\beta}(x,p^{\mathrm{III}})\,,
\end{equation}
\item $\displaystyle\nu_\mathrm{U}=\nu_{\alpha}-\nu_{\beta},\,\nu_\mathrm{L}=\nu_{\beta}\,,$
\begin{equation}
              a=a^{(\alpha-\beta)/\beta}(x,p)\equiv
              a^{\alpha/\beta}(x,p^{\mathrm{IV}})\,,
\end{equation}
          \item $\displaystyle\nu_\mathrm{U}=\nu_{\beta},\, \nu_\mathrm{L}=\nu_{\alpha}-\nu_{\beta}\,,$
\begin{equation}
              a=a^{\beta/(\alpha-\beta)}(x,p)\equiv
              a^{\alpha/\beta}(x,p^{\mathrm{V}})\,,
\end{equation}
         \item $\displaystyle\nu_\mathrm{U}=\nu_{\alpha}+\nu_{\beta},\,\nu_\mathrm{L}=\nu_{\alpha}-\nu_{\beta}\,,$
\begin{equation}
              a=a^{(\alpha+\beta)/(\alpha-\beta)}(x,p)\equiv
              a^{\alpha/\beta}(x,p^{\mathrm{VI}})\,.
\end{equation}
        \end{enumerate}

Simple combinational resonances of the~type CS occur at the~same radii as the~direct resonances D1, D2, D3. Therefore, it is enough to relate the~ratios of the~simple combinational and direct resonances. For example, the~frequency ratio $\nu_{\alpha}:\nu_{\beta} = 5:4$ implies $(\nu_{\alpha}+\nu_{\beta}):\nu_{\alpha} = 9:5$, $\nu_{\alpha}:(\nu_{\alpha}-\nu_{\beta}) = 5:1$, etc.

\section{Simple combinational resonances of three frequencies $\nu_{\mathrm{K}}$, $\nu_{\theta}$, $\nu_{r}$}\label{apendix-B}

We consider resonances of oscillations with one simple frequency of the~orbital frequencies and a~simple combination (beat) of the~other two frequencies:

\begin{enumerate}[{CT}1.]\setlength{\itemsep}{1.5ex plus 1ex}
    \item 
        $\displaystyle \nu_\mathrm{U}=\nu_{\mathrm{K}},\,\nu_\mathrm{L}=\nu_{\theta}-\nu_{r}\,.$

       The resonance function $a^{\mathrm{K}/(\theta-{r})}(x,p)$ is given by
    \begin{equation}
       \left(\alpha_{\theta}-\alpha_{{r}}\right)^2-2p\left(\alpha_{\theta}+\alpha_{{r}}\right)+p^2=0\,.
    \end{equation}
       In the~explicit form the~solution reads as
     \begin{eqnarray}\label{aCT1}
      \lefteqn{a=a^{\mathrm{K}/(\theta-{r})}(x,p) \equiv \sqrt{x}+\frac{1}{2\cdot 3^{5/6}}}\\
       \lefteqn{\times\left[\sqrt{\frac{A^{2/3}+B}{A^{1/3}}}-\sqrt{A^{1/3} \left(\frac{4 \sqrt{3} p
        x^{5/2}}{\sqrt{A+A^{1/3} B}}-1\right)-\frac{B}{A^{1/3}}}\,\right],}\nonumber
    \end{eqnarray}
       where
       \begin{eqnarray}
         \lefteqn{A=6 p^2 x^5+\sqrt{36 p^4 x^{10}-B^3}\,,}\\
         \lefteqn{B=3^{1/3} p x^3 \left[4+(p-4) x\right]\,.}
       \end{eqnarray}
\item 
        $\displaystyle \nu_\mathrm{U}=\nu_{\theta},\,\nu_\mathrm{L}=\nu_{\mathrm{K}}-\nu_{r}\,.$

The resonance function $a^{\theta/(\mathrm{K}-{r})}(x,p)$ is given by
\begin{equation}
        p^2\alpha_{\theta}^2-2p\alpha_{\theta}\left(1+\alpha_{{r}}\right)+\left(1-\alpha_{{r}}\right)^2=0\,.
\end{equation}
\item 
        $\displaystyle
        \nu_\mathrm{U}=\nu_{r},\,\nu_\mathrm{L}=\nu_{\mathrm{K}}-\nu_{\theta}\,.$

        The resonance function $a^{{r}/(\mathrm{K}-\theta)}(x,p)$  is given by
\begin{equation}
        \left(1-\alpha_{\theta}\right)^2-2p\alpha_{{r}}\left(1+\alpha_{\theta}\right)+p^2\alpha_{{r}}^2=0\,.
\end{equation}
\item 
        $\displaystyle
        \nu_\mathrm{U}=\nu_{\theta}+\nu_{r},\,\nu_\mathrm{L}=\nu_{\mathrm{K}}\,.$

        The resonance function $a^{(\theta+{r})/\mathrm{K}}(x,p)$ is given by
\begin{equation}
        p^2\left(\alpha_{{r}}-\alpha_{\theta}\right)^2-2p\left(\alpha_{{r}}+\alpha_{\theta}\right)+1=0\,.
\end{equation}
        The~condition $m<n<2m$ has to be satisfied.
    \item 
        $\displaystyle
        \nu_\mathrm{U}=\nu_{\mathrm{K}}+\nu_{r},\, \nu_\mathrm{L}=\nu_{\theta}\,.$

        The resonance function $a^{(\mathrm{K}+{r})/\theta}(x,p)$  is given by
\begin{equation}
        \left(\alpha_{\theta}-p\right)^2-2p\alpha_{{r}}\left(\alpha_{\theta}+p\right)+p^2\alpha_{{r}}^2=0\,.
\end{equation}
    \item 
        $\displaystyle
        \nu_\mathrm{U}=\nu_{\mathrm{K}}+\nu_{\theta},\,\nu_\mathrm{L}=\nu_{r}\,.$

        The resonance function $a^{(\mathrm{K}+\theta)/{r}}(x,p)$  is given by
        \begin{equation}
        p^2\left(\alpha_{\theta}-1\right)^2-2p\alpha_{{r}}\left(\alpha_{\theta}+1\right)+
        \alpha_{{r}}^2=0\,.
        \end{equation}
        The~condition $n>2m$ has to be satisfied.
\end{enumerate}

\begin{figure*}
\begin{center}
\includegraphics[width=.91\hsize]{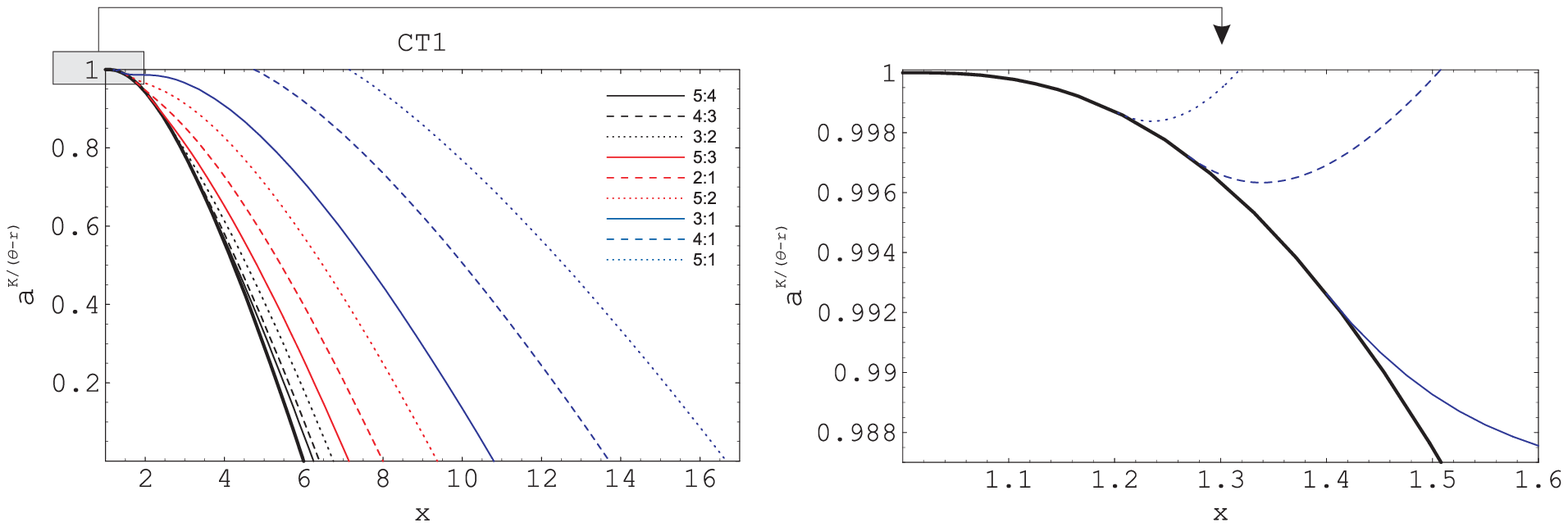}
\end{center}
\begin{center}
\begin{minipage}[h]{.48\hsize}
\includegraphics[width=\hsize]{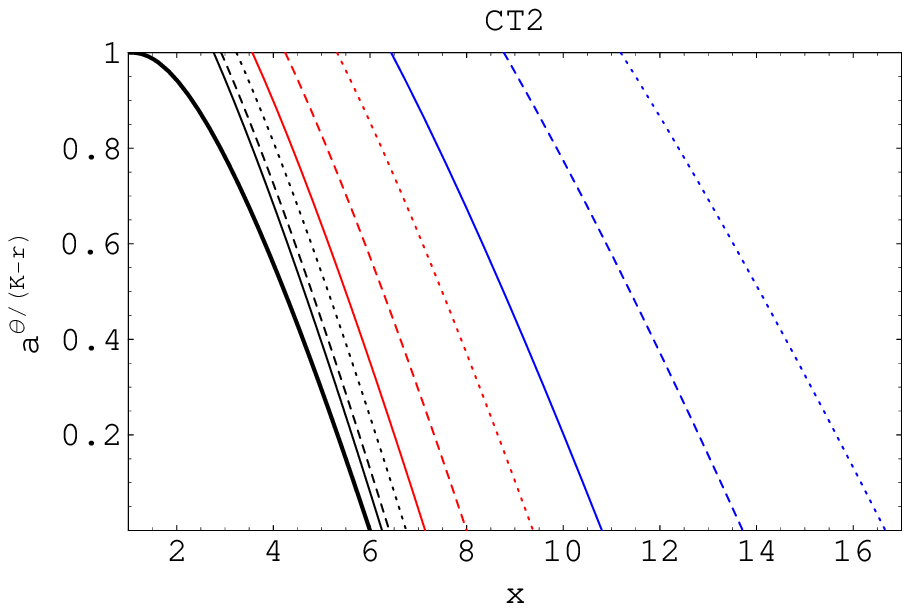}
\end{minipage}
\begin{minipage}[h]{.48\hsize}
\includegraphics[width=\hsize]{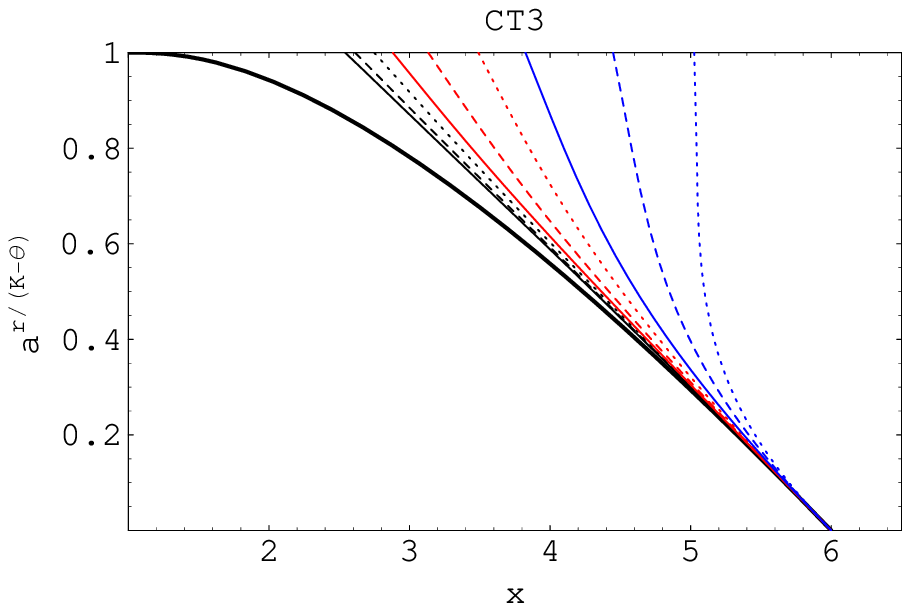}
\end{minipage}
\end{center}
\begin{center}
\begin{minipage}[h]{.48\hsize}
\includegraphics[width=\hsize]{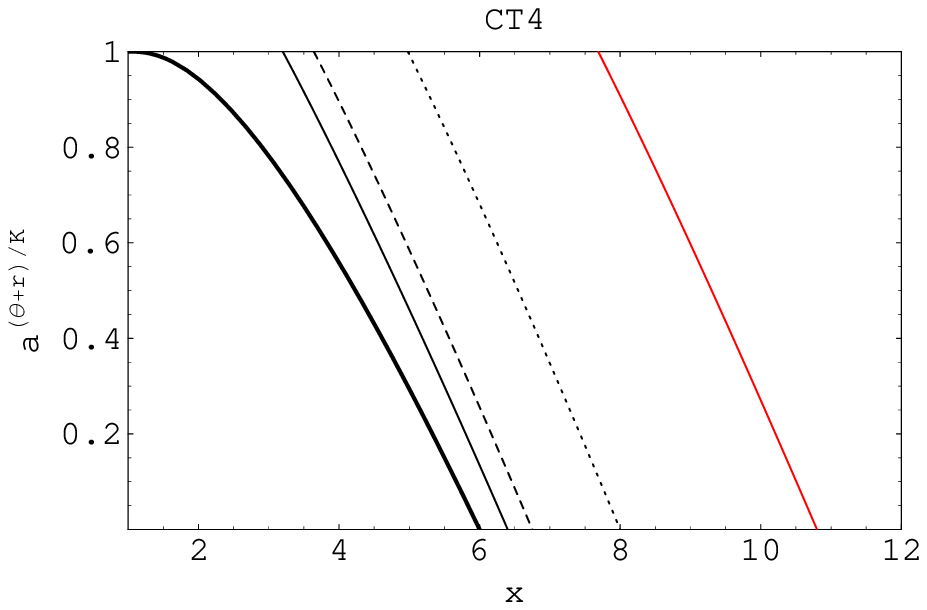}
\end{minipage}
\begin{minipage}[h]{.48\hsize}
\includegraphics[width=\hsize]{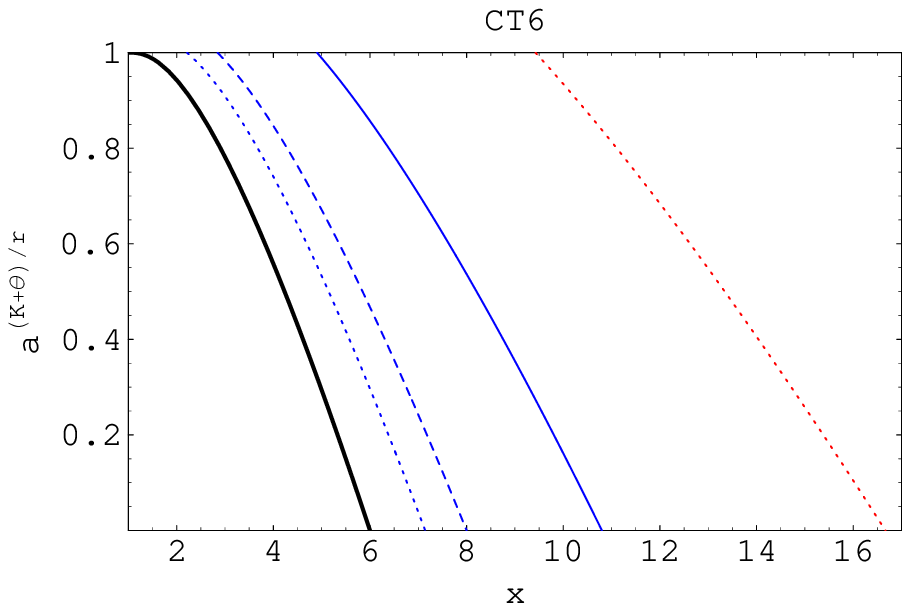}
\end{minipage}
\end{center}
\begin{center}
\includegraphics[width=.91\hsize]{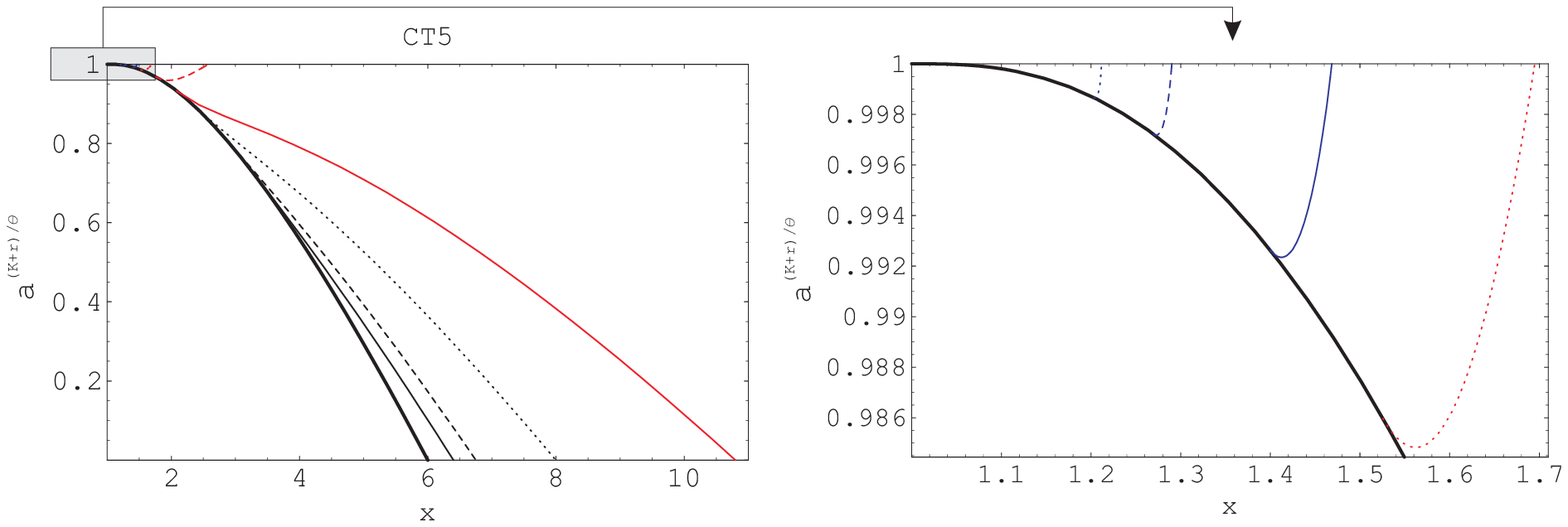}
\end{center}
\caption{The~spin resonance functions $a^{\nu_\mathrm{U}/\nu_\mathrm{L}}(x,p)$ for the~simple combinational resonances CT1\,--\,CT6 and the~frequency ratios $n$\,:\,$m$\,=\,5\,:\,4 (black
solid line), 4\,:\,3 (black dashed line), 3\,:\,2 (black dotted
line), 5\,:\,3 (red solid line), 2\,:\,1 (red dashed line),
5\,:\,2 (red dotted line), 3\,:\,1 (blue solid line), 4\,:\,1
(blue dashed line), 5\,:\,1 (blue dotted line). Black thick line
represents $a_{\mathrm{ms}}(x)$, which implicitly determines the~radius of the~marginally stable orbit $x_{\mathrm{ms}}$.
\label{aCT}}
\end{figure*}

Except for the~case CT1, we give the~resonance functions in an~implicit graphical form, because the~functions are too complex and too long to be written explicitly. The implicit resonance functions are given in Fig.~\ref{aCT}.

The~implicit resonance conditions are polynomials of the~fourth order in the~spin $a$ in all six cases. Only one of the~possible solutions of the~resonance condition is physically relevant. Now the~resonance functions  exhibit more complex behaviour in comparison with the~direct resonance functions. Along with those that are monotonous and do not cross the~stability line, we have found monotonous functions crossing the~stability line at the~stability points $a^{\nu_\mathrm{U}/\nu_\mathrm{L}}_{\mathrm{ms}}(p)$ and non-monotonous functions crossing the~stability line. In some cases the~non-monotonous resonance lines appear to be (seemingly) discontinuous because of entering the~region with spin $a>1$, corresponding to naked-singularity spacetimes.

\section{Double-beat combinational resonances}\label{apendix-C}

We introduce 13 types of the~double-beat combinational resonances:

\noindent
DC1.    $\,\displaystyle
            \frac{\nu_{\mathrm{K}}-\nu_{r}}{\nu_{\theta}-\nu_{r}}\,,\quad$
DC2.    $\,\displaystyle
        \frac{\nu_{\mathrm{K}}+\nu_{r}}{\nu_{\theta}-\nu_{r}}\,,\quad$
DC3.    $\,\displaystyle
        \frac{\nu_{\mathrm{K}}-\nu_{\theta}}{\nu_{\theta}-\nu_{r}}\,,$\\
DC4.    $\,\displaystyle
        \frac{\nu_{\mathrm{K}}+\nu_{\theta}}{\nu_{\theta}-\nu_{r}}\,,\quad$
DC5.    $\,\displaystyle
        \frac{\nu_{\mathrm{K}}-\nu_{r}}{\nu_{\theta}+\nu_{r}}\,,\quad$
DC6.    $\,\displaystyle
        \frac{\nu_{\mathrm{K}}+\nu_{r}}{\nu_{\theta}+\nu_{r}}\,,$\\
DC7.    $\,\displaystyle
        \frac{\nu_{\mathrm{K}}-\nu_{\theta}}{\nu_{\theta}+\nu_{r}}\,,\quad$
DC8.    $\,\displaystyle
        \frac{\nu_{\mathrm{K}}+\nu_{\theta}}{\nu_{\theta}+\nu_{r}}\,,\quad$
DC9.    $\,\displaystyle
        \frac{\nu_{\mathrm{K}}-\nu_{r}}{\nu_{\mathrm{K}}-\nu_{\theta}}\,,$\\
DC10.   $\,\displaystyle
        \frac{\nu_{\mathrm{K}}+\nu_{r}}{\nu_{\mathrm{K}}-\nu_{\theta}}\,,\quad$
DC11.   $\,\displaystyle
        \frac{\nu_{\theta}-\nu_{r}}{\nu_{\mathrm{K}}-\nu_{\theta}}\,,$\\
DC12.   $\,\displaystyle
        \frac{\nu_{\theta}+\nu_{r}}{\nu_{\mathrm{K}}-\nu_{\theta}}\,,\quad$
DC13.   $\,\displaystyle
        \frac{\nu_{\theta}+\nu_{r}}{\nu_{\mathrm{K}}-\nu_{r}}\,.$

We present the~implicit graphical form of the~solution of the~resonance conditions related to the~double-beat combinational resonances DC1\,--\,DC13. We also give the~overview of the~allowed ranges of the~direct and combinational resonances limited by the~stability points that are explicitly given and by the~local extrema of the~resonance functions in Table~\ref{TabLimitniSpiny}.

\begin{figure*}
\begin{center}
\includegraphics[width=.89\hsize]{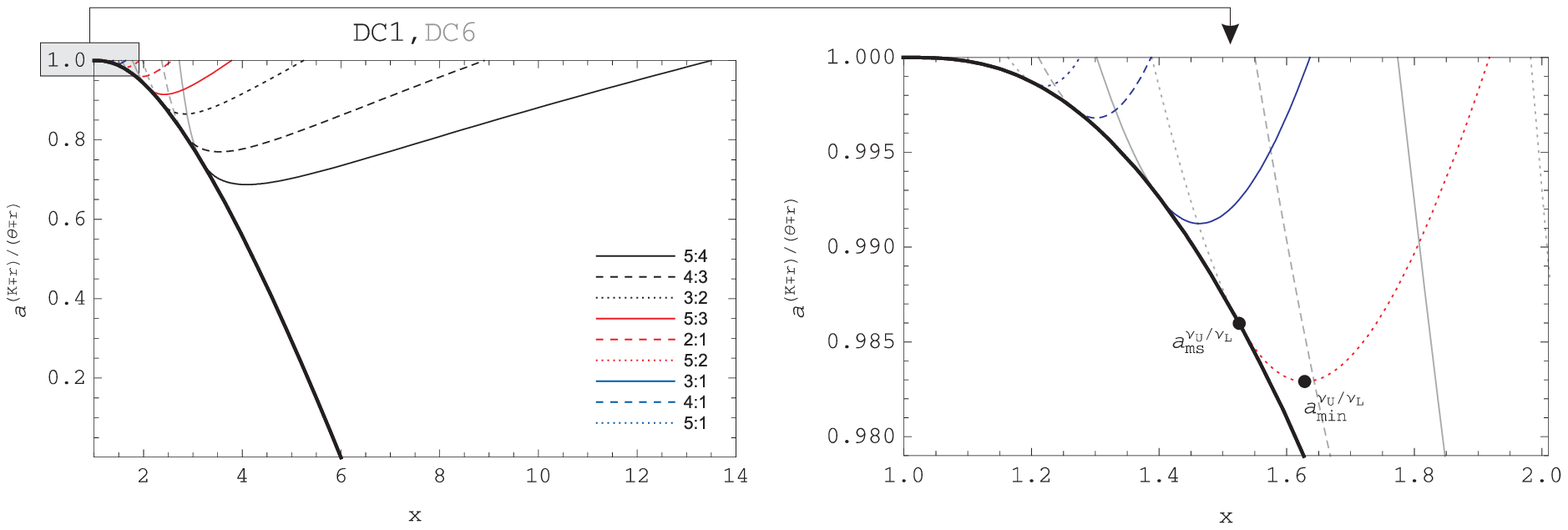}
\end{center}
\begin{center}
\includegraphics[width=.89\hsize]{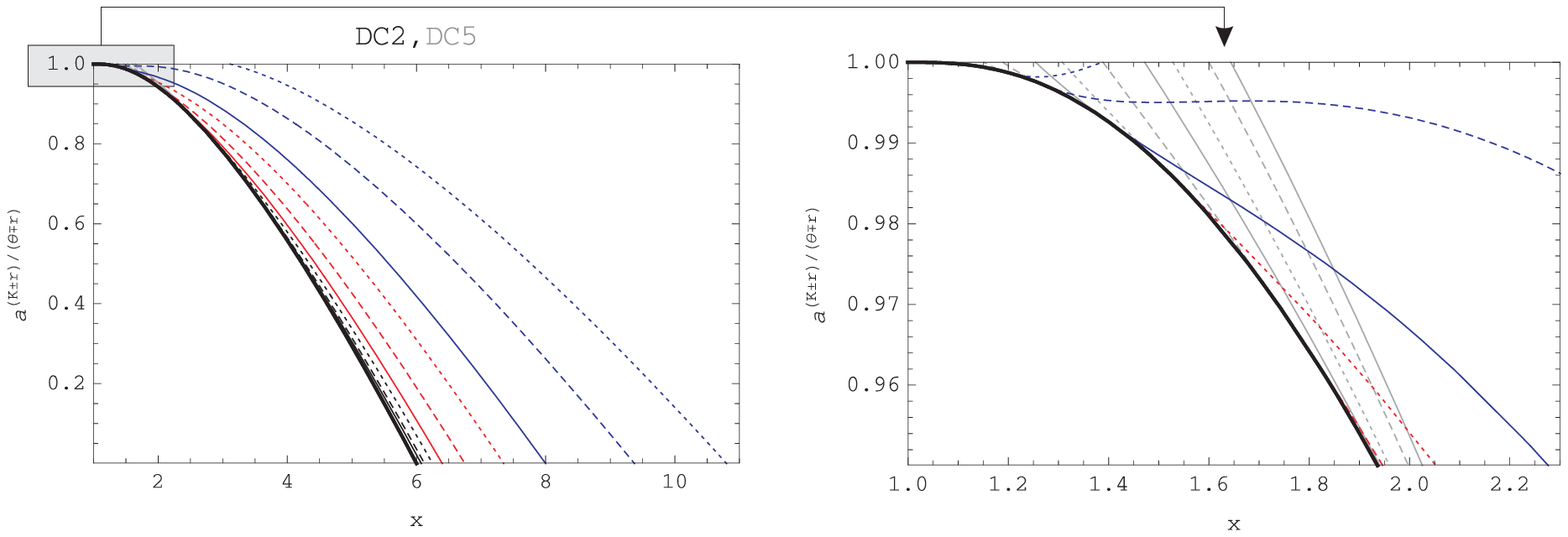}
\end{center}
\begin{center}
\begin{minipage}[h]{.44\hsize}
\includegraphics[width=\hsize]{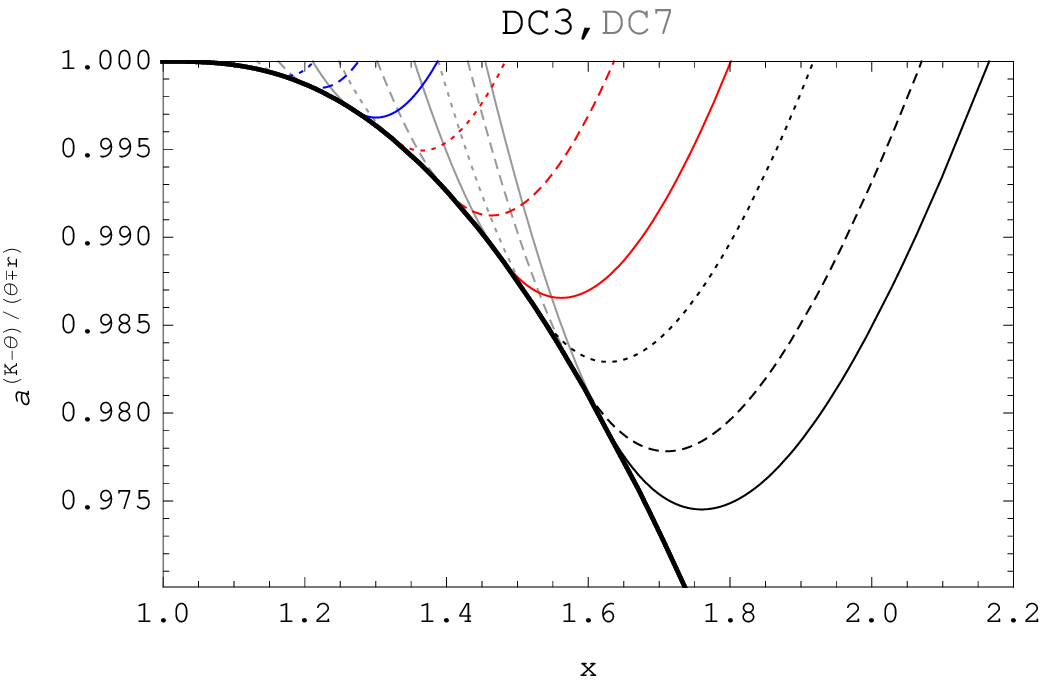}
\end{minipage}
\hspace{4ex}
\begin{minipage}[h]{.41\hsize}
\includegraphics[width=\hsize]{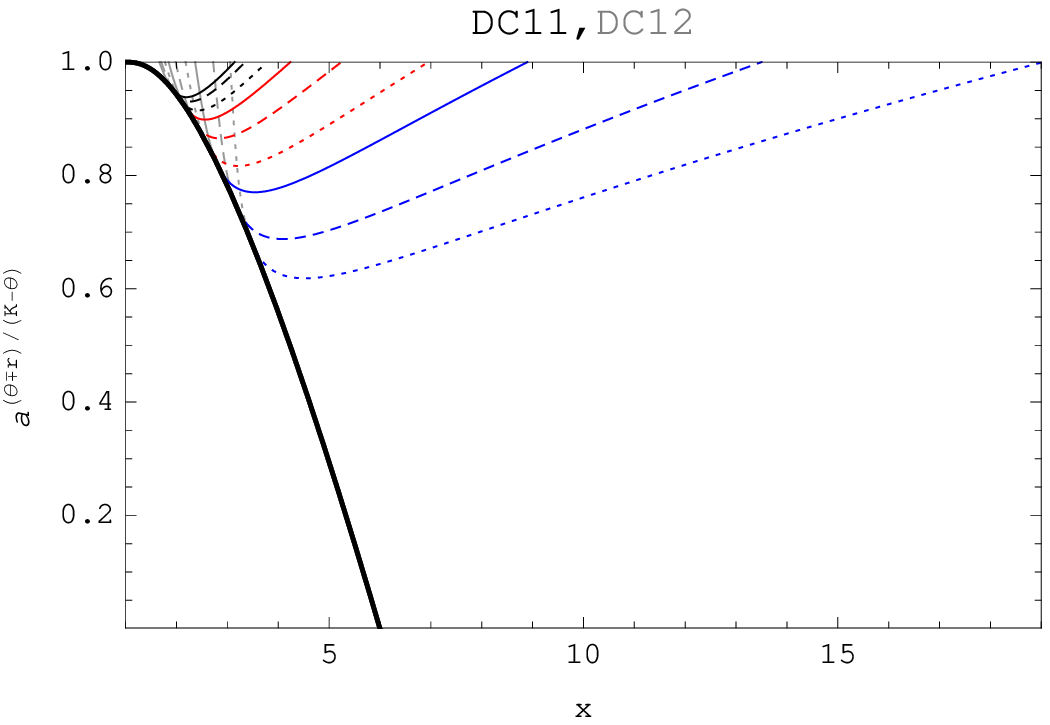}
\end{minipage}
\end{center}
\begin{center}
\includegraphics[width=.89\hsize]{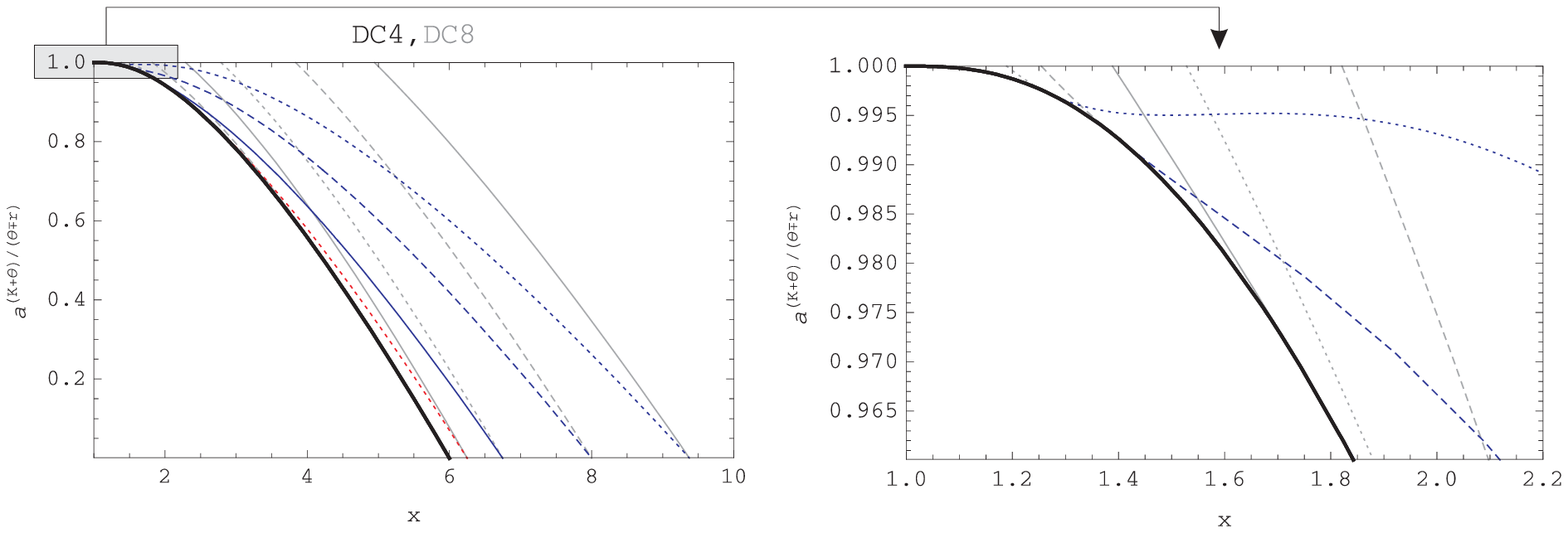}
\end{center}
\caption{The~spin resonance functions $a^{\nu_\mathrm{U}/\nu_\mathrm{L}}(x,p)$ for the~double combinational resonances of the~types DC1 -- DC4, DC11 and the~frequency ratios $n$\,:\,$m$\,=\,5\,:\,4 (black solid line), 4\,:\,3
(black dashed line), 3\,:\,2 (black dotted line), 5\,:\,3 (red
solid line), 2\,:\,1 (red dashed line), 5\,:\,2 (red dotted line),
3\,:\,1 (blue solid line), 4\,:\,1 (blue dashed line), 5\,:\,1
(blue dotted line). The~grey lines represent the~solutions of the~same resonance equations that are relevant for the~corresponding types of resonances (DC5 -- DC8, DC12). The~related solutions are separated by the~line of marginally stable orbits $a_{\mathrm{ms}}(x)$ (black thick line), they have a~common stability point $a^{\nu_\mathrm{U}/\nu_\mathrm{L}}_{\mathrm{ms}}(p)$.\label{aDC1}}
\end{figure*}

\begin{figure*}
\begin{center}
\includegraphics[width=.89\hsize]{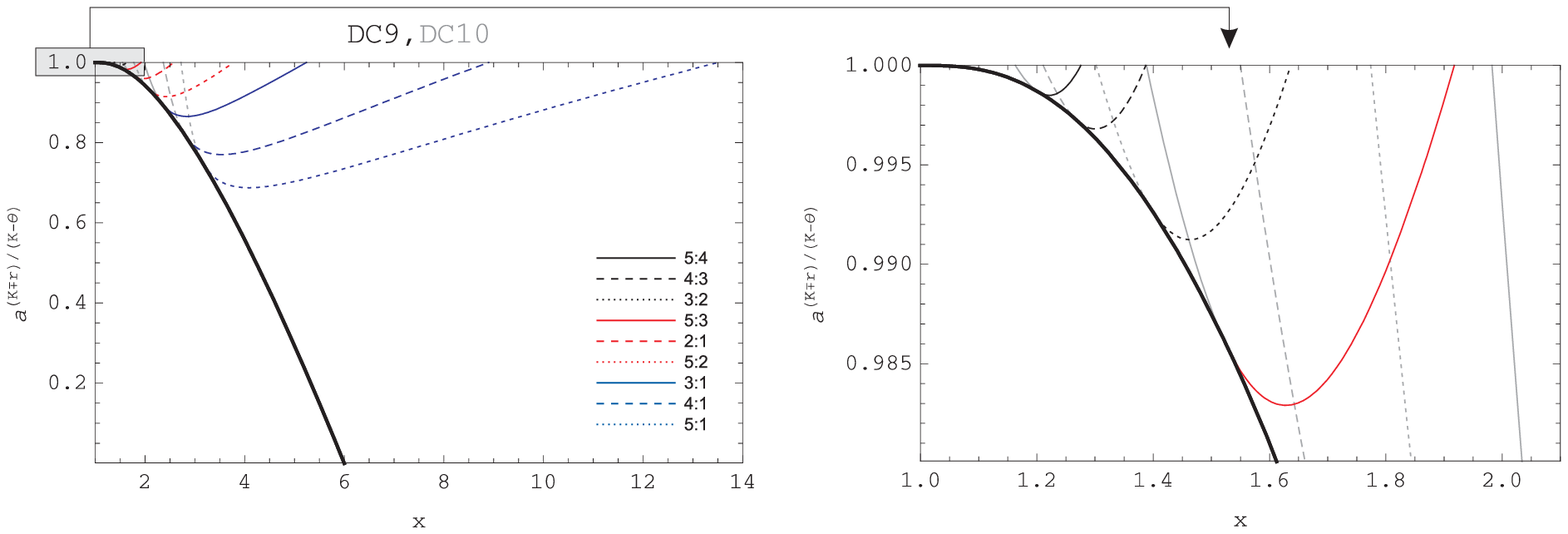}
\end{center}
\begin{center}
\begin{minipage}[h]{.43\hsize}
\includegraphics[width=\hsize]{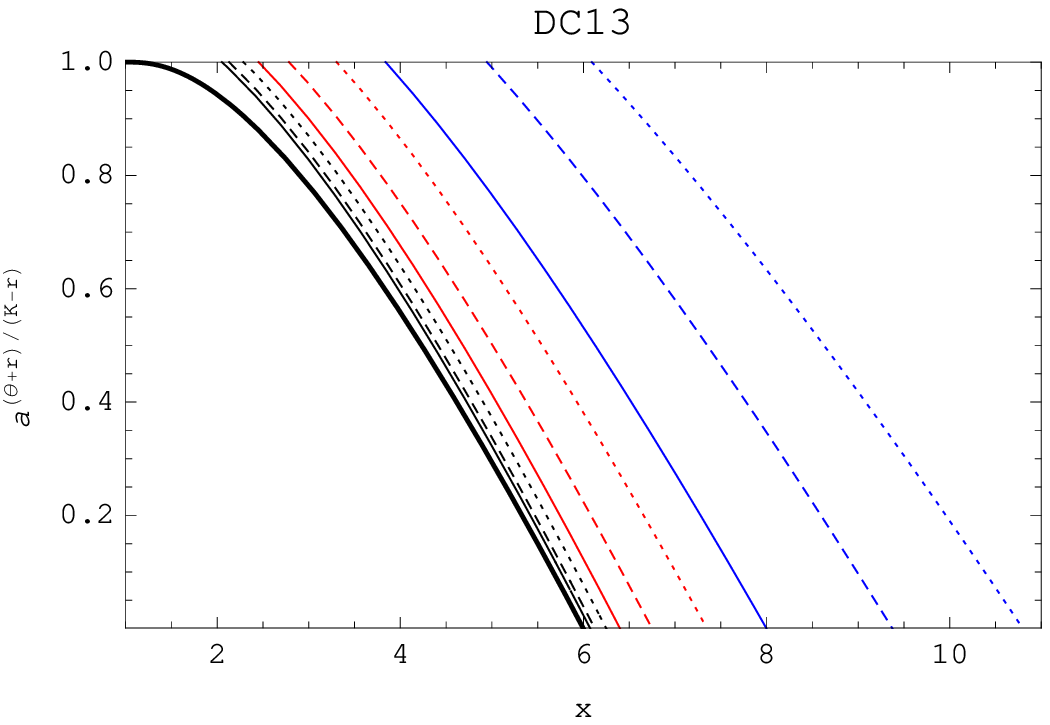}
\end{minipage}
\hspace{4ex}
\begin{minipage}[h]{.43\hsize}
\includegraphics[width=\hsize]{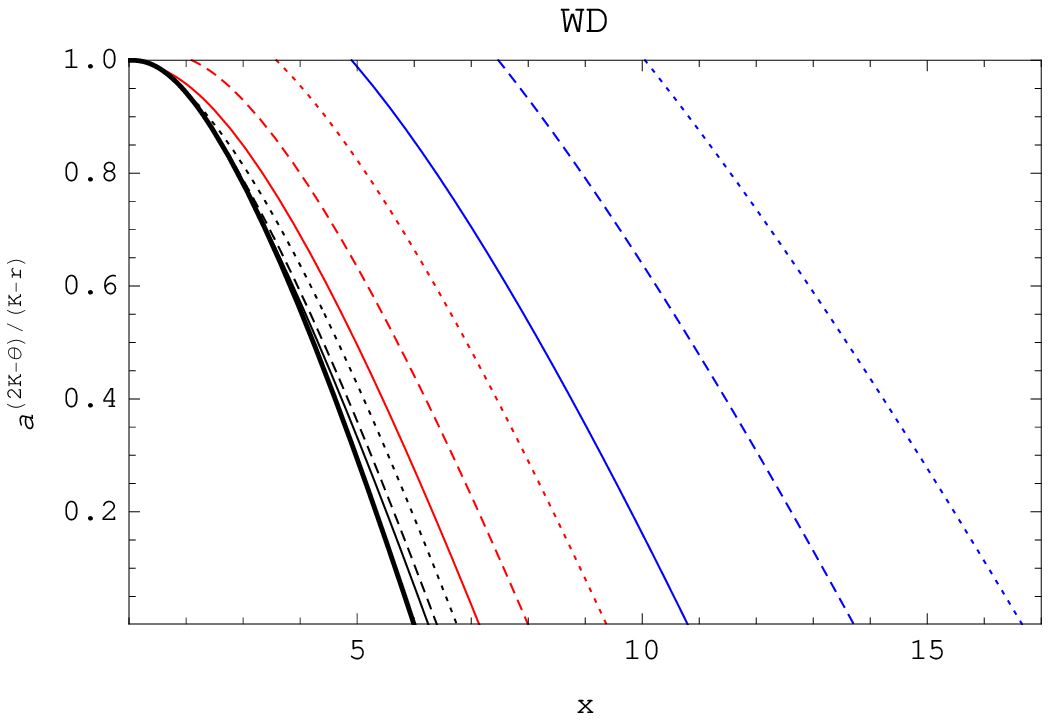}
\end{minipage}
\end{center}
\caption{The~spin resonance functions $a^{\nu_\mathrm{U}/\nu_\mathrm{L}}(x,p)$ for the~double combinational resonances of the~types DC9, DC10, DC13 and the~warped disc oscillation model (WD) for the~frequency ratios $n$\,:\,$m$\,=\,5\,:\,4 (black
solid line), 4\,:\,3 (black dashed line), 3\,:\,2 (black dotted
line), 5\,:\,3 (red solid line), 2\,:\,1 (red dashed line),
5\,:\,2 (red dotted line), 3\,:\,1 (blue solid line), 4\,:\,1
(blue dashed line), 5\,:\,1 (blue dotted line). Black thick line
represents $a_{\mathrm{ms}}(x)$, which implicitly determines the~radius of the~marginally stable orbit
$x_{\mathrm{ms}}$. The~grey lines are related to the~resonance DC10 -- for details see Fig.~\ref{aDC1}\label{aDC2-WD}.}
\end{figure*}

The results obtained for the~relevant resonance functions are given in Figs.~\ref{aDC1} and \ref{aDC2-WD}.
In order to keep the~discussion straightforward, we assume the~frequency ratio $n/m > 1$ in all 13 cases. Therefore, we treat the~frequency relations allowing for ratios higher and lower than 1 as separate (inverse) cases. For each relation of the~beat frequencies, we give the~relation between $\nu_{\theta}/\nu_{r}$ and $\nu_{\mathrm{K}}/\nu_{r}$ in terms of the~ratio $n/m$ that uniquely determines the~spin ``resonance'' function $a^{\nu_{\mathrm{U}}/\nu_{\mathrm{L}}}(x,p)$. Then we determine ``resonance equations'' relating the~functions $\alpha_{\theta}(x,a)$, $\alpha_{{r}}(x,a)$ to the~frequency ratio $p$ that are used for numerical determination of the~spin ``resonance'' functions; however, the~``resonance equations'' do not give unique results and usually determine two spin ``resonance'' functions that must be separated during the~numerical procedure. The~separation is given by the~stability line, i.e., the~function $a_{\mathrm{ms}}(x)$ giving the~marginally stable orbits. 
The resonance conditions are given in the~following form.
\begin{enumerate}[{DC}1.]\setlength{\itemsep}{1.5ex plus 1ex}
 \item 
        $\displaystyle
        \frac{\nu_{\theta}}{\nu_{r}}=1 + \frac{m}{n}\left(\frac{\nu_{\mathrm{K}}}{\nu_{r}}-1\right)\,,$
        \begin{eqnarray}
        \lefteqn{a^{\left(\mathrm{K}-{r}\right)/\left(\theta-{r}\right)}(x,p) \mbox{ is solution of the~equation}}\nonumber\\
         \lefteqn{\left(\alpha_{\theta}-p\right)^2-2\left(\alpha_{\theta}+p\right)\left(1-\sqrt{p}\right)^2\alpha_{{r}}+\left(1-\sqrt{p}\right)^4\alpha_{{r}}^2=0\,.}\nonumber
        \end{eqnarray}
\item 
        $\displaystyle
                \frac{\nu_{\theta}}{\nu_{r}}=1 + \frac{m}{n}\left(\frac{\nu_{\mathrm{K}}}{\nu_{r}}+1\right)\,,$
        \begin{eqnarray}
        \lefteqn{a^{\left(\mathrm{K}+{r}\right)/\left(\theta-{r}\right)}(x,p) \mbox { is solution of the~equation}}\nonumber\\
        \lefteqn{
        \left(\alpha_{\theta}-p\right)^2-2\left(\alpha_{\theta}+p\right)\left(1+\sqrt{p}\right)^2\alpha_{{r}}+\left(1+\sqrt{p}\right)^4\alpha_{{r}}^2=0\,.}\nonumber
        \end{eqnarray}
\item 
        $\displaystyle
                \frac{\nu_{\theta}}{\nu_{r}}=\left(1 + \frac{m}{n}\frac{\nu_{\mathrm{K}}}{\nu_{r}}\right)
        \left(1+\frac{m}{n}\right)^{-1}\,,$
        \begin{eqnarray}
        \lefteqn{a^{\left(\mathrm{K}-\theta\right)/\left(\theta-{r}\right)}(x,p) \mbox{ is solution of the~equation}}\nonumber\\
        \lefteqn{
        \alpha_{\theta}^2\left(1+\sqrt{p}\right)^4-2\alpha_{\theta}\left(1+\sqrt{p}\right)^2\left(\alpha_{{r}}+p\right)+
        \left(\alpha_{{r}}-p\right)^2=0\,.}\nonumber
        \end{eqnarray}
\item 
        $\displaystyle
                \frac{\nu_{\theta}}{\nu_{r}}=\left(1 + \frac{m}{n}\frac{\nu_{\mathrm{K}}}{\nu_{r}}\right)
        \left(1-\frac{m}{n}\right)^{-1}\,,$
        \begin{eqnarray}
        \lefteqn{a^{\left(\mathrm{K}+\theta\right)/\left(\theta-{r}\right)}(x,p) \mbox{ is solution of the~equation}}\nonumber\\
        \lefteqn{
        \alpha_{\theta}^2\left(1-\sqrt{p}\right)^4-2\alpha_{\theta}\left(1-\sqrt{p}\right)^2\left(\alpha_{{r}}+p\right)
        +\left(\alpha_{{r}}-p\right)^2=0\,.}\nonumber
        \end{eqnarray}
\item 
        $\displaystyle
                \frac{\nu_{\theta}}{\nu_{r}}= -1 + \frac{m}{n}\left(\frac{\nu_{\mathrm{K}}}{\nu_{r}}-1\right)\,,$
        \begin{eqnarray}
        \lefteqn{a^{\left(\mathrm{K}-{r}\right)/\left(\theta+{r}\right)}(x,p) \mbox{ is solution of the~equation}}\nonumber\\
        \lefteqn{
        \left(\alpha_{\theta}-p\right)^2-2\left(\alpha_{\theta}+p\right)\left(1+\sqrt{p}\right)^2
        \alpha_{{r}}+\left(1+\sqrt{p}\right)^4\alpha_{{r}}^2=0\,.}\nonumber
        \end{eqnarray}
\item 
        $\displaystyle
                \frac{\nu_{\theta}}{\nu_{r}}=-1 + \frac{m}{n}\left(\frac{\nu_{\mathrm{K}}}{\nu_{r}}+1\right)\,,$
        \begin{eqnarray}
        \lefteqn{a^{\left(\mathrm{K}+{r}\right)/\left(\theta+{r}\right)}(x,p) \mbox{ is solution of the~equation}}\nonumber\\
        \lefteqn{
        \left(\alpha_{\theta}-p\right)^2-2\left(\alpha_{\theta}+p\right)\left(1-\sqrt{p}\right)^2
        \alpha_{{r}}+\left(1-\sqrt{p}\right)^4\alpha_{{r}}^2=0\,.}\nonumber
        \end{eqnarray}
\item 
         $\displaystyle
                \frac{\nu_{\theta}}{\nu_{r}}=\left(-1 + \frac{m}{n}\frac{\nu_{\mathrm{K}}}{\nu_{r}}\right)
        \left(1+\frac{m}{n}\right)^{-1}\,,$
        \begin{eqnarray}
        \lefteqn{a^{\left(\mathrm{K}-\theta\right)/\left(\theta+{r}\right)}(x,p) \mbox{ is solution of the~equation}}\nonumber\\
        \lefteqn{
        \alpha_{\theta}^2\left(1+\sqrt{p}\right)^4-2\alpha_{\theta}\left(1+\sqrt{p}\right)^2\left(\alpha_{{r}}+p\right)+
        \left(\alpha_{{r}}-p\right)^2=0\,.}\nonumber
        \end{eqnarray}
\item 
         $\displaystyle
                \frac{\nu_{\theta}}{\nu_{r}}=\left(-1 + \frac{m}{n}\frac{\nu_{\mathrm{K}}}{\nu_{r}}\right)
        \left(1-\frac{m}{n}\right)^{-1}\,,$
        \begin{eqnarray}
        \lefteqn{a^{\left(\mathrm{K}+\theta\right)/\left(\theta+{r}\right)}(x,p) \mbox{ is solution of the~equation}}\nonumber\\
        \lefteqn{
        \alpha_{\theta}^2\left(1-\sqrt{p}\right)^4-2\alpha_{\theta}\left(1-\sqrt{p}\right)^2
        \left(\alpha_{{r}}+p\right)+\left(\alpha_{{r}}-p\right)^2=0\,.}\nonumber
        \end{eqnarray}
\item 
         $\displaystyle
                \frac{\nu_{\theta}}{\nu_{r}}=\left(1-\frac{m}{n}\right)
        \frac{\nu_{\mathrm{K}}}{\nu_{r}}+ \frac{m}{n}\,,$
        \begin{eqnarray}
        \lefteqn{a^{\left(\mathrm{K}-{r}\right)/\left(\mathrm{K}-\theta\right)}(x,p) \mbox{ is solution of the~equation}}\nonumber\\
        \lefteqn{
        \left[\alpha_{\theta}-\left(1-\sqrt{p}\right)^2\right]^2-2p\left[\alpha_{\theta}+\left(1-\sqrt{p}\right)^2\right]
        \alpha_{{r}}+p^2\alpha_{{r}}^2=0\,.}\nonumber
        \end{eqnarray}
\item 
       $\displaystyle
                \frac{\nu_{\theta}}{\nu_{r}}=\left(1-\frac{m}{n}\right)
        \frac{\nu_{\mathrm{K}}}{\nu_{r}}- \frac{m}{n}\,,$
        \begin{eqnarray}
        \lefteqn{a^{\left(\mathrm{K}+{r}\right)/\left(\mathrm{K}-\theta\right)}(x,p) \mbox{ is solution of the~equation}}\nonumber\\
        \lefteqn{
        \left[\alpha_{\theta}-\left(1-\sqrt{p}\right)^2\right]^2-
        2p\left[\alpha_{\theta}+\left(1-\sqrt{p}\right)^2\right]
        \alpha_{{r}}+p^2\alpha_{{r}}^2=0\,.}\nonumber
        \end{eqnarray}
\item 
        $\displaystyle
        \frac{\nu_{\theta}-\nu_{r}}{\nu_{\mathrm{K}}-\nu_{\theta}}=\frac{n}{m}
        \quad \Rightarrow \quad
        \frac{\nu_{\theta}}{\nu_{r}}=\left(\frac{m}{n}+\frac{\nu_{\mathrm{K}}}{\nu_{r}}\right)
        \left(1 + \frac{m}{n}\right)^{-1}\,,$
        \begin{eqnarray}
        \lefteqn{a^{\left(\theta-{r}\right)/\left(\mathrm{K}-\theta\right)}(x,p) \mbox{ is solution of the~equation}}\nonumber\\
        \lefteqn{
        \alpha_{\theta}^2\left(1+\sqrt{p}\right)^4-2\alpha_{\theta}\left(1+\sqrt{p}\right)^2
        \left(1+p\alpha_{{r}}\right)+\left(1-p\alpha_{{r}}\right)^2=0\,.}\nonumber
        \end{eqnarray}
\item 
        $\displaystyle
       \frac{\nu_{\theta}}{\nu_{r}}=\left(-\frac{m}{n}+\frac{\nu_{\mathrm{K}}}{\nu_{r}}\right)
        \left(1 + \frac{m}{n}\right)^{-1}\,,$
        \begin{eqnarray}
        \lefteqn{a^{\left(\theta+{r}\right)/\left(\mathrm{K}-\theta\right)}(x,p) \mbox{ is solution of the~equation}}\nonumber\\
        \lefteqn{
        \alpha_{\theta}^2\left(1+\sqrt{p}\right)^4-
        2\alpha_{\theta}\left(1+\sqrt{p}\right)^2\left(1+p\alpha_{{r}}\right)+\left(1-p\alpha_{{r}}\right)^2
        =0\,.}\nonumber
        \end{eqnarray}
\item 
        $\displaystyle
                \frac{\nu_{\theta}}{\nu_{r}}=\frac{m}{n}
        \left(\frac{\nu_{\mathrm{K}}}{\nu_{r}} - 1\right) - 1\,,$
        \begin{eqnarray}
        \lefteqn{a^{\left(\theta+{r}\right)/\left(\mathrm{K}-{r}\right)}(x,p) \mbox{ is solution of the~equation}}\nonumber\\
        \lefteqn{
        \left(\alpha_{\theta}-p\right)^2-2\left(\alpha_{\theta}+p\right)\left(1+\sqrt{p}\right)^2
        \alpha_{{r}}+\left(1+\sqrt{p}\right)^4\alpha_{{r}}^2=0\,.}\nonumber
        \end{eqnarray}
\end{enumerate}

\begin{table*}[!]
\caption{\label{TabLimitniSpiny}Allowed ranges of the~Kerr geometry spin for the~direct orbital resonance  model D3 and the~combinational resonances of the~types CT1, CT5, DC1, DC2, DC5, DC6 for given frequency ratio $n\!:\!m$. The~special values of spin $a^{\nu_\mathrm{U}/\nu_\mathrm{L}}_{\mathrm{ms}}(p)$ represent the~limiting (minimum or maximum) values of spin for given frequency ratio $n\!:\!m$ and given type of resonance, along with the~minima of non-monotonic resonance lines $a^{\nu_\mathrm{U}/\nu_\mathrm{L}}_{\mathrm{min}}(p)$ ($a^{\nu_\mathrm{U}/\nu_\mathrm{L}}_{\mathrm{min}}(p)\lesssim a^{\nu_\mathrm{U}/\nu_\mathrm{L}}_{\mathrm{ms}}(p)$, see Figs.~\ref{aCT},~\ref{aDC1}).}
\begin{center}
\small
\begin{tabular}{ccccccccc}
\hline\hline
 $n\!:\!m$ & $a_{\mathrm{ms}}^{\nu_\mathrm{U}/\nu_\mathrm{L}}(n\!:\!m)$ &
    \rule[-2.5ex]{0sp}{6ex}
$\displaystyle \frac{\nu_{\mathrm{K}}}{\nu_{\theta}}$ &
$\displaystyle \frac{\nu_{\mathrm{K}}}{\nu_{\theta}-\nu_{r}}$ &
$\displaystyle \frac{\nu_{\mathrm{K}}+\nu_{r}}{\nu_{\theta}}$ &
$\displaystyle \frac{\nu_{\mathrm{K}}-\nu_{r}}{\nu_{\theta}-\nu_{r}}$ &
$\displaystyle \frac{\nu_{\mathrm{K}}+\nu_{r}}{\nu_{\theta}-\nu_{r}}$ &
$\displaystyle \frac{\nu_{\mathrm{K}}-\nu_{r}}{\nu_{\theta}+\nu_{r}}$ &
$\displaystyle \frac{\nu_{\mathrm{K}}+\nu_{r}}{\nu_{\theta}+\nu_{r}}$ \\
 \hline
  $5:4$ & $0.744446$ & $a_{\mathrm{ms}}^{\nu_\mathrm{U}/\nu_\mathrm{L}}\leq a \leq 1 $ & $0 \leq a \leq a_{\mathrm{ms}}^{\nu_\mathrm{U}/\nu_\mathrm{L}}$ & $0 \leq a \leq a_{\mathrm{ms}}^{\nu_\mathrm{U}/\nu_\mathrm{L}}$ & $a_{\mathrm{min}}^{\nu_\mathrm{U}/\nu_\mathrm{L}}\leq a \leq 1 $ & $0 \leq a \leq a_{\mathrm{ms}}^{\nu_\mathrm{U}/\nu_\mathrm{L}}$ & $a_{\mathrm{ms}}^{\nu_\mathrm{U}/\nu_\mathrm{L}}\leq a \leq 1 $ & $a_{\mathrm{ms}}^{\nu_\mathrm{U}/\nu_\mathrm{L}}\leq a \leq 1 $\\
  $4:3$ & $0.815817$ & $a_{\mathrm{ms}}^{\nu_\mathrm{U}/\nu_\mathrm{L}}\leq a \leq 1 $ & $0 \leq a \leq a_{\mathrm{ms}}^{\nu_\mathrm{U}/\nu_\mathrm{L}}$ & $0 \leq a \leq a_{\mathrm{ms}}^{\nu_\mathrm{U}/\nu_\mathrm{L}}$ & $a_{\mathrm{min}}^{\nu_\mathrm{U}/\nu_\mathrm{L}}\leq a \leq 1 $ & $0 \leq a \leq a_{\mathrm{ms}}^{\nu_\mathrm{U}/\nu_\mathrm{L}}$ & $a_{\mathrm{ms}}^{\nu_\mathrm{U}/\nu_\mathrm{L}}\leq a \leq 1 $ & $a_{\mathrm{ms}}^{\nu_\mathrm{U}/\nu_\mathrm{L}}\leq a \leq 1 $\\
  $3:2$ & $0.893977$ & $a_{\mathrm{ms}}^{\nu_\mathrm{U}/\nu_\mathrm{L}}\leq a \leq 1 $ & $0 \leq a \leq a_{\mathrm{ms}}^{\nu_\mathrm{U}/\nu_\mathrm{L}}$ & $0 \leq a \leq a_{\mathrm{ms}}^{\nu_\mathrm{U}/\nu_\mathrm{L}}$ & $a_{\mathrm{min}}^{\nu_\mathrm{U}/\nu_\mathrm{L}}\leq a \leq 1 $ & $0 \leq a \leq a_{\mathrm{ms}}^{\nu_\mathrm{U}/\nu_\mathrm{L}}$ & $a_{\mathrm{ms}}^{\nu_\mathrm{U}/\nu_\mathrm{L}}\leq a \leq 1 $ & $a_{\mathrm{ms}}^{\nu_\mathrm{U}/\nu_\mathrm{L}}\leq a \leq 1 $\\
  $5:3$ & $0.932874$ & $a_{\mathrm{ms}}^{\nu_\mathrm{U}/\nu_\mathrm{L}}\leq a \leq 1 $ & $0 \leq a \leq a_{\mathrm{ms}}^{\nu_\mathrm{U}/\nu_\mathrm{L}}$ & $0 \leq a \leq a_{\mathrm{ms}}^{\nu_\mathrm{U}/\nu_\mathrm{L}}$ & $a_{\mathrm{min}}^{\nu_\mathrm{U}/\nu_\mathrm{L}}\leq a \leq 1 $ & $0 \leq a \leq a_{\mathrm{ms}}^{\nu_\mathrm{U}/\nu_\mathrm{L}}$ & $a_{\mathrm{ms}}^{\nu_\mathrm{U}/\nu_\mathrm{L}}\leq a \leq 1 $ & $a_{\mathrm{ms}}^{\nu_\mathrm{U}/\nu_\mathrm{L}}\leq a \leq 1 $\\
  $2:1$ & $0.967775$ & $a_{\mathrm{ms}}^{\nu_\mathrm{U}/\nu_\mathrm{L}}\leq a \leq 1 $ & $0 \leq a \leq a_{\mathrm{ms}}^{\nu_\mathrm{U}/\nu_\mathrm{L}}$ & $a_{\mathrm{min}}^{\nu_\mathrm{U}/\nu_\mathrm{L}}\leq a \leq 1 $ & $a_{\mathrm{min}}^{\nu_\mathrm{U}/\nu_\mathrm{L}}\leq a \leq 1 $ & $0 \leq a \leq a_{\mathrm{ms}}^{\nu_\mathrm{U}/\nu_\mathrm{L}}$ & $a_{\mathrm{ms}}^{\nu_\mathrm{U}/\nu_\mathrm{L}}\leq a \leq 1 $ & $a_{\mathrm{ms}}^{\nu_\mathrm{U}/\nu_\mathrm{L}}\leq a \leq 1 $\\
  $5:2$ & $0.985975$ & $a_{\mathrm{ms}}^{\nu_\mathrm{U}/\nu_\mathrm{L}}\leq a \leq 1 $ & $0 \leq a \leq a_{\mathrm{ms}}^{\nu_\mathrm{U}/\nu_\mathrm{L}}$ & $a_{\mathrm{min}}^{\nu_\mathrm{U}/\nu_\mathrm{L}}\leq a \leq 1 $ & $a_{\mathrm{min}}^{\nu_\mathrm{U}/\nu_\mathrm{L}}\leq a \leq 1 $ & $0 \leq a \leq a_{\mathrm{ms}}^{\nu_\mathrm{U}/\nu_\mathrm{L}}$ & $a_{\mathrm{ms}}^{\nu_\mathrm{U}/\nu_\mathrm{L}}\leq a \leq 1 $ & $a_{\mathrm{ms}}^{\nu_\mathrm{U}/\nu_\mathrm{L}}\leq a \leq 1 $\\
  $3:1$ & $0.992634$ & $a_{\mathrm{ms}}^{\nu_\mathrm{U}/\nu_\mathrm{L}}\leq a \leq 1 $ & $0 \leq a \leq a_{\mathrm{ms}}^{\nu_\mathrm{U}/\nu_\mathrm{L}}$ & $a_{\mathrm{min}}^{\nu_\mathrm{U}/\nu_\mathrm{L}}\leq a \leq 1 $ & $a_{\mathrm{min}}^{\nu_\mathrm{U}/\nu_\mathrm{L}}\leq a \leq 1 $ & $0 \leq a \leq a_{\mathrm{ms}}^{\nu_\mathrm{U}/\nu_\mathrm{L}}$ & $a_{\mathrm{ms}}^{\nu_\mathrm{U}/\nu_\mathrm{L}}\leq a \leq 1 $ & $a_{\mathrm{ms}}^{\nu_\mathrm{U}/\nu_\mathrm{L}}\leq a \leq 1 $\\
  $4:1$ & $0.997216$ & $a_{\mathrm{ms}}^{\nu_\mathrm{U}/\nu_\mathrm{L}}\leq a \leq 1 $ & $0 \leq a \leq 1\phantom{_{\mathrm{ms}}^{\nu_\mathrm{U}/\nu_\mathrm{L}}}$ & $a_{\mathrm{min}}^{\nu_\mathrm{U}/\nu_\mathrm{L}}\leq a \leq 1 $ & $a_{\mathrm{min}}^{\nu_\mathrm{U}/\nu_\mathrm{L}}\leq a \leq 1 $ & $0 \leq a \leq a_{\mathrm{ms}}^{\nu_\mathrm{U}/\nu_\mathrm{L}}$ & $a_{\mathrm{ms}}^{\nu_\mathrm{U}/\nu_\mathrm{L}}\leq a \leq 1 $ & $a_{\mathrm{ms}}^{\nu_\mathrm{U}/\nu_\mathrm{L}}\leq a \leq 1 $ \\
  $5:1$ & $0.998659$ & $a_{\mathrm{ms}}^{\nu_\mathrm{U}/\nu_\mathrm{L}}\leq a \leq 1 $ & $0 \leq a \leq 1\phantom{_{\mathrm{ms}}^{\nu_\mathrm{U}/\nu_\mathrm{L}}}$ & $a_{\mathrm{min}}^{\nu_\mathrm{U}/\nu_\mathrm{L}}\leq a \leq 1 $ & $a_{\mathrm{min}}^{\nu_\mathrm{U}/\nu_\mathrm{L}}\leq a \leq 1 $ & $0 \leq a \leq 1\phantom{_{\mathrm{ms}}^{\nu_\mathrm{U}/\nu_\mathrm{L}}}$ & $a_{\mathrm{ms}}^{\nu_\mathrm{U}/\nu_\mathrm{L}}\leq a \leq 1 $ & $a_{\mathrm{ms}}^{\nu_\mathrm{U}/\nu_\mathrm{L}}\leq a \leq 1 $ \\
  \hline
\end{tabular}
\end{center}
\end{table*}

\vspace{0.4cm}

Generally, the~resonant radii corresponding to specific spin ``resonance'' functions are determined by the~related ``resonance equations'' representing fourth-order polynomials in terms of the~black hole spin. The~``resonance equations'' are identical in the~following twin cases: DC1 $\equiv$ DC6, DC2 $\equiv$ DC5, DC3 $\equiv$ DC7, DC4 $\equiv$ DC8, DC9 $\equiv$ DC10, DC11 $\equiv$ DC12. 
The~resonant radii are then determined by different branches of the~``resonance equations'' solutions, separated by the~line of the~marginally stable orbits $a_{\mathrm{ms}}(x)$. Only the~case of DC13 is an~exception, where the~resonant lines do not touch the~marginally stable line. Some of the~``resonance equations'' are related because of the~definition of the~frequency ratio ($p \rightarrow 1/p$), namely DC3~$\leftrightarrow$~DC11, DC5~$\leftrightarrow$~DC13, DC7~$\leftrightarrow$~DC12. Furthermore, there are additional interesting connections between the~resonance lines of the~classes DC1 (DC6) and DC9 (DC10) (see Figs.~\ref{aDC1} and \ref{aDC2-WD}) expressed by the~relations
\begin{eqnarray}
  a^{\left(\mathrm{K}-{r}\right)/\left(\theta-{r}\right)}(x,n\!:\!m) &\equiv& a^{\left(\mathrm{K}-{r}\right)/\left(\mathrm{K}-\theta\right)}(x,k\!:\!l)\,,\label{vztahDC1aDC9}\\
  a^{\left(\mathrm{K}+{r}\right)/\left(\theta+{r}\right)}(x,n\!:\!m) &\equiv& a^{\left(\mathrm{K}+{r}\right)/\left(\mathrm{K}-\theta\right)}(x,k\!:\!l)\,.\label{vztahDC6aDC10}
\end{eqnarray}
The~corresponding frequency ratios $n\!:\!m$ and $k\!:\!l$ are given in Table~\ref{TabVztahDC1aDC9}.

\begin{table}[t]
\caption{\label{TabVztahDC1aDC9}The~corresponding frequency ratios $n\!:\!m$ and $k\!:\!l$ for which the~resonance lines of the~types DC1 (DC6) and DC9 (DC10) are identical and satisfy the~relations (\ref{vztahDC1aDC9}) and (\ref{vztahDC6aDC10}).}
\begin{center}
\small
\begin{tabular}{cccccccccc}
\hline\hline
$n\!:\!m$ & $5\!:\!4$ & $4\!:\!3$ & $3\!:\!2$ & $5\!:\!3$ & $2\!:\!1$ & $5\!:\!2$ & $3\!:\!1$ & $4\!:\!1$ & $5\!:\!1$ \\
$k\!:\!l$ & $5\!:\!1$ & $4\!:\!1$ & $3\!:\!1$ & $5\!:\!2$ & $2\!:\!1$ & $5\!:\!3$ & $3\!:\!2$ & $4\!:\!3$ & $5\!:\!4$ \\
\hline
\end{tabular}
\end{center}
\end{table}

We can see from Figs.~\ref{aDirect}, \ref{aCT}\,--\,\ref{aDC2-WD} that some of the~considered orbital model versions do not work for the~whole range of the~Kerr black hole (neutron star) spin $a$. For example, for the~direct orbital resonance model D3 ($\nu_{\mathrm{K}}\!:\!\nu_{\theta}$) there are minimal values of spin $a_{\mathrm{ms}}^{\mathrm{K}/\theta}(n\!:\!m)$ for the~given frequency ratio $n\!:\!m$ that determines the~stability point for considered resonance functions (see Fig.~\ref{aDirect}). Such stability points $a_{\mathrm{ms}}^{\nu_\mathrm{U}/\nu_\mathrm{L}}(p)$ also represent limiting (minimum or maximum) values of spin for related combinational resonances of the~types CT1, CT5, DC1, DC2, DC5, and DC6. Another limit on spin allowed for resonance of a~given type and frequency ratio occurs in the~case of non-monotonic resonance functions $a^{\nu_\mathrm{U}/\nu_\mathrm{L}}(x,p)$ by their local minimum $a^{\nu_\mathrm{U}/\nu_\mathrm{L}}_{\mathrm{min}}(p)$ (see, e.g., Fig.~\ref{aDC1}). The~limiting values of the~spin, and related allowed spin ranges for given frequency ratio $n\!:\!m$, are presented for given types of resonance in Table~\ref{TabLimitniSpiny}.

When two twin-peak QPOs are observed with frequency ratios $n:m$ and $n':m'$, respectively, we have to find two versions of resonance that could explain both the~ratios and magnitudes of the~observed frequencies and, for a~given range of allowed mass in the~source, they must predict the~same black hole spin $a$, or more precisely, overlapping intervals of the~spin. Therefore, it is clear that, generally, two observed twin-peak QPOs could make the~spin estimates more precise. Two different resonances are necessary when two twin peaks are observed with the~same ratios but different magnitudes.

\clearpage

\onecolumn

\section{Detailed table guide across all the~possible triple frequency ratio sets and related values of the~black hole spin $a$}\label{apendix-D}


\begin{table}[h]
\caption{Notation system of double combinations of both the~direct (D1 -- D3) and some of the~simple triple combinational resonances (CT1 and CT2) with identical top (T), bottom (B), and both types of mixed (M) frequencies. The~``$+$'' symbol denotes the~resonance with the~top frequency and ``$-$'' denotes the~resonance with the~identical bottom frequency.\label{top-id-znaceni}}
\begin{minipage}{.48\linewidth}
\begin{center}

\end{center}

\clearpage
\tabcolsep=3pt
\begin{table}[t]
\caption{Notation system of double combinations of the~direct resonance D1 and all double combinational resonances (DC1 -- DC13) with identical top (T), bottom (B), and both types of mixed (M) frequencies. The~``$+$'' symbol denotes the~resonance with the~top frequency and ``$-$'' denotes the~resonance with the~identical bottom frequency.\label{top-id-D-DC-znaceni}}
\begin{minipage}{.24\linewidth}
\begin{center}
\begin{tabular}{cc}
  \hline
  \hline
 \centering {\rule[-4mm]{0mm}{10mm}
  $+\;+$} & \centering $\displaystyle\frac{\nu_{\theta}}{\nu_{r}}$  \tabularnewline
    \hline
  \centering {\rule[-4mm]{0mm}{10mm}
  $\displaystyle\frac{\nu_{\mathrm{K}}-\nu_{r}}{\nu_{\theta}-\nu_{r}}$}
   & T1(DC1) \tabularnewline
  \centering {\rule[-4mm]{0mm}{10mm}
  $\displaystyle\frac{\nu_{\mathrm{K}}+\nu_{r}}{\nu_{\theta}-\nu_{r}}$}
   & T1(DC2) \tabularnewline
  \centering {\rule[-4mm]{0mm}{10mm}
  $\displaystyle\frac{\nu_{\mathrm{K}}-\nu_{\theta}}{\nu_{\theta}-\nu_{r}}$}
   & T1(DC3) \tabularnewline
  \centering {\rule[-4mm]{0mm}{10mm}
  $\displaystyle\frac{\nu_{\mathrm{K}}+\nu_{\theta}}{\nu_{\theta}-\nu_{r}}$}
   & T1(DC4) \tabularnewline
  \centering {\rule[-4mm]{0mm}{10mm}
  $\displaystyle\frac{\nu_{\mathrm{K}}-\nu_{r}}{\nu_{\theta}+\nu_{r}}$}
   & T1(DC5) \tabularnewline
  \centering {\rule[-4mm]{0mm}{10mm}
  $\displaystyle\frac{\nu_{\mathrm{K}}+\nu_{r}}{\nu_{\theta}+\nu_{r}}$}
   & T1(DC6) \tabularnewline
  \centering {\rule[-4mm]{0mm}{10mm}
  $\displaystyle\frac{\nu_{\mathrm{K}}-\nu_{\theta}}{\nu_{\theta}+\nu_{r}}$}
   & T1(DC7) \tabularnewline
\hline
\end{tabular}
\end{center}
\end{minipage}
\hfill%
\begin{minipage}{.24\linewidth}
\begin{center}
\begin{tabular}{cc}
  \hline
  \hline
 \centering {\rule[-4mm]{0mm}{10mm}
  $-\;-$} & \centering $\displaystyle\frac{\nu_{\theta}}{\nu_{r}}$  \tabularnewline
    \hline
  \centering {\rule[-4mm]{0mm}{10mm}
  $\displaystyle\frac{\nu_{\mathrm{K}}-\nu_{r}}{\nu_{\theta}-\nu_{r}}$}
   & B1(DC1) \tabularnewline
  \centering {\rule[-4mm]{0mm}{10mm}
  $\displaystyle\frac{\nu_{\mathrm{K}}+\nu_{r}}{\nu_{\theta}-\nu_{r}}$}
   & B1(DC2) \tabularnewline
  \centering {\rule[-4mm]{0mm}{10mm}
  $\displaystyle\frac{\nu_{\mathrm{K}}-\nu_{\theta}}{\nu_{\theta}-\nu_{r}}$}
   & B1(DC3) \tabularnewline
  \centering {\rule[-4mm]{0mm}{10mm}
  $\displaystyle\frac{\nu_{\mathrm{K}}+\nu_{\theta}}{\nu_{\theta}-\nu_{r}}$}
   & B1(DC4) \tabularnewline
  \centering {\rule[-4mm]{0mm}{10mm}
  $\displaystyle\frac{\nu_{\mathrm{K}}-\nu_{r}}{\nu_{\theta}+\nu_{r}}$}
   & B1(DC5) \tabularnewline
  \centering {\rule[-4mm]{0mm}{10mm}
  $\displaystyle\frac{\nu_{\mathrm{K}}+\nu_{r}}{\nu_{\theta}+\nu_{r}}$}
   & B1(DC6) \tabularnewline
  \centering {\rule[-4mm]{0mm}{10mm}
  $\displaystyle\frac{\nu_{\mathrm{K}}-\nu_{\theta}}{\nu_{\theta}+\nu_{r}}$}
   & B1(DC7) \tabularnewline
\hline
  \end{tabular}
\end{center}
\end{minipage}
\hfill%
\begin{minipage}{.24\linewidth}
\begin{center}
\begin{tabular}{cc}
  \hline
  \hline
 \centering {\rule[-4mm]{0mm}{10mm}
  $+\;-$} & \centering $\displaystyle\frac{\nu_{\theta}}{\nu_{r}}$  \tabularnewline
    \hline
  \centering {\rule[-4mm]{0mm}{10mm}
  $\displaystyle\frac{\nu_{\mathrm{K}}-\nu_{r}}{\nu_{\theta}-\nu_{r}}$}
   & M(DC1)1 \tabularnewline
  \centering {\rule[-4mm]{0mm}{10mm}
  $\displaystyle\frac{\nu_{\mathrm{K}}+\nu_{r}}{\nu_{\theta}-\nu_{r}}$}
   & M(DC2)1 \tabularnewline
  \centering {\rule[-4mm]{0mm}{10mm}
  $\displaystyle\frac{\nu_{\mathrm{K}}-\nu_{\theta}}{\nu_{\theta}-\nu_{r}}$}
   & M(DC3)1 \tabularnewline
  \centering {\rule[-4mm]{0mm}{10mm}
  $\displaystyle\frac{\nu_{\mathrm{K}}+\nu_{\theta}}{\nu_{\theta}-\nu_{r}}$}
   & M(DC4)1 \tabularnewline
  \centering {\rule[-4mm]{0mm}{10mm}
  $\displaystyle\frac{\nu_{\mathrm{K}}-\nu_{r}}{\nu_{\theta}+\nu_{r}}$}
   & M(DC5)1 \tabularnewline
  \centering {\rule[-4mm]{0mm}{10mm}
  $\displaystyle\frac{\nu_{\mathrm{K}}+\nu_{r}}{\nu_{\theta}+\nu_{r}}$}
   & M(DC6)1 \tabularnewline
  \centering {\rule[-4mm]{0mm}{10mm}
  $\displaystyle\frac{\nu_{\mathrm{K}}-\nu_{\theta}}{\nu_{\theta}+\nu_{r}}$}
   & M(DC7)1 \tabularnewline
\hline
  \end{tabular}
\end{center}
\end{minipage}
\hfill%
\begin{minipage}{.24\linewidth}
\begin{center}
\begin{tabular}{cc}
  \hline
  \hline
 \centering {\rule[-4mm]{0mm}{10mm}
  $-\;+$} & \centering $\displaystyle\frac{\nu_{\theta}}{\nu_{r}}$  \tabularnewline
    \hline
  \centering {\rule[-4mm]{0mm}{10mm}
  $\displaystyle\frac{\nu_{\mathrm{K}}-\nu_{r}}{\nu_{\theta}-\nu_{r}}$}
   & M1(DC1) \tabularnewline
  \centering {\rule[-4mm]{0mm}{10mm}
  $\displaystyle\frac{\nu_{\mathrm{K}}+\nu_{r}}{\nu_{\theta}-\nu_{r}}$}
   & M1(DC2) \tabularnewline
  \centering {\rule[-4mm]{0mm}{10mm}
  $\displaystyle\frac{\nu_{\mathrm{K}}-\nu_{\theta}}{\nu_{\theta}-\nu_{r}}$}
   & M1(DC3) \tabularnewline
  \centering {\rule[-4mm]{0mm}{10mm}
  $\displaystyle\frac{\nu_{\mathrm{K}}+\nu_{\theta}}{\nu_{\theta}-\nu_{r}}$}
   & M1(DC4) \tabularnewline
  \centering {\rule[-4mm]{0mm}{10mm}
  $\displaystyle\frac{\nu_{\mathrm{K}}-\nu_{r}}{\nu_{\theta}+\nu_{r}}$}
   & M1(DC5) \tabularnewline
  \centering {\rule[-4mm]{0mm}{10mm}
  $\displaystyle\frac{\nu_{\mathrm{K}}+\nu_{r}}{\nu_{\theta}+\nu_{r}}$}
   & M1(DC6) \tabularnewline
  \centering {\rule[-4mm]{0mm}{10mm}
  $\displaystyle\frac{\nu_{\mathrm{K}}-\nu_{\theta}}{\nu_{\theta}+\nu_{r}}$}
   & M1(DC7) \tabularnewline
\hline
  \end{tabular}
\end{center}
\end{minipage}
\end{table}

\clearpage

\begin{center}
\begin{longtable}{ccccc}
\caption{Top identity -- characteristic triple frequency ratio sets and related values of the~black hole spin.\label{top-id-D-DC}}\\
\endfirsthead
\multicolumn{5}{c}%
  {\raisebox{1ex}[0sp]{\small\textbf{\tablename\ D.6.} \textit{(continued)}}}\\
\endhead
\endfoot
\endlastfoot
    \hline
    \hline
  $\nu_{\mathrm{U}}$ & $\nu_{\mathrm{M}}$ & $\nu_{\mathrm{L}}$ & T1(DC1) sets & $a$\\
    \cline{1-5}
  $\left(\nu_{\mathrm{K}}-\nu_{r}\right)^{4:3}=\nu_{\theta}^{4:3}$\rule{0sp}{2.5ex} &
  -- & $\nu_{r}^{4:3}=\left(\nu_{\theta}-\nu_{r}\right)^{4:3}$
   & $\textbf{{\phantom{1}}4}:\textbf{3}$ & $0.807331${\phantom{*}}\\
  $\left(\nu_{\mathrm{K}}-\nu_{r}\right)^{4:3}=\nu_{\theta}^{3:2}$ &
  $\left(\nu_{\theta}-\nu_{r}\right)^{4:3}$ & $\nu_{r}^{3:2}$ & $\textbf{12}:9:8$ & $0.783086${\phantom{*}}\\
  $\left(\nu_{\mathrm{K}}-\nu_{r}\right)^{4:3}=\nu_{\theta}^{2:1}$ &
  $\left(\nu_{\theta}-\nu_{r}\right)^{4:3}$ & $\nu_{r}^{2:1}$ & $\textbf{{\phantom{1}}4}:3:2$ & $0.770370${\phantom{*}}\\
  $\left(\nu_{\mathrm{K}}-\nu_{r}\right)^{4:3}=\nu_{\theta}^{3:1}$ &
  $\left(\nu_{\theta}-\nu_{r}\right)^{4:3}$ & $\nu_{r}^{3:1}$ & $\textbf{12}:9:4$ & $0.773163${\phantom{*}}\\
  $\left(\nu_{\mathrm{K}}-\nu_{r}\right)^{4:3}=\nu_{\theta}^{4:1}$ &
  $\left(\nu_{\theta}-\nu_{r}\right)^{4:3}$ & $\nu_{r}^{4:1}$ & $\textbf{{\phantom{1}}4}:3:1$ & $0.775420$* 
  \\
  $\left(\nu_{\mathrm{K}}-\nu_{r}\right)^{3:2}=\nu_{\theta}^{4:3}$ &
  $\nu_{r}^{4:3}$ & $\left(\nu_{\theta}-\nu_{r}\right)^{3:2}$ & $\textbf{12}:9:8$ & $0.925660${\phantom{*}}\\
  $\left(\nu_{\mathrm{K}}-\nu_{r}\right)^{3:2}=\nu_{\theta}^{3:2}$ &
  -- & $\nu_{r}^{3:2}=\left(\nu_{\theta}-\nu_{r}\right)^{3:2}$ & $\textbf{{\phantom{1}}3}:\textbf{2}$ & $0.892290${\phantom{*}}\\
\hline
\end{longtable}
\end{center}

\clearpage

\begin{center}
\begin{longtable}{ccccc}
\caption{Bottom identity -- characteristic triple frequency ratio sets and related values of the~black hole spin.\label{bot-id-D-DC}}\\
\endfirsthead
\multicolumn{5}{c}%
  {\raisebox{1ex}[0sp]{\small\textbf{\tablename\ D.7.} \textit{(continued)}}}\\
\endhead
\endfoot
\endlastfoot
    \hline
    \hline
  $\nu_{\mathrm{U}}$ & $\nu_{\mathrm{M}}$ & $\nu_{\mathrm{L}}$ & B1(DC1) sets & $a$\\
    \cline{1-5}
  $\left(\nu_{\mathrm{K}}-\nu_{r}\right)^{4:3}=\nu_{\theta}^{4:3}$\rule{0sp}{2.5ex} &
  -- &
  $\nu_{r}^{4:3}=\left(\nu_{\theta}-\nu_{r}\right)^{4:3}$ &
  $\textbf{{\phantom{1}}4}:\textbf{3}$ &
  $0.807331${\phantom{*}}\\
  $\nu_{\theta}^{4:1}$ &
  $\left(\nu_{\mathrm{K}}-\nu_{r}\right)^{4:3}$ &
  $\nu_{r}^{4:1}=\left(\nu_{\theta}-\nu_{r}\right)^{4:3}$ &
  $12:4:\textbf{3}$ & $0.805935${\phantom{*}}\\
  $\nu_{\theta}^{3:1}$ &
  $\left(\nu_{\mathrm{K}}-\nu_{r}\right)^{4:3}$ &
  $\nu_{r}^{3:1}=\left(\nu_{\theta}-\nu_{r}\right)^{4:3}$ &
  ${\phantom{1}}9:4:\textbf{3}$ & $0.790993${\phantom{*}}\\
  $\nu_{\theta}^{3:2}$ &
  $\left(\nu_{\mathrm{K}}-\nu_{r}\right)^{4:3}$ &
  $\nu_{r}^{3:2}=\left(\nu_{\theta}-\nu_{r}\right)^{4:3}$ &
  ${\phantom{1}}9:8:\textbf{6}$ & $0.789110${\phantom{*}}\\
  $\nu_{\theta}^{2:1}$ &
  $\left(\nu_{\mathrm{K}}-\nu_{r}\right)^{4:3}$ &
  $\nu_{r}^{2:1}=\left(\nu_{\theta}-\nu_{r}\right)^{4:3}$ &
  ${\phantom{1}}6:4:\textbf{3}$ & $0.780930$* 
  \\
  $\nu_{\theta}^{4:1}$ &
  $\left(\nu_{\mathrm{K}}-\nu_{r}\right)^{3:2}$ &
  $\nu_{r}^{4:1}=\left(\nu_{\theta}-\nu_{r}\right)^{3:2}$ &
  ${\phantom{1}}8:3:\textbf{2}$ & $0.916623${\phantom{*}}\\
  $\left(\nu_{\mathrm{K}}-\nu_{r}\right)^{3:2}$ &
  $\nu_{\theta}^{4:3}$ &
  $\nu_{r}^{4:3}=\left(\nu_{\theta}-\nu_{r}\right)^{3:2}$ &
  ${\phantom{1}}9:8:\textbf{6}$ & $0.915657${\phantom{*}}\\
\hline
\end{longtable}
\end{center}

\clearpage

\begin{center}
\begin{longtable}{ccccc}
\caption{Mixed identity -- characteristic triple frequency ratio sets and related values of the~black hole spin.\label{mid-id-D-DC}}\\
\endfirsthead
\multicolumn{5}{c}%
  {\raisebox{1ex}[0sp]{\small\textbf{\tablename\ D.8.} \textit{(continued)}}}\\
\endhead
\endfoot
\endlastfoot
    \hline
    \hline
  $\nu_{\mathrm{U}}$ & $\nu_{\mathrm{M}}$ & $\nu_{\mathrm{L}}$ & M(DC1)1 sets & $a$\\
  \cline{1-5}
  $\nu_{\theta}^{4:3}$\rule{0sp}{2.5ex} &
  $\nu_{r}^{4:3}=\left(\nu_{\mathrm{K}}-\nu_{r}\right)^{4:3}$ &
  $\left(\nu_{\theta}-\nu_{r}\right)^{4:3}$ &
  $16:\textbf{12}:9$ &
  $0.829397${\phantom{*}}\\
  $\nu_{\theta}^{4:1}$ &
  $\nu_{r}^{4:1}=\left(\nu_{\mathrm{K}}-\nu_{r}\right)^{4:3}$ &
  $\left(\nu_{\theta}-\nu_{r}\right)^{4:3}$ &
  $16:\textbf{4}:3$ &
  $0.828052${\phantom{*}}\\
  $\nu_{\theta}^{3:1}$ &
  $\nu_{r}^{3:1}=\left(\nu_{\mathrm{K}}-\nu_{r}\right)^{4:3}$ &
  $\left(\nu_{\theta}-\nu_{r}\right)^{4:3}$ &
  $12:\textbf{4}:3$ &
  $0.809741${\phantom{*}}\\
  $\nu_{\theta}^{3:2}$ &
  $\nu_{r}^{3:2}=\left(\nu_{\mathrm{K}}-\nu_{r}\right)^{4:3}$ &
  $\left(\nu_{\theta}-\nu_{r}\right)^{4:3}$ &
  ${\phantom{1}}6:\textbf{4}:3$ &
  $0.807099${\phantom{*}}\\
  $\nu_{\theta}^{2:1}$ &
  $\nu_{r}^{2:1}=\left(\nu_{\mathrm{K}}-\nu_{r}\right)^{4:3}$ &
  $\left(\nu_{\theta}-\nu_{r}\right)^{4:3}$ &
  ${\phantom{1}}8:\textbf{4}:3$ &
  $0.796257${\phantom{*}}\\\cline{5-5}
& & & & $0.998895${\phantom{*}}\\
  \raisebox{1.6ex}[0pt]
     {$\nu_{\theta}^{4:1}$} &
      \raisebox{1.6ex}[0pt]
      {$\nu_{r}^{4:1}=\left(\nu_{\mathrm{K}}-\nu_{r}\right)^{3:2}$} &
      \raisebox{1.6ex}[0pt]
      {$\left(\nu_{\theta}-\nu_{r}\right)^{3:2}$} &
      \raisebox{1.6ex}[0pt]
  {$12:\textbf{3}:2 \rlap{\,\,\,\,\,\,\,\Bigg\{}$} &
  $0.956922${\phantom{*}}\\
\hline
\end{longtable}
\end{center}

\clearpage

\section{Detailed table guide for the~triple frequency 3:2:1}\label{apendix-E}

\renewcommand{\arraystretch}{1.4}
\tabcolsep=6pt
\begin{table*}[h]
\caption{\label{3/2/1}The~relevant versions of the~multi-resonant model with assumed observed characteristic frequency ratio set $\nu_{\mathrm{U}}:\nu_{\mathrm{M}}:\nu_{\mathrm{L}}=$
3\,:\,2\,:\,1. The~radius of marginally stable orbit $x_\mathrm{ms}$ and corresponding resonant radii $x_\mathrm{1}$ and $x_\mathrm{2}$ are given. An~asterisk denotes the~special values of the~black hole spin when the~resonance points share the~same radius ($x_\mathrm{1} = x_\mathrm{2}\equiv x_\mathrm{3:2:1}$).}
\begin{center}
\begin{tabular}{llllllll}
 \hline \hline
    Set & {\rule[-2mm]{0mm}{6mm}
$\nu_{\mathrm{U}}$} & $\nu_{\mathrm{M}}$ & $\nu_{\mathrm{L}}$ & $a$ & $x_{\mathrm{ms}}$ & $x_1$ & $x_2$ \\
 \hline
``Magic''  & $\nu_{\mathrm{K}}$ & $\nu_{\theta}^{3:2}$ &
$\nu_{r}^{3:1}$ & $0.983043$* &
    $1.571$ & \multicolumn{2}{c}{$2.395$} \\
T45 & $\nu_{\mathrm{K}}^{3:1}=\nu_{\theta}^{3:2}$ &
$\left(\nu_{\mathrm{K}}-\nu_{r}\right)^{3:2}$ &
$\left(\nu_{\theta}-\nu_{r}\right)^{3:1}$
 & $0.885010$ &
    $2.419$ & $3.720$ & $4.299$ \\
B15 & $\nu_{\theta}^{3:1}$ & $\nu_{\theta}^{2:1}$ &
$\nu_{r}^{3:1}=\left(\nu_{\mathrm{K}}-\nu_{r}\right)^{2:1}$
 & $0.616894$ & $3.758$ & $4.241$ & $5.833$ \\
B15 & $\nu_{\theta}^{3:1}$ & $\nu_{\theta}^{2:1}$ &
$\nu_{r}^{3:1}=\left(\nu_{\mathrm{K}}-\nu_{r}\right)^{2:1}$
 & $0.999667$ & $1.121$ & $1.411$ & $4.250$ \\
 B22 & $\nu_{\mathrm{K}}^{3:1}$ & $\nu_{\mathrm{K}}^{2:1}$ & $\nu_{r}$ &
 $0.913806$ &
    $2.225$ & $2.885$ & $3.935$ \\
 B24 & $\nu_{\mathrm{K}}^{3:1}$ & $\nu_{\mathrm{K}}^{2:1}$ & $\nu_{r}^{2:1}=\left(\nu_{\theta}-\nu_{r}\right)^{3:1}$ &
 $0.980124$ &
    $1.612$ & $2.551$ & $3.519$ \\
 B25 & $\nu_{\mathrm{K}}^{3:1}$ & $\nu_{\theta}^{2:1}$ & $\nu_{r}^{3:1}=\left(\nu_{\mathrm{K}}-\nu_{r}\right)^{2:1}$ &
 $0.475159$ &
    $4.330$ & $4.988$ & $6.359$ \\
 B45 & $\nu_{\mathrm{K}}^{3:1}$ & $\nu_{\theta}^{2:1}$ & $\left(\nu_{\theta}-\nu_{r}\right)^{3:1}=\left(\nu_{\mathrm{K}}-\nu_{r}\right)^{2:1}$ &
 $0.922985$  &
    $2.158$ & $3.794$ & $4.594$ \\
 M14 & $\nu_{\mathrm{K}}^{3:2}$ & $\nu_{\theta}^{2:1}=\left(\nu_{\theta}-\nu_{r}\right)^{3:2}$ & $\nu_{r}^{2:1}$ &
 $0.544870$ &
    $4.055$ & $4.347$ & $5.477$ \\
 M24 & $\nu_{\mathrm{K}}^{3:2}$ & $\nu_{\mathrm{K}}^{2:1}=\left(\nu_{\theta}-\nu_{r}\right)^{3:2}$ & $\nu_{r}^{2:1}$ &
 $0.535413$  &
    $4.093$ & $4.394$ & $5.832$ \\
 M54 & $\nu_{\mathrm{K}}^{3:2}$ & $\nu_{\theta}^{2:1}=\left(\nu_{\theta}-\nu_{r}\right)^{3:2}$ & $\left(\nu_{\mathrm{K}}-\nu_{r}\right)^{2:1}$ &
 $0.336030$ &
    $4.849$ & $5.327$ & $6.857$ \\
T1(DC1) & $\left(\nu_{\mathrm{K}}-\nu_{r}\right)^{3:2}=\nu_{\theta}^{3:1}$ &
   $\left(\nu_{\theta}-\nu_{r}\right)^{3:2}$ & $\nu_{r}^{3:1}$ & & & \multicolumn{2}{c}{} \\
 B1(DC11)
 & $\nu_{\theta}^{3:1}$ &
  $\left(\nu_{\theta}-\nu_{r}\right)^{2:1}$ &
  $\nu_{r}^{3:1}=\left(\nu_{\mathrm{K}}-\nu_{\theta}\right)^{2:1}$ &
 \raisebox{1.6ex}[0pt]
 {$\llap{{\Bigg\}\,\,}} 0.865670$*} &
 \raisebox{1.6ex}[0pt]
 {$2.539$} &
 \multicolumn{2}{c}
 {\raisebox{1.6ex}[0pt]
 {$2.880$}} \\
 T1(DC3)
 & $\left(\nu_{\mathrm{K}}-\nu_{\theta}\right)^{3:2}=\nu_{\theta}^{3:1}$ &
  $\left(\nu_{\theta}-\nu_{r}\right)^{3:2}$ &
  $\nu_{r}^{3:1}$ &
 $0.986666$* &
 $1.514$ & \multicolumn{2}{c}{$1.753$} \\
 T1(DC9)
 &  $\left(\nu_{\mathrm{K}}-\nu_{r}\right)^{3:1}=\nu_{\theta}^{3:2}$ &
  $\nu_{r}^{3:2}$ &
  $\left(\nu_{\mathrm{K}}-\nu_{\theta}\right)^{3:1}$ & & & & \\
 M(DC11)
 & $\nu_{\theta}^{3:2}$ &
  $\nu_{r}^{3:2}=\left(\nu_{\theta}-\nu_{r}\right)^{2:1}$ &
  $\left(\nu_{\mathrm{K}}-\nu_{\theta}\right)^{2:1}$ &
 \raisebox{1.6ex}[0pt]
 {$\llap{{\Bigg\}\,\,}} 0.892290$} &
 \raisebox{1.6ex}[0pt]
 {$2.372$} &
 \raisebox{1.6ex}[0pt]
 {$3.601$} &
 \raisebox{1.6ex}[0pt]
 {$5.034$} \\
 T1(DC11)
 & $\left(\nu_{\theta}-\nu_{r}\right)^{3:1}=\nu_{\theta}^{3:2}$ &
  $\nu_{r}^{3:2}$ &
  $\left(\nu_{\mathrm{K}}-\nu_{\theta}\right)^{3:1}$ &
 $0.772687$ &
 $3.046$ & $3.792$ & $6.036$ \\
 T1(DC11)
 & $\left(\nu_{\theta}-\nu_{r}\right)^{3:2}=\nu_{\theta}^{3:1}$ &
  $\left(\nu_{\mathrm{K}}-\nu_{\theta}\right)^{3:2}$ &
  $\nu_{r}^{3:1}$ &
 $0.927324$ &
 $2.125$ & $2.131$ & $2.419$ \\
 B1(DC9)
 & $\left(\nu_{\mathrm{K}}-\nu_{r}\right)^{3:1}$ &
  $\nu_{\theta}^{2:1}$ &
  $\nu_{r}^{2:1}=\left(\nu_{\mathrm{K}}-\nu_{\theta}\right)^{3:1}$ & & & & \\
 M1(DC1)
 & $\left(\nu_{\mathrm{K}}-\nu_{r}\right)^{3:2}$ &
  $\nu_{\theta}^{2:1}=\left(\nu_{\theta}-\nu_{r}\right)^{3:2}$ &
  $\nu_{r}^{2:1}$ &
  \raisebox{1.6ex}[0pt]
 {$\llap{{\Bigg\}\,\,}} 0.868917$} &
 \raisebox{1.6ex}[0pt]
 {$2.520$} &
 \raisebox{1.6ex}[0pt]
 {$2.648$} &
 \raisebox{1.6ex}[0pt]
 {$3.510$} \\
 B1(DC12)
 & $\left(\nu_{\theta}+\nu_{r}\right)^{3:1}$ &
  $\nu_{\theta}^{2:1}$ &
  $\nu_{r}^{2:1}=\left(\nu_{\mathrm{K}}-\nu_{\theta}\right)^{3:1}$ &
 $0.851581$ &
 $2.623$ & $2.675$ & $3.640$ \\
 M1(DC7)
 & $\left(\nu_{\mathrm{K}}-\nu_{\theta}\right)^{3:2}$ &
  $\nu_{\theta}^{2:1}=\left(\nu_{\theta}+\nu_{r}\right)^{3:2}$ &
  $\nu_{r}^{2:1}$ &
 $0.987594$ &
 $1.498$ & $1.502$ & $2.326$ \\
 \hline
\end{tabular}
\end{center}
\end{table*}

\end{document}